\documentclass[preprint,3p,authoryear]{elsarticle}

\journal{arxiv}
\usepackage{natbib}

\usepackage[utf8]{inputenc}
\usepackage{bm}
\usepackage{amsmath,comment}
\usepackage[hidelinks]{hyperref}
\hypersetup{colorlinks=true,linkcolor=blue,urlcolor=blue}
\usepackage{multicol}
\usepackage{booktabs}
\usepackage{amsthm}
\newtheorem{remark}{Remark}
\usepackage[table,xcdraw,dvipsnames]{xcolor}
\usepackage{graphicx}
\usepackage{float}
\usepackage{subcaption}
\usepackage{algorithm}
\usepackage{algpseudocode}
\usepackage{pdflscape}
\usepackage{array}
\usepackage{multirow}
\usepackage{diagbox}
\usepackage{cleveref}
\usepackage{amsfonts}
\usepackage{amssymb}
\usepackage{ulem}
\usepackage{fourier-orns}
\usepackage[labelfont=bf,labelsep=period]{caption}
\usepackage[labelfont=rm,font=footnotesize]{subcaption}

\numberwithin{equation}{section}
\usepackage{subcaption}
\usepackage{chngcntr}
\usepackage[makeroom]{cancel}
\counterwithout{equation}{section}

\DeclareMathAlphabet{\altmathcal}{OMS}{cmsy}{m}{n}
\DeclareMathAlphabet{\altmathcalb}{OMS}{cmsy}{b}{n}
\DeclareMathAlphabet{\mathcalboondox}{U}{BOONDOX-calo}{m}{n}
\DeclareMathAlphabet{\mathbbmsl}{U}{bbm}{m}{sl}

\newcommand{\Pro}{P_\mathbf{X}(\mathbf{x},t)}
\newcommand{\bx}{\mathbf{x}}
\newcommand{\orcid}[1]{\href{https://orcid.org/#1}{\texorpdfstring{\includegraphics*[width=8pt]{figs/orcid}}~~}{}}

\definecolor{drot}{rgb}{0.7,0,0.1}

\newcommand{\rev}[1]{{\color{black}#1}}
\newcommand{\revv}[1]{\textcolor{black}{{#1}}}

\begin{document}
\baselineskip14pt
\sloppy

\begin{frontmatter}

%\title{SPLIT-PINN: Separable Probability Learning Technique via Physics-Informed Neural Networks in High-Dimensional Probabilistic Dynamics}
\title{SPLIT-PINN: Separable Probability Learning Technique via Physics-Informed Neural Networks for High-Dimensional Probabilistic Modeling}

\author[UW-Math]{Pouria Behnoudfar}
\author[UW-Mech]{Deekshith Naidu Ponnana}
\author[JHU]{Noah J. Schmelzer}
\author[LANL]{Janith Wanni}
\author[LANL]{George T. Gray III}
\author[UW-MSE]{Dan J. Thoma}
\author[UW-Mech]{Curt A. Bronkhorst}
\author[UW-Math]{Nan Chen}
\author[UW-Mech]{Wenxiao Pan\corref{cor}}
\ead{wpan9@wisc.edu}
\cortext[cor]{Corresponding author}
\address[UW-Mech]{{Department of Mechanical Engineering, University of Wisconsin-Madison, Madison, WI, USA}}
\address[UW-Math]{{Department of Mathematics, University of Wisconsin-Madison, Madison, WI, USA}}
\address[JHU]{{Department of Civil Engineering, Johns Hopkins University, Baltimore, MD, USA}}
\address[LANL]{{Materials Physics and Applications Division, Los Alamos National Laboratory, Los Alamos, NM, USA}}
\address[UW-MSE]{Department of Materials Science and Engineering, University of Wisconsin-Madison, Madison, WI, USA}


\begin{abstract}
We present a probabilistic modeling framework for incorporating small-scale spatial heterogeneity into macroscopic descriptions of material behavior for polycrystalline metallic materials. Spatially heterogeneous material state fields are represented using probability density functions (PDFs), providing a principled statistical description of microstructural variability and state evolution across different computational polycrystalline realizations. The framework is built on the inverse identification of a probabilistic transport model, formulated as a Liouville equation with an unknown drift term. To enable accurate, stable, and interpretable inference of this drift field in high-dimensional, transport-dominated settings, we develop a Separable Probability Learning Technique via Physics-Informed Neural Networks (SPLIT-PINN). This method incorporates a marginal-correction drift decomposition, orthogonality constraints, and residual-based adaptive training to enhance well-posedness, numerical stability, and physical consistency without imposing restrictive parametric assumptions. Using SPLIT-PINN, the drift field governing the temporal evolution of joint state PDFs is inferred directly from data. After benchmark validation, the framework is applied to physical computational datasets describing the evolution of polycrystalline microstructural states, including von Mises stress, dislocation density, and equivalent plastic strain rate. The learned Liouville model, trained on a single dataset, is subsequently used in forward predictions of the temporal evolution of joint and marginal PDFs for multiple unseen polycrystal realizations. Quantitative comparisons with reference PDFs demonstrate that the proposed framework yields accurate and robust probabilistic predictions and generalizes effectively across datasets.

\end{abstract}

\end{frontmatter}
\section{Introduction}

Polycrystalline metallic materials are generally aggregate composites composed of single crystals (grains) with a chemical composition of the pure or alloy material. Single crystals are atomistically organized into a specific crystallographic structure, which then dictates their response to mechanical loading. The crystallographic structure and atom type prescribe elastic moduli and locations that are best suited to facilitate the motion of dislocations. Single crystals are then spatially heterogeneous in their response to mechanical loading; the degree to which is very much dependent upon material type. Some materials and alloys are designed to include specific atomic or intermetallic features which provide physical barriers to dislocation motion; adding additional spatial heterogeneity. The spatial heterogeneity of atomic and structural features leads directly to the significant spatially heterogeneous mechanical response of polycrystalline metals to mechanical loading (\cite{BRONKHORST2021102903,SCHMELZER2025104318,DUNHAM2025104258}). During conditions of extreme mechanical loading, the spatial heterogeneity of deformation and stress can be very large and the primary contributor to the initiation of damage and failure processes (\cite{BRONKHORST2021102903,SCHMELZER2025104318}). 

The spatial heterogeneity of polycrystalline metallic materials response to extreme mechanical loading generally involves physical processes (i.e. dislocation glide, intergranular interaction, deformation twinning, structural phase transformation, pore/crack nucleation) occurring at very small length scales (e.g., atom to grain size) in relation to the size of most structural engineering components. Theories and computational tools which represent these small length scale physical processes continue to be developed in concert with our improved understanding (e.g., \cite{BertinNPJ,Bertin_2019,BERTIN2024119884,Arsenlis_2007,DANG2022111786,MADEC2017166,GROGER20085401,GROGER20085412,MadecSci03,DEVINCRE2006741}). These appropriately high fidelity tools at small length scales for metallic materials remain computationally expensive and therefore cannot be fully engaged within large length scale theories of the same physical processes. This is the case even though such high fidelity is necessary to properly describe extreme events such as damage in strongly history dependent materials. In this case, the spatial heterogeneity of material state evolves with deformation (time) and in a way which leads to material damage events (\cite{ZHANG2023105386}). It is important to include such spatially dependent structural evolution even in our high length scale theories of material behavior.

The unresolved small-scale heterogeneities within larger length scale theories have motivated the adoption of probabilistic modeling frameworks in large length scale descriptions of material behaviors. In particular, one path is to introduce stochastic forcing to account for unresolved small-scale variability when the evolution of state variables cannot be fully captured through deterministic modeling. This approach naturally results in the Fokker–Planck equation, which governs the evolution of probability density functions (PDFs) under drift and diffusion terms~(\cite{friedrich1997description,gardiner2004handbook, risken1989fokker}). In this formulation, the drift term represents the mean deterministic behavior, while the diffusion term accounts for fluctuations arising from excitations, collective activities, and other unresolved mechanisms~(\cite{chavanis2003generalized,bontempi1994chaotic}). Therefore, the added noise is often a compensation for the missed dynamics and not a physical description of the fluctuations. As a result, this framework has become a standard tool for representing material-state uncertainties, especially when only partial state information is available~(\cite{kalidindi2015hierarchical,zaiser2006scale}). However, this reliance on added noise can obscure \revv{the true deterministic structure of the system, particularly in materials where the underlying dynamics are strongly driven by internal (e.g., intergranular, dislocation structure) interactions rather than external randomness (e.g., excitation or forcing).}

As the dimensionality of the material state variable space increases, \revv{the effect of the unresolved variables diminishes, and hence the effective dynamics become progressively more deterministic, thereby reducing the need for ad hoc diffusion terms. This viewpoint is consistent with the interpretation of stochasticity in many physical systems as an artifact of dimensional reduction rather than true randomness~(\cite{pavliotis2008multiscale,choromanska2015loss}).} 
Thus, a probabilistic model formulated as a Liouville partial differential equation (PDE) with only a drift term may be sufficient to describe the evolution of system states, particularly their PDFs. In this work, state PDFs are employed to describe the spatially heterogeneous material state fields in polycrystalline metallic materials. The probabilistic model, in the form of the Liouville equation, is inferred inversely to 
predict the temporal evolution of these state PDFs across different polycrystal statistical volume elements. Our proposed framework solves the resulting inverse problem to approximate the drift terms that best explain the observed evolution of the PDF governed by the Liouville equation. 

There have been approaches developed to solve inverse problems, ranging from classical optimization to variational techniques and data assimilation methods~(\cite{kalnay2003atmospheric,tarantola2005inverse,talagrand1987variational}). More recently, physics-informed neural networks (PINNs) have emerged as a powerful tool for combining data with governing physical laws to model complex systems (\cite{raissi2019physics,karniadakis2021physics,Zhao2024comprehensive,Shukla2022Scalable,Cuomo2022Scientific,poulet2023physics, hasan2025pc}). While PINNs have shown remarkable success in forward simulations, their application to inverse problems, especially in high-dimensional settings, remains challenging. Inverse estimation of drift terms governing the evolution of joint PDFs is particularly ill-conditioned, and directly applying PINNs often suffers from instability or poor generalization~(\cite{Zhao2024comprehensive}).

In this work, we propose \textbf{SPLIT-PINN} (Separable Probability Learning via Physics-Informed Neural Networks), a systematic framework for reconstructing high-dimensional drift fields directly from PDF data. The approach leverages a marginal-correction drift decomposition, adaptive residual-based training, and additional orthogonality constraints to improve the well-posedness and robustness of the inverse solution. After reconstructing the drift term, the Liouville equation can be used in forward simulations to predict the temporal evolution of state PDFs (joint and marginal) across different polycrystal statistical volume elements. Moreover, we equip SPLIT-PINN with artificial viscosity~(\cite{coutinho2023physics}) and residual-based adaptive distribution~(\cite{wu2023comprehensive}) to allow capture of sharp fronts commonly observed in drift terms.  

\rev{It is worth noting that the Liouville equation is primarily suited to systems where randomness originates from initial conditions or system parameters, precisely the regime of the present framework. In our case, the randomness arises from spatial heterogeneity in the polycrystalline microstructure, which can be naturally captured by variability in the initial  conditions of state variables. Like other PDF-governing PDEs, traditional mesh-based approaches for joint PDFs remain susceptible to the curse of dimensionality as their dimension grows. The proposed SPLIT-PINN framework mitigates this challenge through two complementary mechanisms: the mesh-free nature of PINNs avoids exponential grid scaling, while the marginal-correction decomposition reduces the learning complexity by firstly handling a set of 1D marginal problems before correcting for joint dependencies via a residual drift term.}

\rev{
A related line of efforts in probabilistic modeling include the  globally-evolving-based generalized density evolution equation (GE-GDEE)~(\cite{chen2022globally}), its extension to double-randomness in velocity or drift terms (GV-GDEE), and the more general dimension-reduced probability density evolution equation (DR-PDEE)~(\cite{chen2024stochastic}). These approaches address the challenge of high-dimensional probability evolution by leveraging known governing equations to derive reduced-order PDEs for the PDF of selected response quantities, typically low-dimensional or scalar. More recently, PINNs were further incorporated to solve DR-PDEE formulations and, in some cases, to identify unknown coefficients from simulation data (e.g., DR-PDE-Net~(\cite{hao2024multi}) and MO-MPINN~(\cite{hao2025dr})). 
While SPLIT-PINN shares the overarching goal of making the prediction of high-dimensional PDF evolution tractable, it differs in several key aspects. First, no governing equations of the underlying physical system are assumed; instead, the drift field in the Liouville equation is entirely unknown and inferred directly from data, resulting in an inverse problem rather than a forward modeling approach. Second, the decomposition of the joint PDF into marginal components is introduced as a computational strategy to improve well-posedness and numerical stability of the learning problem, rather than as a physics-based dimensionality reduction derived from system dynamics. Third, the primary objective is the reconstruction and evolution of the joint PDF over multiple state variables, whereas DR-PDEE-based methods focus on marginal PDFs of selected scalar quantities. }

The main contributions of this paper are summarized as follows:
\begin{itemize}
%\item We consider a deterministic approach to predict the evolution of the PDF with focus on the microstructure-sensitive materials. 
\item Probabilistic descriptions, in particular the PDFs, are used to account for the spatially heterogeneous material state fields in polycrystalline metallic materials.

\item The probabilistic model, in the form of the Liouville equation, is inferred by SPLIT-PINN and can effectively predict the temporal evolution of state PDFs.

\item SPLIT-PINN, tailored for high-dimensional inverse problems, incorporates a marginal-correction decomposition that stabilizes learning and ensures physical consistency in high-dimensional PDF dynamics.
\item The efficacy of SPLIT-PINN is demonstrated on real datasets of polycrystal state evolution, including the von Mises stress, dislocation density, and equivalent plastic strain rate, showing accurate reconstruction of their joint PDFs and predictive capability of the Liouville model for new polycrystal statistical volume elements.
\end{itemize}

The rest of this paper is organized as follows. In Section~\ref{sec:formulation}, we describe the Liouville model and formulation, and then propose the framework of SPLIT-PINN in Section~\ref{sec:SPLIT-PINN}. Next, Section~\ref{sec:bench} validates the performance of SPLIT-PINN over a benchmark problem. We provide the details of probabilistic modeling of polycrystal state evolution in Section~\ref{sec:polycrystal} and explain how SPLIT-PINN infers the PDF evolution of different polycrystal realizations. The concluding remarks are given in Section~\ref{sec:conclude}. 

\section{Liouville Model and Formulation}
\label{sec:formulation}

In the probabilistic modeling, the state variables ($n_d$-dimension) are assumed to follow an $n_d$-dimensional stochastic dynamical system with random initial conditions and a deterministic vector operator, as: 
\begin{equation}
\dot{\mathbf{X}}(t) = \mathbf{A}(\mathbf{X}(t),t)\;,
\label{eq:state_sys}
\end{equation}
where 
$
\mathbf{X}(t) = (X_1(t), \dots, X_{n_d}(t))^T
$
denotes the time-dependent $n_d \times 1$ vector of state variables, varying over the temporal domain $T=[t_0,t_f]$; and 
$
\mathbf{A} = (A_1, \dots, A_{n_d})^T
$
is the deterministic vector operator~(\cite{khalil2002nonlinear}). The initial condition for this stochastic system is given by
\begin{equation}
\mathbf{X}(t)\big|_{t=t_0} = \mathbf{X^0},
\label{eq:initial}
\end{equation}
where $\mathbf{X^0}$ is a \textit{stochastic} vector, capturing the distributional variability of the state variables at the initial time. 

Next, we denote the joint PDF of $\mathbf{X}(t)$ as $\Pro$, with $
\mathbf{x} = (x_1, x_2, \dots, x_{n_d})^T \in \Omega
$ representing realizations of $\mathbf{X}$ over the domain $
\Omega = \Omega_1 \times \Omega_2 \times \cdots \times \Omega_{n_d},
$ which is constructed from each one-dimensional spatial domains $\Omega_i$ associated with each state variable $X_i$. Thus, the governing transport equation for $\Pro$ follows (\cite{lasota2013chaos,lin1995probabilistic}):
\begin{equation}
\frac{\partial \Pro}{\partial t} + \sum_{i=1}^{n_d} \frac{\partial}{\partial x_i} \big[ A_i(\mathbf{x},t)\Pro \big] = 0,
\label{eq:liouville}
\end{equation}
which is a first-order PDE convecting the probabilistic uncertainty via the deterministic operator $\mathbf{A}$ in the drift term over the spatiotemporal domain $\Omega_T=\Omega \times T$. Eq. \eqref{eq:liouville}, known as the Liouville equation~(\cite{ehrendorfer1994liouville}), describes the spatiotemporal evolution of the joint PDF while preserving the total probability. The initial condition for Eq. \eqref{eq:liouville} is: 
\begin{equation}
    P_\mathbf{X}(\mathbf{x},0) = P_\mathbf{X^0}(\mathbf{x^0}) \;,    
\end{equation}
where $P_\mathbf{X^0}(\mathbf{x^0})$ is the joint PDF of $\mathbf{X^0}$. In this work, $P_\mathbf{X^0}(\mathbf{x^0})$ corresponds to the initial joint PDF of the state variables considered, obtained from the polycrystal simulation (\S\ref{sec:polycrystal}), and varies across different polycrystal realizations.

\section{SPLIT-PINN Methodology}\label{sec:SPLIT-PINN}

The probabilistic model considered in Eq.~\eqref{eq:liouville} conserves the total probability in the absence of diffusion or source terms. Therefore, our objective is to infer the drift vector field $\mathbf{A}(\mathbf{x},t)$ from 
$\Pro$ observed at discrete spatial-temporal points. Let $P_{X,j}=P_X(\mathbf{x}_j, t_j)$ denote the probability density of the observed stochastic state variable $\mathbf{x}_j \in \mathbb{R}^{n_d}$ at time $t_j$, with $j=1, \cdots, N_O$. Once $\mathbf{A}(\mathbf{x},t)$ is determined, we can use the probabilistic model in Eq.~\eqref{eq:liouville} to predict the temporal evolution of the joint PDF $\Pro$ with different initial conditions.

To alleviate the significant overhead incurred in balancing multiple competitive loss functions~(\cite{Bischof2025Multi}), we first construct a neural network (NN) to approximate $\Pro$ using the available data samples, before using PINNs for learning $\mathbf{A}(\mathbf{x},t)$. The resulting NN model, $\mathcal{NN}_{P_{X}}$, is subsequently employed to estimate $P^\mathcal{NN}_X(\mathbf{x},t)$ across the domain, as the input for PINNs to infer $\mathbf{A}(\mathbf{x},t)$. The architectural details and training procedures are described below. In \ref{app:joint}, we present an example demonstrating the superior performance achieved through sequential training of the data-driven and physics-based networks.

The NN model, $\mathcal{NN}_{P_{X}}(\mathbf{x},t;\theta_{P_X})$, is trained from the available data samples $\{x_j, t_j, P_{X,j}\},\, j=1, \cdots, N_O$, to approximate the underlying probability density field $\Pro$. The network has a two-dimensional input for the space-time points and four hidden layers. Its output is enforced to satisfy normalization and non-negativity constraints:
\begin{equation}
       \Pro \geq 0, 
        \qquad 
        \int_{\Omega} \Pro \, dx = 1, \quad \forall t,
        \label{eq:prob_constraints}
\end{equation}
%%%%%%%%%%%%%%%%%
These constraints are incorporated directly into the network architecture. Specifically, the positivity preservation is enforced by employing a sigmoid activation function in the last layer, while the normalization condition, ensuring that the integral of estimated probabilities over the domain equals one at each time, is imposed softly through a residual term in the loss function. More specifically, the data loss is defined as:
\begin{equation}
\mathcal{L}_{\mathrm{data}}
=
\frac{1}{N_O}
\sum_{j}
\left[
P_X^{\mathcal{NN}}(x_j,t_j)
-
P_{X,j}
\right]^2\;.
\label{eq:data_loss}
\end{equation}
%%%%%%%%%%%%%%%%%%%%
The normalization condition 
$\int_{\Omega} P_X^{\mathcal{NN}}(x,t)\,dx = 1$ at time $t_j$ for the $N_{t_j}$ uniformly distributed spatial points is imposed softly through the residual:
\begin{equation}
\mathcal{R}_{\mathrm{norm}}(t_j)
=
\Delta x \sum_{q=1}^{N_{t_j}} 
P_X^{\mathcal{NN}}(x_q,t_j)
- 1,
\end{equation}
where $\Delta x$ is $\prod_i^{n_d}  \Delta_i$ with $\Delta_i$ being the distance of two points in direction $i$. Accordingly, the corresponding loss term reads:
\begin{equation}
\mathcal{L}_{\mathrm{norm}}
=
\frac{1}{N_t}
\sum_{j=1}^{N_t}
\left(
\mathcal{R}_{\mathrm{norm}}(t_j)
\right)^2.
\end{equation}
%%%%%%%%%%%%%%%%%%%
The resulting network model $\mathcal{NN}_{P_{X}}$ yields the estimated $P^\mathcal{NN}_X(\mathbf{x},t)$ across the domain, which is subsequently utilized as the input of PINNs for inferring $\mathbf{A}(\mathbf{x},t)$.

%The second NN, denoted as 
To proceed, denote $\mathcal{NN}_A(\mathbf{x},t;\theta_A)$ the NN model that approximates the drift vector field $\mathbf{A}(\mathbf{x},t)$. Although defining a physics-informed loss function using the PDE~\eqref{eq:liouville} for training $\mathcal{NN}_A(\mathbf{x},t;\theta_A)$ is conceptually straightforward, directly estimating $\mathbf{A}$ through a PINN becomes increasingly challenging as the complexity and dimensionality of the joint PDF grow, especially with insufficient data about $\mathbf{A}$, such as boundary and initial conditions. This challenge is even more pronounced if $\Pro$ is highly localized or exhibits discontinuities within the domain, which can adversely affect training stability and accuracy. Therefore, we next introduce a decomposition approach to infer $\mathbf{A}$, as detailed below.

\subsection{Decomposition via Conditional Drift}

First, we obtain the dynamics of the marginal PDF of $x_i$ by integrating variables $\{x_k\}_{k \neq i}$ in \eqref{eq:liouville} over the spatial domain $\Omega$:
\begin{equation}
\int_{\Omega_{\setminus i}} \left( \frac{\partial \Pro}{\partial t} + \sum_{k=1}^{n_d} \frac{\partial}{\partial x_k} \big[A_k(\mathbf{x},t) \Pro \big] \right) d\mathbf{x}_{\setminus i} = 0 \;,
\end{equation}
where $d\mathbf{x}_{\setminus i} = \prod_{k \neq i} dx_k$ and $\Omega_{\setminus i} = \prod_{k \neq i} \Omega_k$.
Given $\Pro \to 0$ sufficiently fast as $x_k \to \pm\infty$, all boundary terms vanish except for the derivative with respect to $x_k$, yielding the following:

\begin{equation}
\label{eq:marg_der}
 \frac{\partial }{\partial t}\int_{\Omega_{\setminus i}} \Pro  d\mathbf{x}_{\setminus i} + \sum_{k=1}^{n_d} \frac{\partial}{\partial x_k} \int_{\Omega_{\setminus i}} \left[ A_k(\mathbf{x},t)\Pro \right]  d\mathbf{x}_{\setminus i} = 0 \;.
\end{equation}
Then, substituting the marginal PDF:
\begin{equation}
p_i(x_i, t) = \int_{\Omega_{\setminus i}} \Pro \, d\mathbf{x}_{\setminus i}\;,
\label{eq:marginal_PDF}
\end{equation}
into~\eqref{eq:marg_der} and assuming $p_i(x_i, t)\neq 0$, we rewrite the equation of the marginal PDF as:
\begin{equation}
\label{eq:marginal_pde_general}
\frac{\partial p_i(x_i,t)}{\partial t}
+\frac{\partial}{\partial x_i}\!\left(\bar A_i(x_i,t)\,p_i(x_i,t)\right)
+\sum_{\substack{k=1\\k\neq i}}^{n_d}
\int_{\partial\Omega_{\setminus i}}
A_k(\mathbf{x},t)\,p(\mathbf{x},t)\, n_k \, dS_{\setminus i}
=0 \;.
\end{equation}
Here, \(n_k\) denotes the \(k\)-th component of the outward unit normal vector
\(\mathbf{n}=(n_1,\ldots,n_{n_d})\) on \(\partial\Omega_{\setminus i}\), and
\(dS_{\setminus i}\) denotes the boundaries on \(\partial\Omega_{\setminus i}\).
Under sufficiently fast decay at infinity, or no-flux boundary conditions on
\(\partial\Omega_{\setminus i}\) (very mild assumptions valid in probabilistic distributions), the boundary integral vanishes, recovering the closed 1D marginal
transport equation.
Thus, we define the effective marginal drift $\bar{A}_i(\mathbf{x}, t)$ as:
\begin{equation}
\label{eq:effective_A_general}
\bar{A}_i(x_i, t) = \frac{1}{p_i(x_i, t)} \int_{\Omega_{\setminus i}} A_i(\mathbf{x}, t) P(\mathbf{x}, t) \, d\mathbf{x}_{\setminus i}, 
\quad \text{for } p_i(x_i, t) \neq 0.
\end{equation}
This allows us to decompose each drift component as:
\begin{equation}
\label{eq:correction}
A_i(\mathbf{x}, t) = \bar{A}_i(x_i, t) \otimes \mathbf{1}_{\Omega_{\setminus i}} + R_i(\mathbf{x}, t),
\end{equation}
with $R_i$ being a correction term that captures the interaction and dependency of $A_i(\mathbf{x}, t)$ on the remaining variables $\mathbf{x}_{\neg i}$. Substituting~\eqref{eq:correction} into \eqref{eq:effective_A_general} gives:
\begin{align}
\int_{\Omega_{\setminus i}} A_i(\mathbf{x}, t) P(\mathbf{x}, t) \, d\mathbf{x}_{\setminus i} 
&= \int_{\Omega_{\setminus i}}  \left( \bar{A}_i(x_i, t) + R_i(\mathbf{x}, t) \right) P(\mathbf{x}, t) \, d\mathbf{x}_{\setminus i} \\
&= \bar{A}_i(x_i, t) p_i(x_i, t) + \underbrace{\int_{\Omega_{\setminus i}}  R_i(\mathbf{x}, t) P(\mathbf{x}, t) \, d\mathbf{x}_{\setminus i}}_{=0}.
\end{align}
Therefore, $R_i$ satisfies the orthogonality condition:
\begin{equation}
\label{eq:ri_constraint}
\int_{\Omega_{\setminus i}}  R_i(\mathbf{x}, t) P(\mathbf{x}, t) \, d\mathbf{x}_{\setminus i} = 0 
\quad \text{for each fixed } x_i.
\end{equation}
This constraint ensures that the correction term $R_i$ does not contribute to the marginal drift and is used in the training of the PINN model to enforce consistency between the dynamics of joint and marginal PDFs.

\begin{remark}
     The Liouville equation with only partial information does not provide sufficient constraints to uniquely identify cross-dimensional dependencies, leading to severe non-uniqueness and instability in high dimensions. Moreover, enforcing the full transport equation using physics-informed neural networks suffers from the curse of dimensionality, resulting in weak constraint enforcement and poor generalization away from sampled trajectories.
The SPLIT-PINN framework addresses these challenges by decomposing the learning task into marginal and residual components. First, low-dimensional marginal PDFs are learned using reduced Liouville equations, which admit robust PINN training. The remaining high-dimensional structure is then recovered via a residual correction that is conditioned on the learned marginals and constrained to be orthogonal to them. From a functional perspective, the orthogonality constraint acts as a condition removing null-space components of the Liouville operator associated with marginal invariances, and regularizing the inverse problem~(\cite{gardiner2009stochastic}).
 As a result, SPLIT-PINN transforms the original ill-posed joint learning problem into a hierarchically constrained and identifiable formulation, significantly improving stability and data efficiency.
\end{remark}
\subsection{Physics-informed learning of marginal and correction terms} 
Therefore, the next step of SPLIT-PINN is to estimate the marginal drift term for each dimension. 
To this end, we first compute each marginal PDF $p_i(x_i, t)$ using Eq. \eqref{eq:marginal_PDF} and the NN model $\mathcal{NN}_{\Pro}$ previously trained to approximate $\Pro$ from its observed data $P_o=P_X(\mathbf{x}_o, t_o)$.

Subsequently, we train a PINN based on the governing equation~\eqref{eq:marginal_pde_general} to inversely infer the marginal drift term $\bar{A}_i(x_i, t)$, following the same procedure described in the previous section for the one-dimensional case. 
The corresponding physics-informed loss function is defined as
\begin{equation}
\mathcal{L}_{\text{marg}} =
\frac{1}{N_c} \sum_{j=1}^{N_c} 
\left|
\mathcal{R}_{\text{marg}}(x_i^j, t_i^j; \theta_{\Pro}, \theta_{\bar{A}_i})
\right|^2,
\label{eq:loss_marg}
\end{equation}
where $\mathcal{R}_{\text{marg}}$ denotes the residual of the PDE~\eqref{eq:marginal_pde_general} for the $i$-th marginal PDF over $N_c$ collocation points.

After repeating this process for all $i = 1, \ldots, n_d$, we train a second-stage neural network to learn the correction term $\bm{R}(\mathbf{x}, t)$ by minimizing the residual of the full Liouville equation, rewritten as
\begin{equation}
\label{eq:pde_multi}
    \frac{\partial \Pro}{\partial t} 
    + \nabla \cdot \left( \left( \bar{\mathbf{A}}(\mathbf{x}, t) + \bm{R}(\mathbf{x}, t) \right) \Pro \right) = 0\;,
\end{equation}
where $\bar{\mathbf{A}} = [\bar{A}_i]$ and $\bm{R} = [R_i]$ for $i = 1, \ldots, n_d$. The physics-informed loss function for training the correction term incorporates both the Liouville residual \eqref{eq:pde_multi} and the orthogonality constraint~\eqref{eq:ri_constraint}, and is defined as
\begin{equation}
\mathcal{L}_{\text{corr}} =
\frac{1}{N_c} \sum_{j=1}^{N_c} 
\left|
\mathcal{R}_{\text{corr}}(\mathbf{x}_c^j,t_c^j; \theta_{\Pro}, \theta_{\bm{R}})
\right|^2
+ 
\lambda_{\text{orth}}
\sum_{i=1}^{n_d}
\left|
\int_{\Omega_{\setminus i}}  
R_i(\mathbf{x}, t) \, \Pro \, d\mathbf{x}_{\setminus i}
\right|^2,
\label{eq:loss_corr}
\end{equation}
where $\mathcal{R}_{\text{corr}}$ denotes the residual of the Liouville equation involving both the estimated marginal drift $\bar{\mathbf{A}}$ and the learned correction $\bm{R}$; $\lambda_{\text{orth}}$ is a weighting coefficient that enforces the orthogonality constraint.

\begin{figure}[H]
    \centering
    \includegraphics[width=0.95\linewidth]{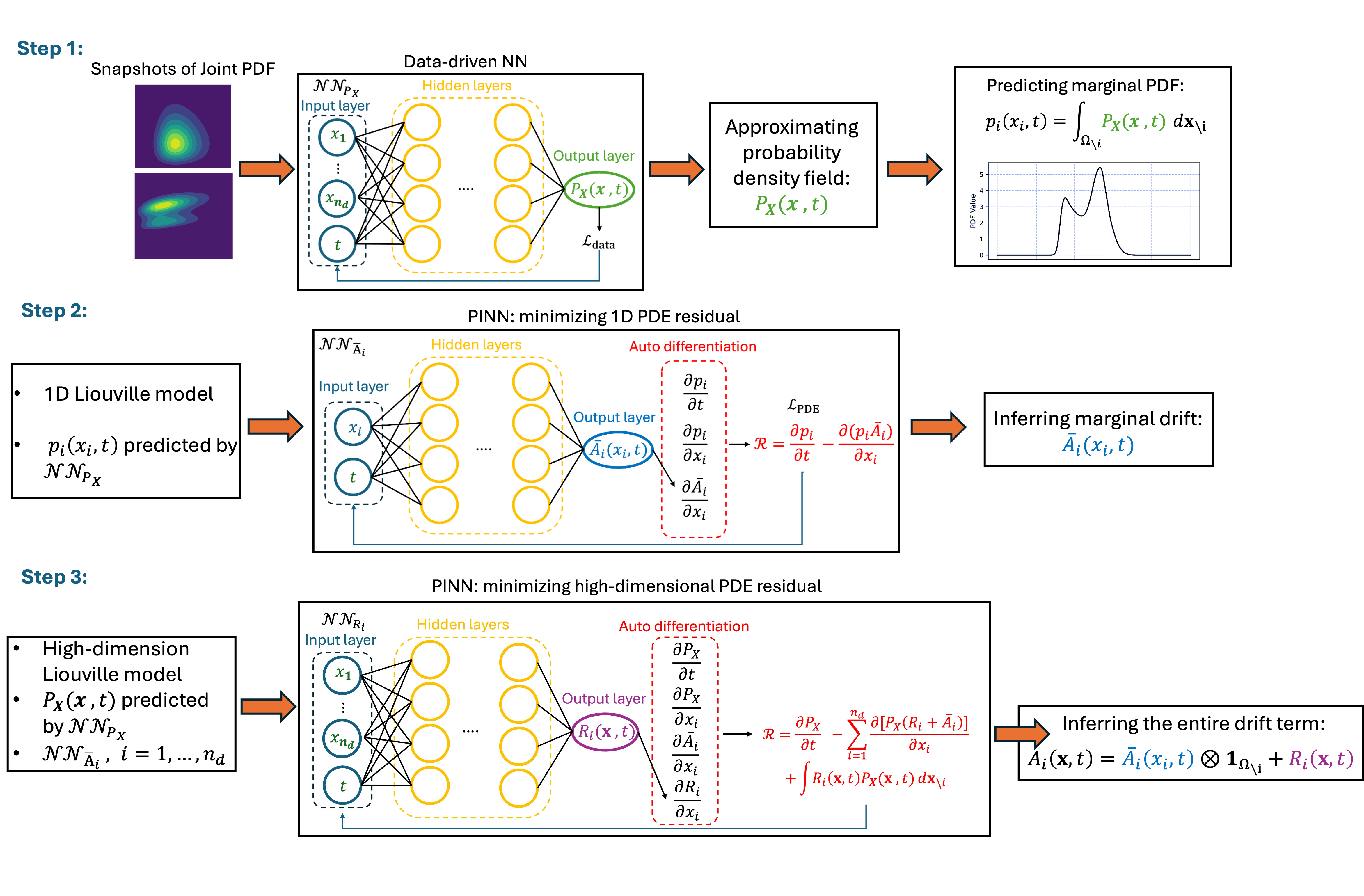}
    \vspace{-11pt}
    
    \caption{Overview of the SPLIT-PINN framework.}
    \label{fig:over_nd}
\end{figure}

Finally, by estimating $\bar{A}_i$ and $R_i$ for all $i = 1, \ldots, n_d$, we can recover the full drift structure of the system in an interpretable and physically consistent manner. We provide an overview of the SPLIT-PINN framework in Fig.~\ref{fig:over_nd}. 

\begin{remark}
The initial and boundary conditions of $\Pro$ are enforced during the training of the first NN. However, in the second NN, no explicit boundary term is included because no information is available about $\mathbf{A}(x,t)$ on the boundaries or at the initial time. This lack of boundary information renders the inverse problem increasingly ill-posed, particularly as the dimensionality grows~(\cite{kabanikhin2011inverse, poulet2024slip}). Addressing this challenge is one of our motivations for developing the SPLIT-PINN.

SPLIT-PINN mitigates the ill-posedness of learning the full high-dimensional Liouville equation by first learning marginal transport equations and then recovering the remaining joint structure through a residual constrained to be orthogonal to the marginals, effectively removing null-space ambiguities that appear in the high-dimensional Liouville model. 

\end{remark}

\subsection{Interpretability of SPLIT-PINN}

The proposed SPLIT-PINN, which utilizes the decomposition in Eq.~\eqref{eq:correction}, offers several key advantages in terms of accuracy, structure-awareness, and interpretability.  
First, by separating the marginal drift $\bar{A}_i$ from the correction term $R_i$, the one-dimensional components of the drift can be estimated with higher accuracy. This is because training PINNs in one dimension is more stable and precise, enabling finer resolution and better enforcement of physical constraints.  

Second, the correction term $R_i(\mathbf{x}, t)$ plays a crucial structural role: it isolates and captures interdimensional interactions that are not captured by the marginal drift. As a result, the model explicitly distinguishes between local, dimension-wise effects and global, coupled dynamics, thereby improving the overall structure-awareness of the learned representation.  

Finally, this decomposition enhances interpretability. The learned correction term $R_i(\mathbf{x}, t)$ quantifies how strongly the drift in direction $i$ depends on other variables beyond $x_i$. Examining the spatial and temporal patterns of $R_i$ can thus reveal the coupling strength and interaction mechanisms between state variables, providing valuable physical insight into the underlying system dynamics.

\subsection{SPLIT-PINN with Learnable Global Artificial Viscosity}

For a PDE without any diffusion term, its numerical solution through PINN (or other numerical techniques) may develop spurious oscillations, discontinuities, or instability. To mitigate these issues and stabilize training, an artificial viscosity is introduced into the PINN loss formulation, following ~(\cite{coutinho2023physics}). 

Specifically, an adaptive scalar value for the artificial viscosity coefficient $\nu$ is learned during the SPLIT-PINN training procedure. Accordingly, the Liouville residual used in the physics-based loss functions is augmented with an artificial diffusion term. Particularly, for the marginal PDFs, we have:
\begin{equation}
\label{eq:residual1}
\mathcal{R}_{\text{marg},\nu}(\mathbf{x}, t; \theta_{\Pro}, \theta_{\bar{A}_i}), \nu)
= \mathcal{R}_{\text{marg}}(x_i^j, t_i^j; \theta_{\Pro}, \theta_{\bar{A}_i})- \nu \frac{\partial ^2 p_i }{\partial x_i^2}\;,
\end{equation}
%%%%%%%%%%%%%%%%%%
and similarly, for the correction term:
%%%%%%%%%%%%%%%%%%%
\begin{equation}
\label{eq:residual2}
\mathcal{R}_{\text{corr}, \nu}(\mathbf{x}, t; \theta_{\Pro}, \theta_{\mathbf{A}}, \nu)
= \mathcal{R}_{\text{corr}}(x_i^j, t_i^j; \theta_{\Pro}, \theta_{\mathbf{A}})- \nu \nabla^2 \Pro\;,
\end{equation}
where the artificial viscosity coefficient $\nu$ is treated as a trainable parameter and updated via gradient descent (using the Adam optimizer) together with the NN parameters (weights and biases). The residual term~\eqref{eq:residual1} results in the loss function:
\begin{equation}
\mathcal{L}_{\text{marg}} =
\frac{1}{N_c} \sum_{j=1}^{N_c} 
\left|
\mathcal{R}_{\text{marg},\nu}(\mathbf{x}, t; \theta_{\Pro}, \theta_{\bar{A}_i}), \nu)
\right|^2+ \alpha_{\text{visc}} \mathcal{L}_{\text{visc}}(\nu) \;,
\label{eq:loss_marg_vis}
\end{equation}
%%%%%%%%%%%%%%%
where $\alpha_{\text{visc}}$ is a regularization coefficient for the artificial viscosity penalty term, which we define here as $\mathcal{L}_{\text{visc}}(\nu) = \nu^2$. This term penalizes large values of $\nu$ to encourage minimal artificial dissipation. The penalty coefficient $\alpha_{\text{visc}}$ can also be learned as part of the training. Next, for the correction term, we have the following loss function:
\begin{equation}
\mathcal{L}_{\text{corr}} =
\frac{1}{N_c} \sum_{j=1}^{N_c} 
\left|
\mathcal{R}_{\text{corr}, \nu}(\mathbf{x}, t; \theta_{\Pro}, \theta_{\mathbf{A}}, \nu)
\right|^2
+ 
\lambda_{\text{orth}}
\sum_{i=1}^{n_d}
\left|
\int_{\Omega_{\setminus i}}  
R_i(\mathbf{x}, t) \, \Pro \, d\mathbf{x}_{\setminus i}
\right|^2+ \alpha_{\text{visc}} \mathcal{L}_{\text{visc}}(\nu) \;.
\label{eq:loss_corr_vis}
\end{equation}

 This combined formulation allows the network to simultaneously infer the marginal drifts (or correction terms) and the optimal artificial viscosity coefficient $\nu$, ensuring both physical consistency and numerical stability.

\begin{remark}\revv{[Role of the artificial viscosity term in the context of inverse modeling]}

\revv{The introduction of the artificial viscosity (diffusion) term serves primarily as a numerical regularization mechanism to stabilize training in transport-dominated regimes, rather than as a modification of the underlying physical model. To ensure consistency with the original Liouville equation, the viscosity coefficient $\nu$ is treated as a learnable parameter and penalized through 
$\mathcal{L}_{\text{visc}}(\nu) = \nu^2$.}

\revv{From an inverse modeling perspective, the inferred drift corresponds to the drift of the original Liouville equation in the limit as $\nu \to 0$. In practice, when a small but nonzero viscosity is learned, the identified drift should be interpreted as an effective drift associated with weakly regularized dynamics. Empirically, we observe that the learned viscosity remains small (order of $10^{-5}$) due to the imposed penalty, indicating that the inferred dynamics remain close to the diffusion-free Liouville model.} 
\end{remark}

\subsection{Residual-Based Adaptive Distribution (RAD)}
We incorporate adaptivity as the final component of the SPLIT-PINN framework. In learning a full joint PDF, regions of the state space with nonzero probabilities are the important ones. The residual of the governing equation reveals where the model is struggling, i.e., where the drift is misrepresented, and/or where sharp structures are forming~(\cite{liu2024discontinuity}). Residual-Based Adaptive Distribution (RAD) leverages this information in the residual and reallocates sampling points toward the locations where the physics is least satisfied. This adaptive mechanism prevents the model from wasting resolution on already well-captured regions and stabilizes learning in high-dimensional settings where uniform sampling fails to resolve essential features. Following~(\cite{wu2023comprehensive}), RAD defines a new PDF as:
\begin{equation}
   \pi(\bx) \propto \frac{\varepsilon^k(\bx)}{\mathbb{E}[\varepsilon^k(\bx)] + c}\;,
    \label{eq:rad_pdf}
\end{equation}
where $\varepsilon(\bx)$ is the value of the loss function ($\mathcal{L}_{\text{marg}}$ or $\mathcal{L}_{\text{corr}}$) at spatial-temporal collocation points $\bx$;
    $k \geq 0$ and $c \geq 0$ are two hyperparameters that control the emphasis on higher residual regions and the smoothing effect, respectively. Here, we set $k=2$ and $c=0$ following discussions in~(\cite{wu2023comprehensive}). 
    $\mathbb{E}[\varepsilon^k(\bx)]$ denotes the expected value of the $k$-th power of the residual, which can be approximated numerically using Monte Carlo integration:
    \begin{equation}
        \mathbb{E}[\varepsilon^k(\bx)] \approx \frac{1}{N} \sum_{j=1}^N \varepsilon^k(\bx_j)\;,
    \end{equation}
    for a sample $\{\bx_j\}_{j=1}^N$ drawn from current distribution.

We begin by training the NN on an initial set of collocation points denoted by $\mathcal{S}_0$. After an initial training phase, we compute the adaptive sampling distribution $\pi(\bx)$ over a new set of candidate points. Based on this distribution, we collect a new set of points $\mathcal{S}_1$, where the cardinality of $\mathcal{S}_1$ is smaller than that of $\mathcal{S}_0$. We then replace a subset of $\mathcal{S}_0$ with points from $\mathcal{S}_1$ to form an updated set of collocation points. This strategy helps maintain training stability by avoiding drastic changes in the collocation set. After a fixed number of training iterations, we repeat the procedure to refine the sampling toward regions with higher residuals.

Furthermore, to emphasize more on the regions with large pointwise residuals, we weight the loss formulation. For this, let $\mathcal{\tilde{R}}(x,t)$ denote the pointwise residual (marginal or correction). We then define a normalized weight:
\begin{equation}
    w(\bx,t) = \frac{\mathcal{\tilde{R}^2}(\bx,t)}{\overline{\mathcal{\tilde{R}^2}(\bx,t)}} + \delta \;,
\end{equation}
where $\overline{\mathcal{\tilde{R}}(\bx,t)}$ is the mean of ${\mathcal{R}(\bx,t)}$ over the training batch; $\delta$ is a small constant added for numerical stability (e.g., $10^{-5}$). The final weighted loss becomes:

\[
\mathcal{L}_{\text{weighted}} = \frac{1}{N} \sum_{j=1}^N w(\bx_j, t_j) \cdot \tilde{R}(\bx_j, t_j) \;.
\]
This weighted loss, together with the adaptive distribution, dynamically emphasizes high-error regions, encouraging SPLIT-PINN to resolve localized features and/or sharp transitions more effectively.

\section{Validation on a Benchmark Problem}
\label{sec:bench}
To validate the performance of SPLIT-PINN, we first apply it to a benchmark problem. In particular, we consider the two-dimensional  system: 
\begin{equation}
\label{eq:bench}
    d 
    \begin{pmatrix} 
        X_t \\ 
        Y_t 
    \end{pmatrix} 
    =
    \begin{pmatrix} 
        \partial_{X_t} \Phi(X_t, Y_t) \\ 
        \partial_{Y_t} \Phi(X_t, Y_t) 
    \end{pmatrix} 
    dt \;,
\end{equation}
where the potential function is defined as:
\[
    \Phi(x,y) = - (x + \lambda_0)^2 (y + \lambda_1)^2 - (x + \lambda_2)^2 (y + \lambda_3)^2,
\]
with parameters $\lambda_0 = \lambda_1 = 1$ and $\lambda_2 = \lambda_3 = -0.5$. In the absence of stochastic noises in~\eqref{eq:bench}, the Liouville equation describes the PDF evolution of the state variables using the drift terms defined as: $A_x = \frac{\partial\Phi(x,y)}{\partial x}$ and $A_y = \frac{\partial\Phi(x,y)}{\partial y}$. 

The initial distribution of state variables is set to $\mathcal{N}(0, I)$. Sampling from the initial distribution as ensembles, we approximate $(X_t, Y_t)$ over the time interval $[0,1]$ to construct a reference PDF. To determine the ensemble size required for accurate approximation, we compare PDFs generated using several ensemble sizes. The reference solution herein uses $15{,}000$ ensembles, against which we compare the PDFs obtained using $2{,}000$, $5{,}000$, and $10{,}000$ ensembles (more details are provided in \ref{app:benchmark}). 

This system creates a double-well energy landscape with two stable regions separated by saddle structures. As a result, the dynamics exhibit strong cross-dimensional coupling in the drift direction, particularly in regions connecting the two wells. Although the state space is two-dimensional, this setup results in a nontrivial inverse problem because the dynamics and the underlying drift field have different scales. That is, trajectory data constrain the dynamics primarily along sampled paths, while the underlying drift field must be inferred globally. Accurately recovering the full vector-valued drift field in such systems is therefore significantly more challenging than predicting scalar quantities or individual trajectories.

Previous work (e.g., \cite{chen2021solving}) has shown that standard PINNs struggle in approximating the Liouville equation, especially when the objective is to recover the drift field rather than only the trajectories or probability densities. The difficulty arises from two related issues. First, enforcing the Liouville or transport equation through a global residual loss becomes insufficient and diverges from observed trajectories, leading to a poor generalization of the learned vector field. Second, the solutions to inverse problems and the use of PINNs in this application are inherently non-unique. Therefore,  multiple drift fields can be consistent with the same partial probability evolution, unless additional structural constraints are imposed. Existing approaches often address this by prescribing a specific functional form for the drift (e.g., low-order polynomials) and estimating only a finite set of coefficients~(\cite{chen2021solving}). However, this limits the model's expressiveness and restricts its applicability.

The SPLIT-PINN framework addresses these challenges by introducing a structured decomposition of the learning task. As we mentioned before, the high-dimensional structure is recovered through a residual correction that is explicitly constrained to be orthogonal to the learned marginals. From a functional perspective, this orthogonality condition removes null-space components of the Liouville operator associated with marginal invariances, thereby regularizing the inverse problem and restoring identifiability without imposing \textit{a priori} assumptions on the functional form of the drift.

\begin{figure}[H] 
    \centering
    \includegraphics[width=.9\linewidth]{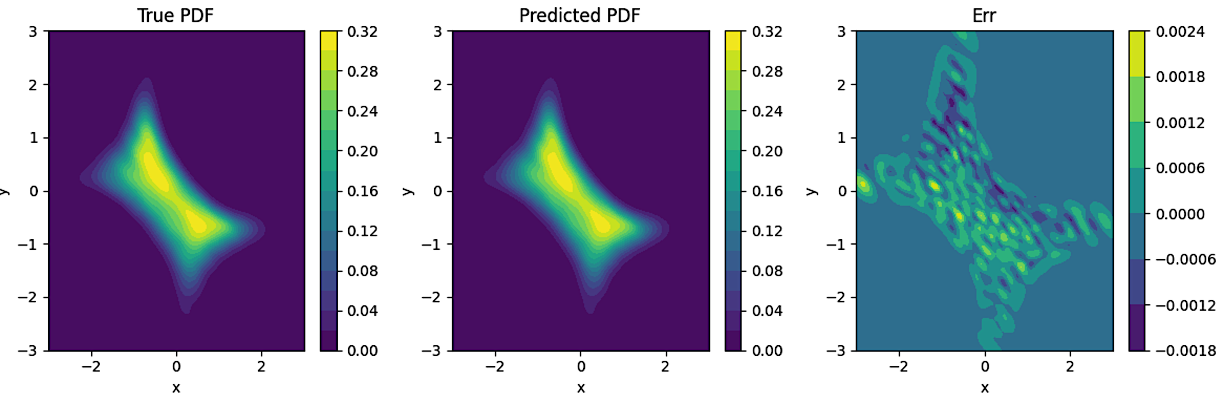}
  \caption{Benchmark problem: Comparison of the predicted PDF \(P(x,y,t)\) against the ground truth. The reported error corresponds to the absolute pointwise error.}
          \label{fig:bench_pdf}
\end{figure}

\begin{figure}[H] 
    \centering
    \includegraphics[width=.7\linewidth]{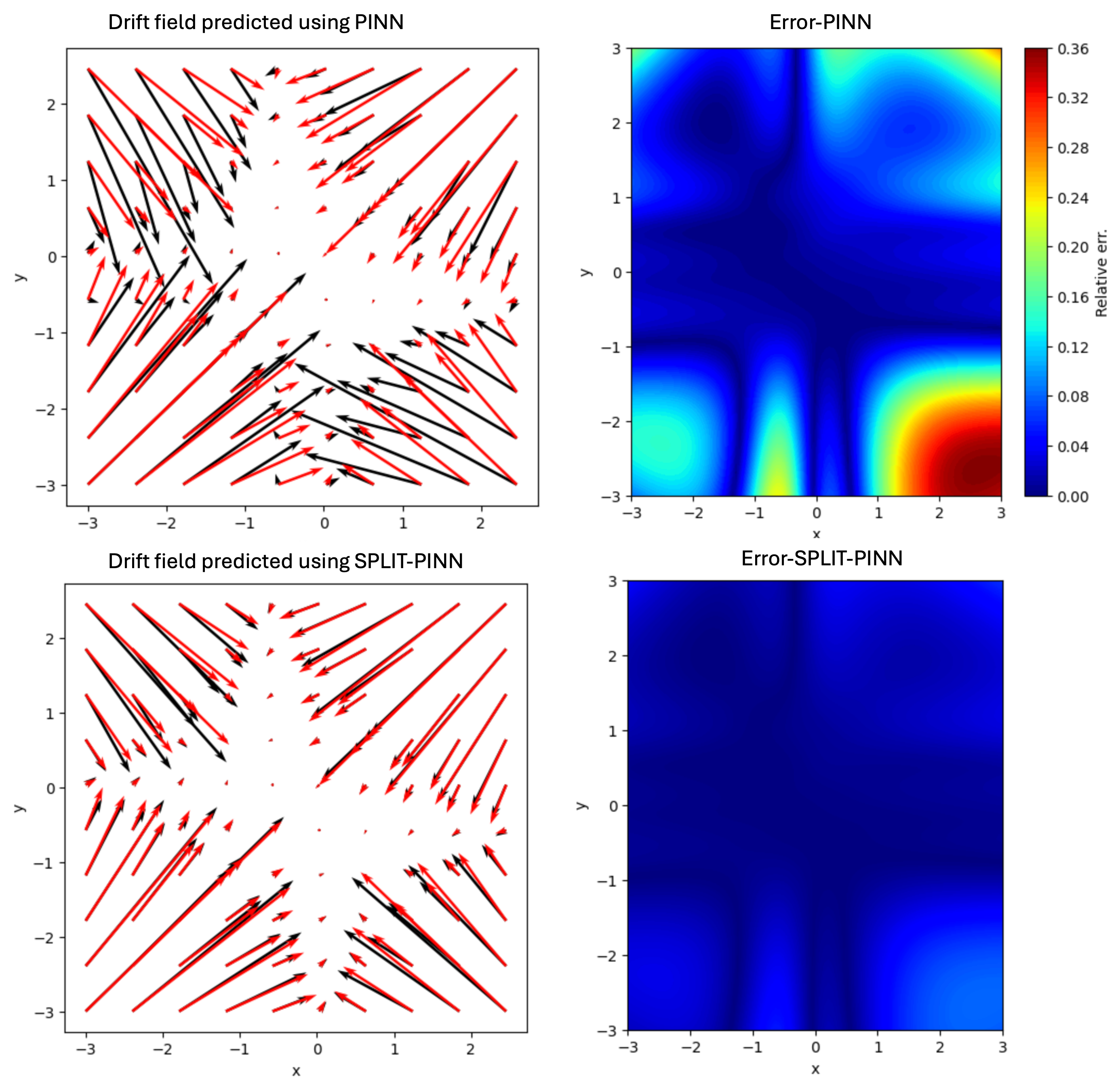}
  \caption{Benchmark problem: Recovered drift vector fields. Black: true drift; red: predicted drift. 
The generic PINN exhibits significant errors, whereas the SPLIT-PINN achieves markedly improved accuracy across the domain.}
    \label{fig:bench_err}
\end{figure}

To validate the effectiveness of SPLIT-PINN, we compare the predicted PDFs with the ground truth, as shown in Fig.~\ref{fig:bench_pdf}. The PDF predicted by SPLIT-PINN faithfully preserves the bimodal structure and accurately matches the true PDF, validating Step 1 of SPLIT-PINN. Next, we compare SPLIT-PINN with a generic PINN in both PDF reconstruction and drift-field recovery, thereby validating Steps 2 and 3 of SPLIT-PINN. As shown in Fig.~\ref{fig:bench_err}, the generic PINN produces drift fields that only qualitatively follow the true dynamics but fail to quantitatively recover the underlying drift field, particularly in the absence of drift-field observations. In contrast, SPLIT-PINN yields a substantially more accurate approximation of the drift vector field across the entire domain, ensuring consistency with the true dynamics. Quantitatively, this improvement corresponds to approximately a fourfold reduction in relative error relative to the generic PINN.

Therefore, this benchmark demonstrates that inverse drift identification can be severely ill-posed if not approached with additional care, for example, by directly using PINNs. By incorporating hierarchical constraints via marginal-residual decomposition and orthogonality enforcement, SPLIT-PINN offers a principled and effective solution that enables accurate and stable recovery of vector-valued drift fields without restrictive parametric assumptions. \rev{For completeness, we present the sensitivity of the drift-term inference to the distribution of the collocation points, considering both uniform and adaptive sampling, as well as to the total number of collocation points in~\ref{sec:collocation}.} \revv{Moreover, we discuss the role of orthogonality enforcement, as also presented in~\ref{sec:collocation}.}

\section{Probabilistic Modeling of Polycrystal State Evolution}
\label{sec:polycrystal}
After validating the effectiveness of SPLIT-PINN on a benchmark problem, we next assess its performance in inferring a predictive statistical model for the temporal evolution of material state variables in deforming polycrystalline metallic materials. The resulting statistical model, in the form of the Liouville equation, predicts the joint PDF and its temporal evolution for state variables such as the von Mises stress, dislocation density, and equivalent plastic strain rate. Prior to applying SPLIT-PINN, we outline the physical model and the methodology used to generate the state variable datasets.

\subsection{Polycrystal state evolution}

It is well understood that most metallic single crystals are atomically structured, which produces an anisotropic elastic-plastic mechanical response to loading. Polycrystalline metallic materials then reflect this anisotropy in single crystal behavior by spatially heterogeneous deformation and material state fields due to significant intergranular interaction, and local material structure evolution. The aggregate heterogeneity evolves with deformation and contributes significantly to the nucleation of damage during extreme loading histories. We provide an overview of a theory of the mechanical behavior of single-crystal body-centered cubic materials applied to pure iron in this work (\cite{cho_anomalous_2018,lee_deformation_2023,SCHMELZER2025104318}). 

\subsection{Single crystal theory overview}

\begin{figure}
    \centering
    \includegraphics[width=0.5\linewidth]{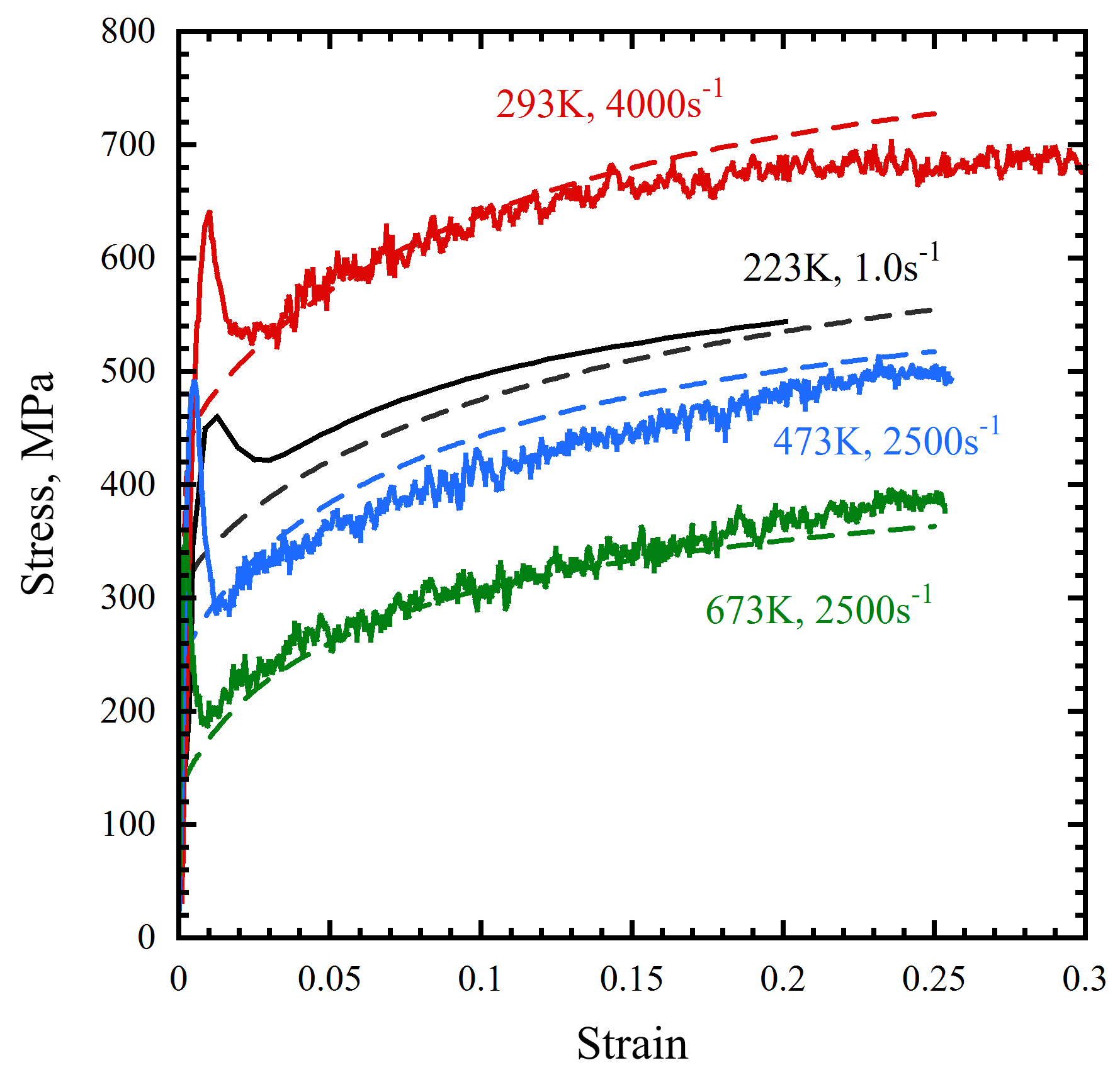}
    \caption{Single crystal model used with a 1000 grain polycrystal simulation in comparison with results of simple compression experiments conducted on polycrystalline pure iron for varying initial temperatures and applied strain rates. The broken lines represent the numerical simulation results.}
    \label{fig:Fe_CrystalModel}
\end{figure}

\begin{figure}
    \centering
    \includegraphics[width=1.00\linewidth]{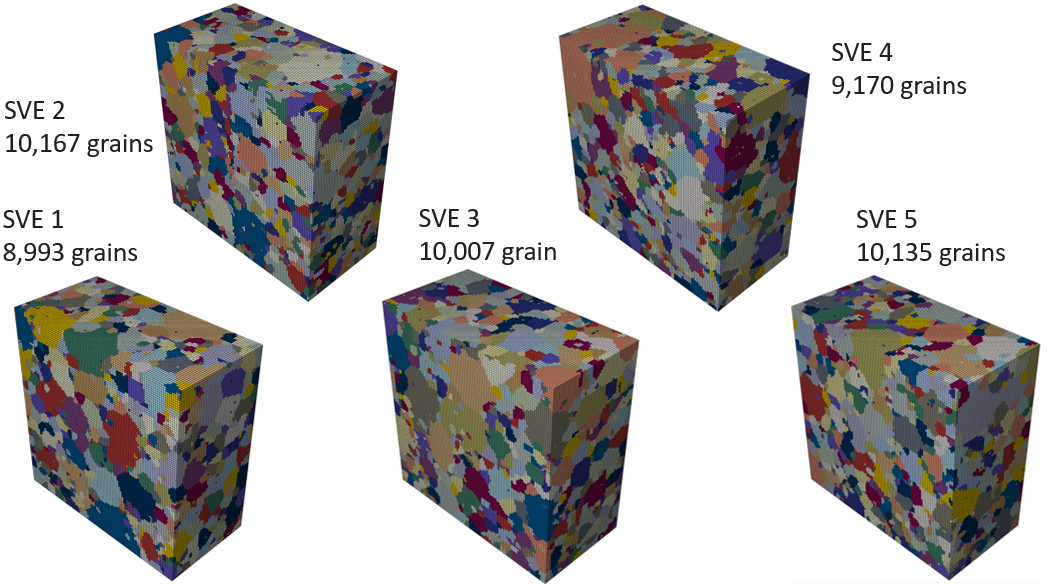}
    \caption{Five polycrystal statistical volume elements of a quenched and tempered low carbon steel plate. Each numerical model contains $100\times100\times100$ hexahedral elements with dimensions of 50 $\mu$m on a side. Each colored region represents a different single crystal. The colorization scheme is random to enable differentiation of different grains. For each SVE realization, each grain is assigned a specified crystallographic orientation to be consistent with experimental results.}
    \label{fig:Poly_Reals}
\end{figure}

\begin{figure}
    \centering
    \includegraphics[width=0.8\linewidth]{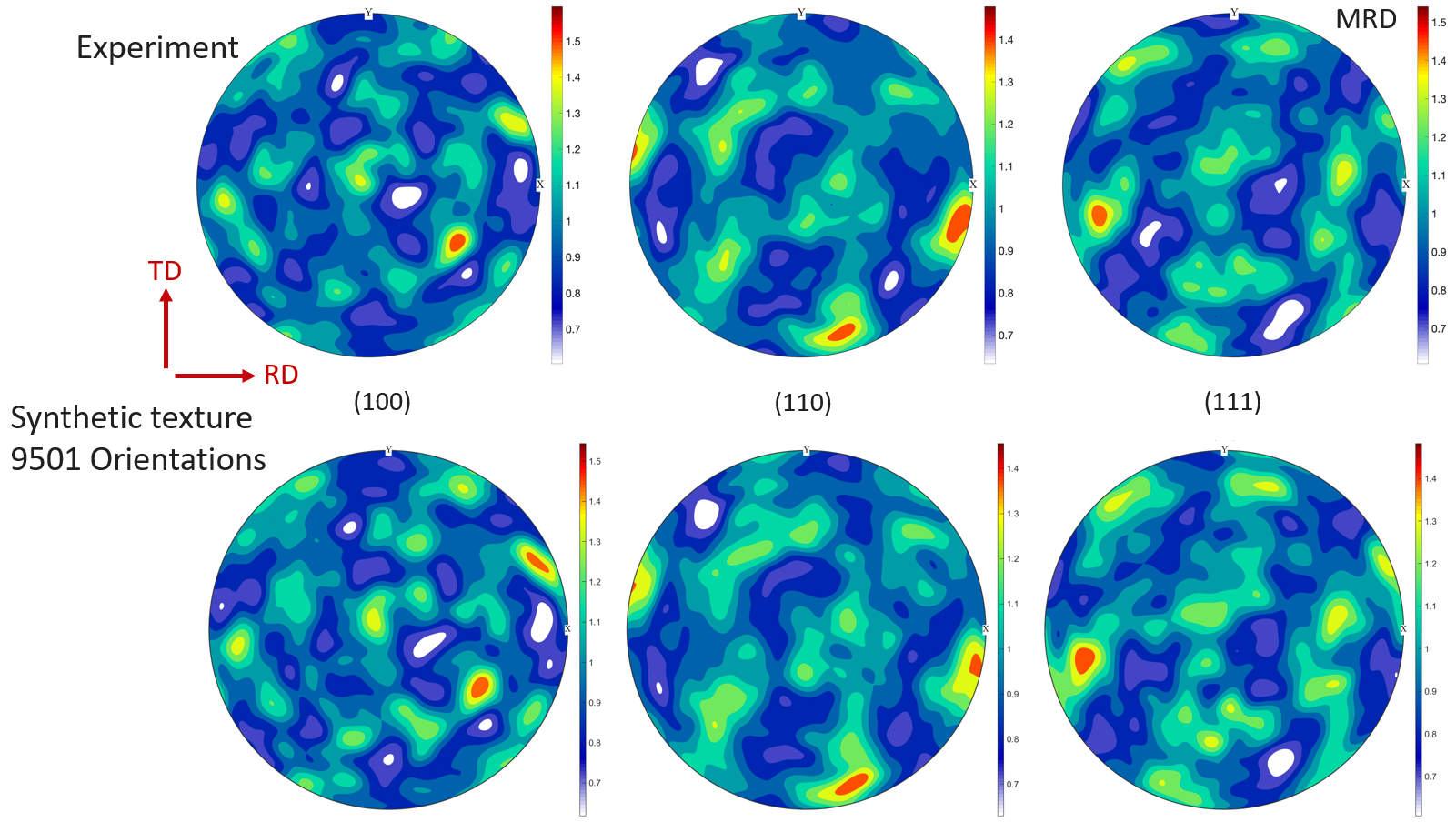}
    \caption{Equal-area pole figures for measured and reconstructed crystallographic texture using 9,501 orientations. These orientations are assigned randomly to grains in each statistical volume element model displayed in Fig.~\ref{fig:Poly_Reals}. }
    \label{fig:CrysTex}
\end{figure}

Finite deformation is represented by a multiplicatively decomposed deformation gradient
\begin{equation}
    \bf{F}=Grad \space \bf{x},\quad \bf{x}=\hat{\bf{x}}(\bf{X},t),\quad \bf{F}=\bf{F}^{e}\bf{F}^{p},
\end{equation}
where "Grad" denotes the spatial gradient in the reference configuration of motion $\hat{\bf{x}}$. The spatial velocity gradient represents the rate of deformation in the current configuration as

\begin{equation}
    \bf{L}=grad \space \bf{v}=\dot{\bf{F}}\bf{F}^{-1}=\bf{L}^{e}+\bf{F}^{e}\bf{L}^{p}\bf{F}^{e-1}, \quad \bf{L}^{e}=\dot{\bf{L}}^{e}\bf{L}^{e-1}, \quad \bf{L}^{p}=\dot{\bf{F}}^{p}\bf{F}^{p-1}.
\end{equation}
The plastic velocity gradient is then defined as
\begin{equation}
    \bf{L}^{p}=\sum^{N}_{\alpha=1} \dot{\gamma}^{\alpha}\bf{S}^{\alpha}_{0}, \quad \bf{S}^{\alpha}_{0}=\bf{m}^{\alpha}_{0} \otimes \bf{n}^{\alpha}_{0},
\end{equation}
where $\dot{\gamma}^{\alpha}_{p}$ is plastic slip rate to be defined shortly and $\bf{m}^{\alpha}_{0}$, $\bf{n}^{\alpha}_{0}$ are slip direction and plane normal for slip system $\alpha$ in the reference configuration. The elastic second Piola stress is defined by the hyperelastic relationship
\begin{equation}
    \bf{T}^{e}=\mathbb{C}[\bf{E}^{e}-\bf{A}(\theta-\theta_{0})],\quad \bf{E}^{e}=\frac{1}{2}(\bf{F}^{eT}\bf{F}^e-\bf{1}),
\end{equation}
where $\mathbb{C}$ is the fourth-order elastic moduli tensor and $\bf{A}$ is the second-order thermal expansion tensor. The plastic slip rate on the $\alpha$th slip system is described by the thermally-activated motion of dislocations expression 
\begin{equation}
    \dot{\gamma}^{\alpha}=\dot{\gamma}_{0}\text{exp}\left(-\frac{\Delta G}{k_{B}\theta}\left<1-\left( \frac{\tau^\alpha_{eff}}{\tilde{s}_{l}}\right)^{p}\right>^{q}\right),\quad \tau^{\alpha}_{eff}=|\tau^{\alpha}|-s^{\alpha}, \quad \tilde{s}^{l}=s_{l}\frac{\mu}{\mu_0}.
\end{equation}
The quantities, $\dot{\gamma}_{0}$ reference slip rate, $\Delta G$ activation energy, $k_B$ Boltzmann's constant, $p$ and $q$ energy barrier shape exponents, $\mu,\mu_0$ current and 0K shear moduli, $s_l$ intrinsic lattice resistance, and $<\bullet>=1/2(|\bullet|+(\bullet))$. The dislocation structural resistance is given by
\begin{equation}
    s^\alpha = s_0 +\mu b\sqrt{\sum^N_{\beta=1}a^{\alpha\beta}\rho^{\beta}}.
\end{equation}
where $s_0$ is the far-field resistance to slip, $b$ is the Burgers vector magnitude, $a^{\alpha\beta}$ is interaction matrix, and $\rho^{\beta}$ is dislocation density on system $\beta$. The evolution of dislocation density follows the following relationship
\begin{equation}
    \dot{\rho}^\alpha=\frac{1}{b}\left( \frac{1}{\pounds^\alpha}-2y^\alpha_c\rho^\alpha\right) |\dot{\gamma^\alpha_p}|, \quad  \frac{1}{\pounds^\alpha}=\sqrt{\sum^{N}_{\beta=1}d^{\alpha\beta}\rho^\beta}, \quad y^\alpha_c=y_{c0}\left( 1-\frac{k_B\theta}{G_{rec}}\text{ln}\left| \frac{\dot{\gamma}^\alpha_{p}}{\dot{\gamma_0}}\right|\right).
\end{equation}
where $\pounds^{\alpha}$ is mean free path of dislocations, $y^\alpha_c$ is annihilation capture radius, $d^{\alpha\beta}=\frac{a^{\alpha\beta}}{k^{2}_{1}}$ for self or coplanar interaction, $d^{\alpha\beta}=\frac{a^{\alpha\beta}}{k^{2}_{2}}$ for other interactions, $k_1,k_2$ mean free path coefficients, $y_{c0}$ a length scale in numbers of Burgers vector magnitudes, and $G_{rec}$ is annihilation activation energy. Details of the theory and methodology for material parameter evaluation can be found in (\cite{lee_deformation_2023,SCHMELZER2025104318}). The material parameters and their values for pure iron can be found in 
\autoref{tab:Params}. 
\begin{table}[H]
\begin{center}
\renewcommand{\arraystretch}{1.2}
\caption{ \label{tab:Params} Material parameters for pure iron.}
\begin{tabular}{c c c c c c c c}
    \hline
    $\rho_M$ [kg/m$^3$]         & 7870                     & \hspace{0.3cm} & $a_{\text{colli }30^{\circ}}$ & 0.5112    & \hspace{0.3cm} & $\Delta G$ [J]    & $2.10 \times 10^{-19}$\\
    $c$ [kJ/kg-K]                 & 460                       & \hspace{0.3cm} & $a_{\text{colli }60^{\circ}}$ & 0.7744    & \hspace{0.3cm} & $s_L$ [MPa]       & 420.0 \\
    $\alpha$ [$\mu$m/m-K]       & 0.12                       & \hspace{0.3cm} & $a_{\text{colli }90^{\circ}}$ & 0.9025    & \hspace{0.3cm} & $y_{c0}$ & $6b$ \\
    $k_B$ [J/K]                 & 1.38 $\times 10^{-23}$    & \hspace{0.3cm} & $a_\text{copl}$               &0.06       & \hspace{0.3cm} & $\rho_0^\alpha$ [m$^{-2}$] & $0.1 \times 10^{12}$\\
    $\mathcal{C}_{11,0}$ [GPa]  & 246.5                     & \hspace{0.3cm} &  $a_{\text{J}} $             &0.05       & \hspace{0.3cm} & $G_\text{rec}$ [J] & $3.0 \times 10^{-20} $\\
    $\mathcal{C}_{12,0}$ [GPa]  & 139.7                     & \hspace{0.3cm} & $a_{\text{XJ}} $& 0.04            & \hspace{0.3cm} & $p$& 0.33\\
    $\mathcal{C}_{44,0}$ [GPa]  & 123.3                      &  \hspace{0.3cm} & $k_1$ & 8.0               & \hspace{0.3cm} & $q$& 1.34\\
    $m_{11}$ [MPa/K]            & -59.7                     &  \hspace{0.3cm} & $k_2$ & 150.0             & \hspace{0.3cm} &$\dot \gamma_0 $ [s$^{-1}$] &$10^7$ \\
    $m_{12}$ [MPa/K]            & -21.5                     & \hspace{0.3cm} & $\theta_r$ [K]                & 100        & \hspace{0.3cm} &$b$ [nm]  & 0.248\\
    $m_{44}$ [MPa/K]            & -22.8                     & \hspace{0.3cm} & $s_0$ [MPa]                   & 0        & \hspace{0.3cm} &  &  \\
    \hline
\end{tabular}
\end{center}
\end{table}
The performance of the single crystal model representing polycrystalline pure iron simple compression experiments is demonstrated in Fig.~\ref{fig:Fe_CrystalModel}. Overall, the single crystal model performs reasonably well in representing the range of initial temperatures and applied strain rates relevant for the polycrystal simulations conducted here. This single crystal model is used to conduct large polycrystal microstructure simulations of a quenched and tempered low carbon steel plate material which was characterized by electron backscatter diffraction (EBSD) imaging (\cite{WanniIJP}). This EBSD data was used within the Dream3D (\cite{Dream3D}) software to produce voxelized microstructures replicating the quenched and tempered low carbon steel plate material. Five statistical volume element realizations with side lengths of 50 $\mu$m were produced for this study and can be viewed in Fig.~\ref{fig:Poly_Reals} composed of $100\times100\times100$ Abaqus C3D8 hexahedral elements. Over the sum of the five statistical volume elements (SVEs), the equivalent sphere diameter grain size is 1.95 $\mu$m, with a minimum of 0.62 $\mu$m and maximum of 17.35 $\mu$m. This is in comparison to the EBSD results of equivalent sphere diameter grain size of 1.91 $\mu$m, minimum of 0.41 $\mu$m and maximum of 16.46 $\mu$m. Note that the colors used to differentiate grains within the five images in Fig.~\ref{fig:Poly_Reals} are arbitrary. Each grain however was assigned a crystallographic orientation based upon the Dream3D derived replication of the crystallographic texture of the steel plate using a pool of 9,501 orientations. This collection of orientations was assigned randomly to the individual grains within each of the five polycrystal aggregate models. The replicated crystallographic orientation is provided in Fig.~\ref{fig:CrysTex} in comparison to experimental data. 

\begin{figure}
    \centering
    \includegraphics[width=0.8\linewidth]{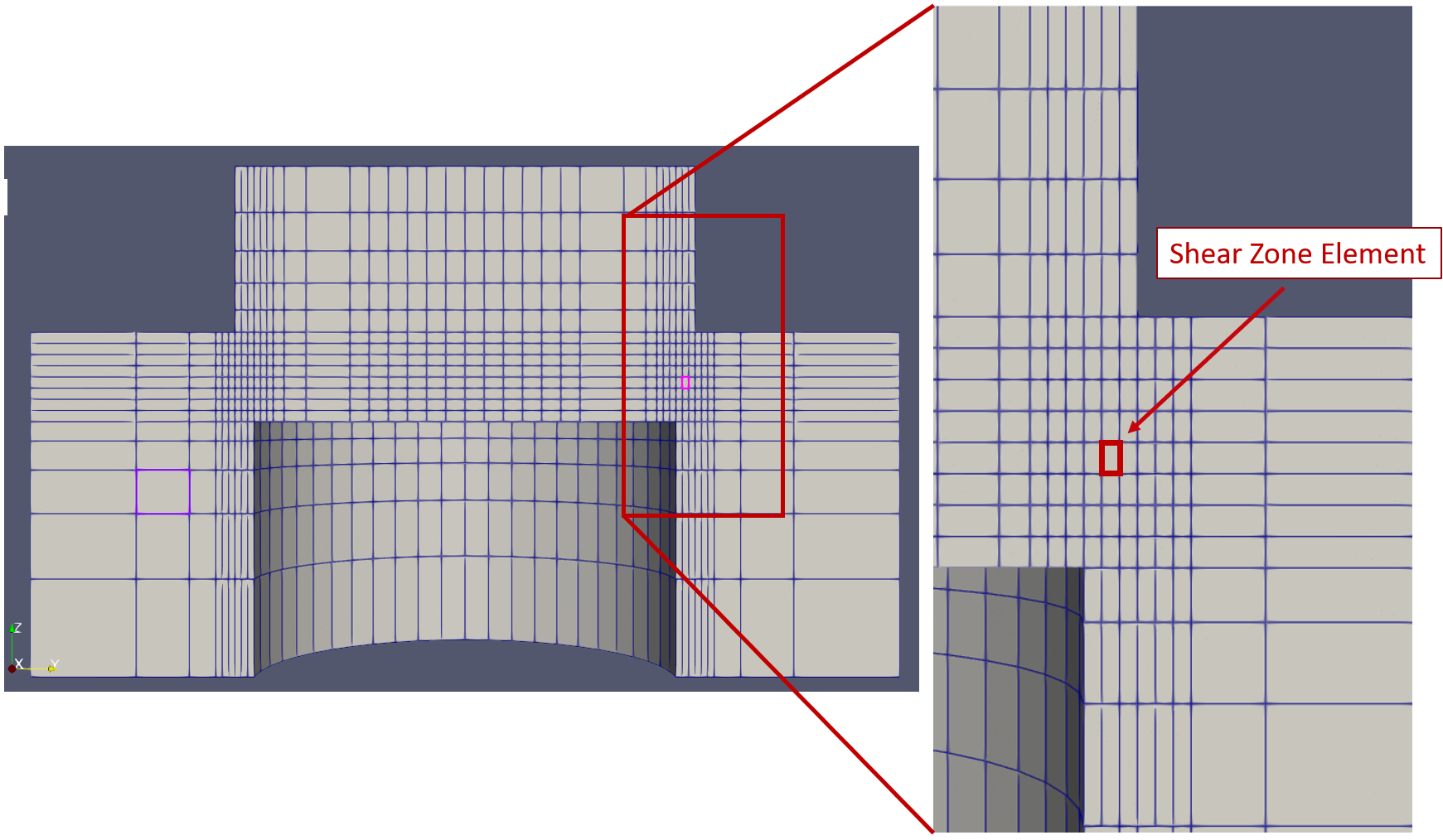}
    \caption{Location within the shear zone of a forced shear sample simulation (\cite{LIEOU2019171} of a steel material from which the time evolution of the stress tensor was extracted up to the time at which the simulation indicated adiabatic shear band initiation}
    \label{fig:ShearZone}
\end{figure}

\begin{figure}
    \centering
    \includegraphics[width=0.9\linewidth]{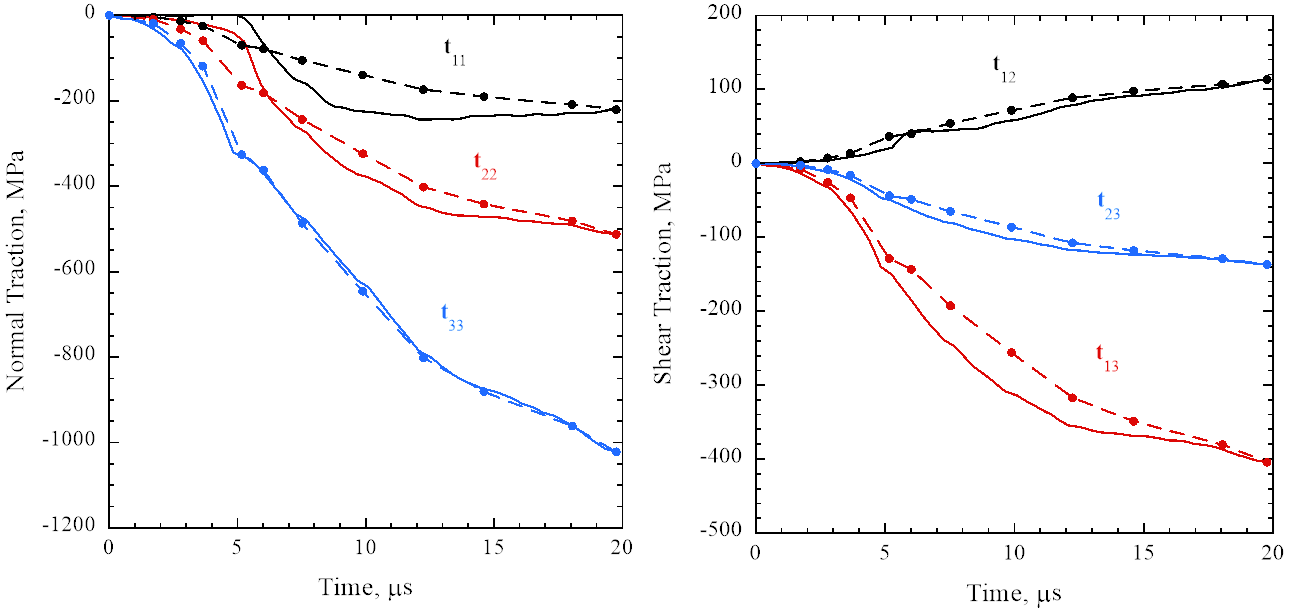}
    \caption{Traction vs. time profiles extracted from the computational element in the forced shear simulation of (\cite{LIEOU2019171}) are shown by the solid lines. The broken lines are the traction profiles applied to the faces of each statistical volume element in Fig.~\ref{fig:Poly_Reals}.}
    \label{fig:tracprof}
\end{figure}

The synthetic numerical microstructure models (Fig.~\ref{fig:Poly_Reals}) are used to approximate features of microstructural evolution leading to adiabatic shear banding in steel materials. These simulation results will be used to develop a methodology for representing the microstructural state evolution in macroscale calculations of adiabatic shear banding (\cite{MOURAD20171,JIN201911,JIN2019416}). The computational results of (\cite{LIEOU2019171}) are used to quantify the stress state evolution within the shear zone of a forced shear sample simulation up to the point at which the simulation predicts adiabatic shear band nucleation. This is similar to the method used by (\cite{BRONKHORST2021102903,SCHMELZER2025104318}) to enable high-fidelity polycrystal simulations of important regions of a much larger sample. For the forced shear simulation, the stress tensor was extracted from an element within the shear zone as illustrated in Fig.~\ref{fig:ShearZone}. The resulting traction profiles and the corresponding representation of each profile applied to the polycrystal realizations in Fig.~\ref{fig:Poly_Reals} are shown in Fig.~\ref{fig:tracprof}. These were applied as surface tractions to each surface of each of the polycrystal cubes. 

\begin{figure}
    \centering
    \includegraphics[width=0.7\linewidth]{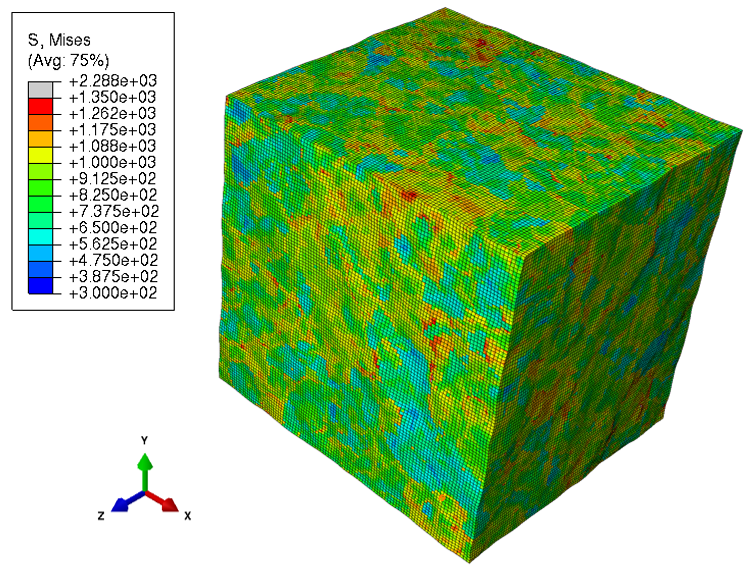}
    \caption{Deformed polycrystal cube at the final simulation time of 19.77 $\mu$s. The surface displays the contour of von Mises stress in units of MPa.}
    \label{fig:singlepolycontour}
\end{figure}

\begin{figure}
    \centering
    \includegraphics[width=1.0\linewidth]{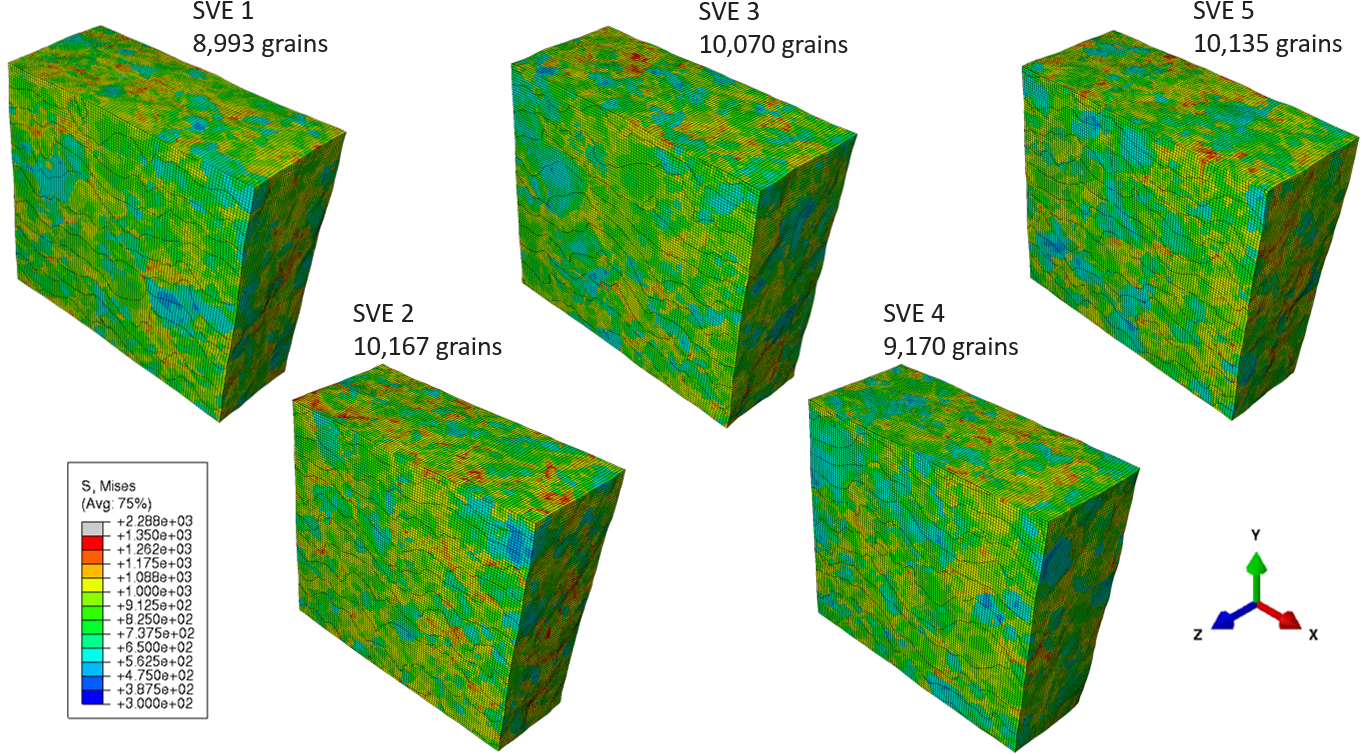}
    \caption{Deformed cross-sectional von Mises stress contour plots of each of the five polycrystal realizations in units of MPa. Each image was taken from the simulation at a time of 19.77 $\mu$s. }
    \label{fig:Mises5}
\end{figure}

\begin{figure}
    \centering
    \includegraphics[width=0.7\linewidth]{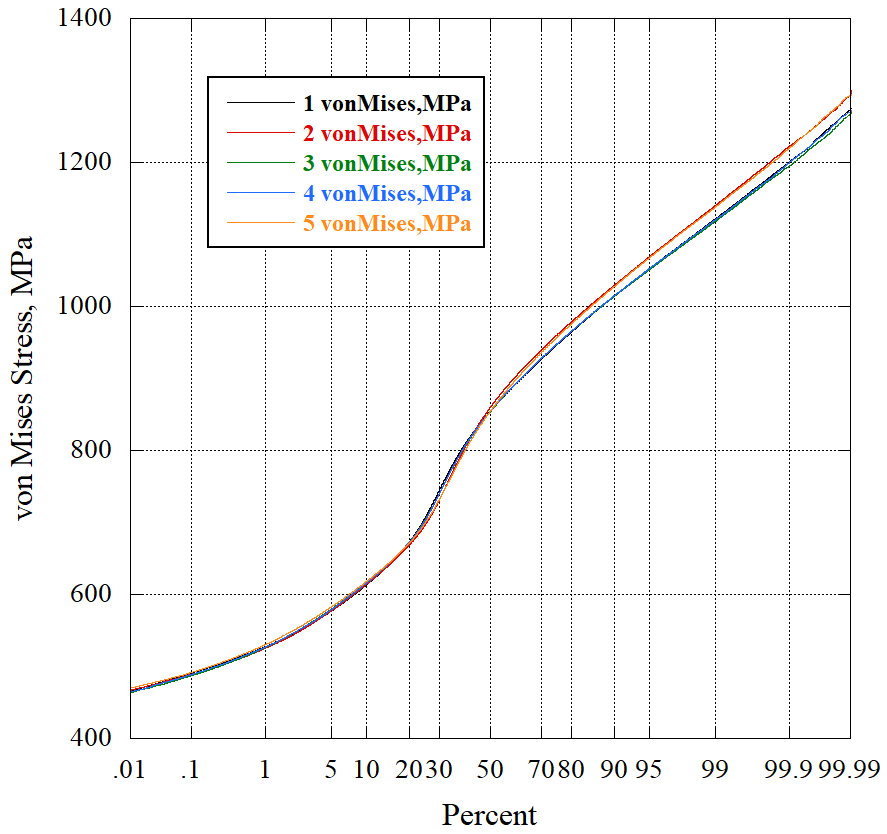}
    \caption{Cumulative probability distribution function for von Mises stress for all integration points within all five SVE simulations at a time of 19.77 $\mu$s.}
    \label{fig:5MisesCPDF}
\end{figure}

At the final simulation time of 19.77 $\mu$s, a deformed polycrystal cube showing spatial distribution of von Mises stress is provided in Fig.~\ref{fig:singlepolycontour}. Note the significant spatial variability in von Mises stress state over the entire cube surface. The cross-sectional von Mises stress contour for each of the five polycrystal realizations is provided in Fig.~\ref{fig:Mises5} at a final simulation time of 19.77 $\mu$s. The comparison between von Mises stress cumulative probability distribution functions for each of the five polycrystal realization simulations is provided in Fig.~\ref{fig:5MisesCPDF}. At this final simulation time, the differences between each simulation are modest but noticeable. Other state variables may realize greater differences in statistical response as it is not known a priori the size necessary to reach a statistically relevant result from a single simulation. In general, when the interest is damage as an extreme event process, the interest is in accurately capturing tails of distributions which is far less well understood.

Note that five polycrystal realization simulations were conducted and presented here despite analysis conducted on only the first four computational datasets. There is no deliberate reason for not conducting analysis on all five datasets as at the start of this study, five polycrystal realizations were constructed and five datasets produced. This complete dataset is presented in its entirety here to enable community access. As the results presented in Fig.~\ref{fig:5MisesCPDF} attest, the first four datasets well represent the physics and statistics of interest in presenting the new mathematical methodologies as is the primary goal of this work.\color{black}

\subsection{PDFs of State Variables}
The simulations described above compute many material state variables. Here we limited our examination to von Mises stress, dislocation density, and equivalent plastic strain rate at discrete time steps of 0.3295 microseconds, covering a total simulation duration of 19.77 microseconds (i.e., 60 time steps). Each state variable was first normalized to the range of $[0, 1]$ to ensure comparability and numerical stability across the analysis. We then computed the PDFs of each normalized variable over the entire spatial domain at each time step. As such, we obtained four sets of time-dependent PDFs, denoted as Datasets 1-4, respectively, corresponding to the SVE 1 through SVE 4 polycrystal realizations shown in Figs. \ref{fig:Poly_Reals} and \ref{fig:Mises5}.

\begin{figure}[H]
    \centering
  
        \includegraphics[width=1\linewidth]{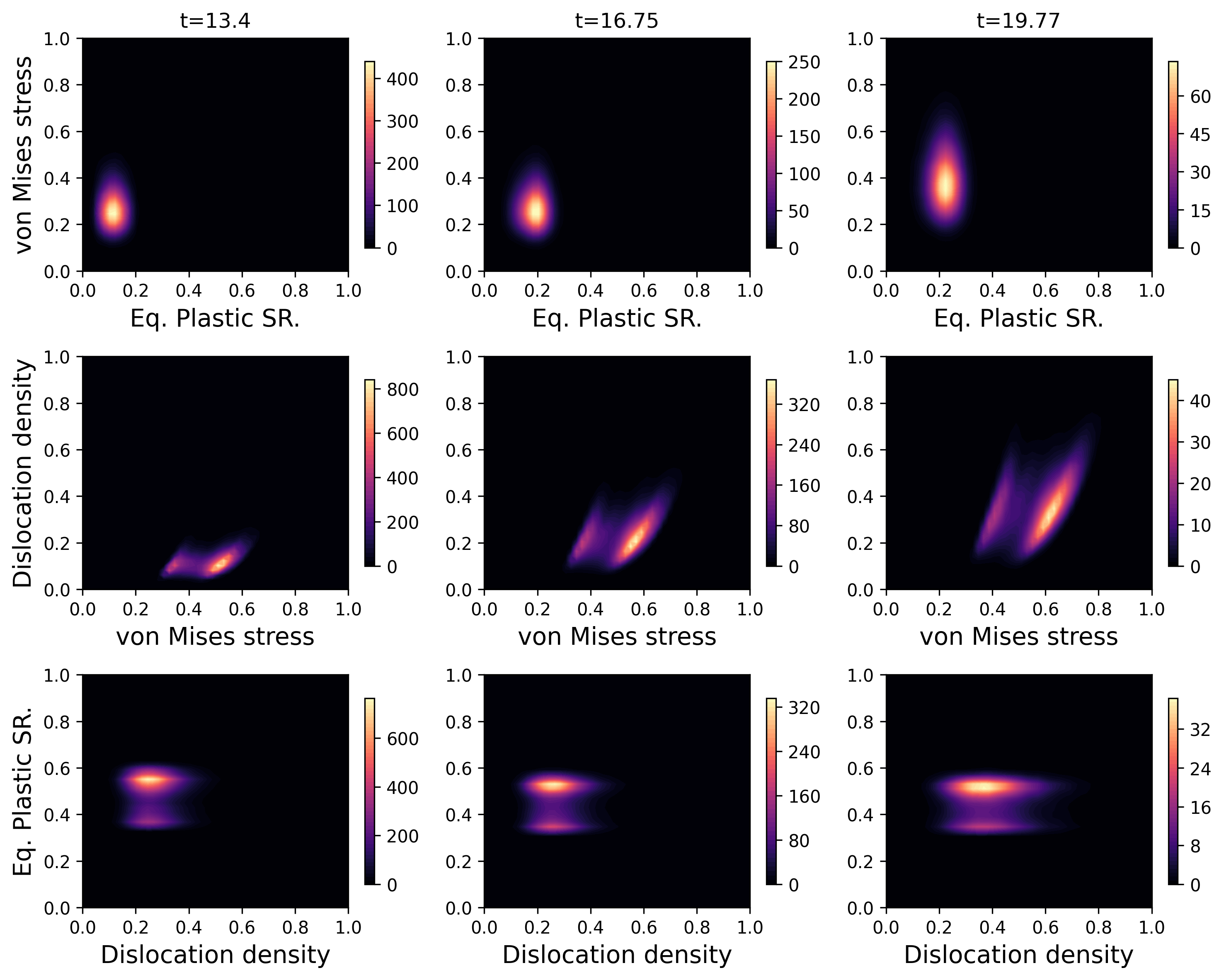}
    \caption{The joint PDF evolution of dataset $\mathcal{D}^1$ used for training SPLIT-PINN at three representative time instants: 
$t = 9.88~\mu\text{s}$, $t \approx 12.19~\mu\text{s}$, and $t = 19.77~\mu\text{s}$. 
Each row corresponds to a 2D cross-section of the joint PDF fixing one state variable (from top to bottom rows: the PDFs with dislocation density=$0.18$, equivalent plastic strain rate =$0.07$, and von Mises stress= $0.1$, respectively). Time is reported in milliseconds $\mu s$.}
    \label{fig:PDF_over_time}
\end{figure}

Next, we split the datasets: we use one dataset, denoted by $\mathcal{D}^1$, for training SPLIT-PINN to infer the drift term in the Liouville model and subsequently apply the learned model to the other three datasets, denoted by $\mathcal{D}^2$--$\mathcal{D}^4$, to predict the time evolution of their joint PDFs. In Fig.~\ref{fig:PDF_over_time}, we show the joint PDF evolution of the dataset $\mathcal{D}^1$ at three representative time instances: $t = 13.4~\mu$s, $t \approx 16.75~\mu$s, and $t = 19.77~\mu$s.

\subsection{Learning the probabilistic model via SPLIT-PINN }

\begin{figure}[H]
   \centering
\includegraphics[width=\linewidth]{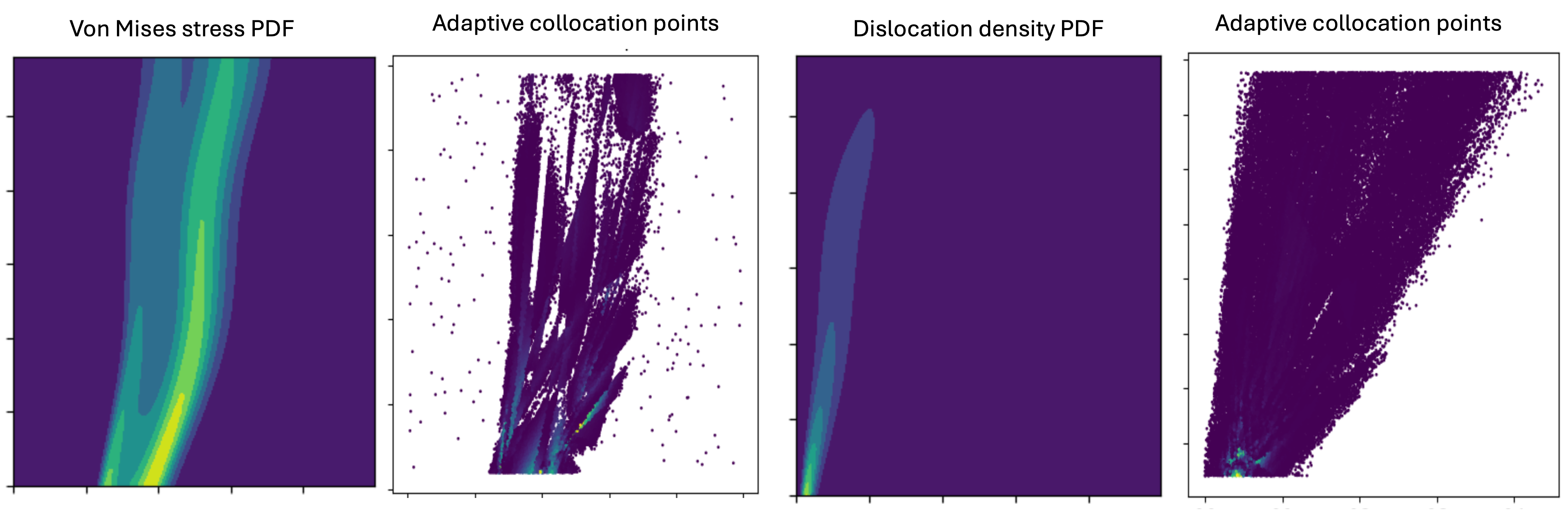}
                           
                        \caption{Adaptive redistribution of collocation points guided by the marginal PDFs of the von Mises stress and dislocation density. RAD strategy concentrates points in regions of high probability and sharp gradients, improving the resolution of the evolving statistical structure in space and time.}
                             \label{fig:RAD}
\end{figure}

Using $\mathcal{D}^1$ dataset of the measured PDF evolution as the training data, we apply the SPLIT-PINN framework to learn a statistical model, in the form of Liouville equation \eqref{eq:liouville}, that describes the temporal evolution of the joint PDF of von Mises stress, dislocation density, and equivalent plastic strain rate. As explained in \S\ref{sec:SPLIT-PINN}, we first consider the marginal PDFs of each state variable, i.e., infer the marginal drift field. Given the localized behaviour of PDFs with sharp interfaces, RAD plays an essential role during the training. Figure~\ref{fig:RAD} shows the adaptive redistribution of collocation points from an initially uniform space–time sampling toward a distribution that conforms to the dominant structure of the PDF. 

Then, we learn the correction drift term to recover the full three-dimensional (3D) drift field. The resulting 3D Liouville PDE is subsequently employed to predict the evolution of the 3D joint PDFs for Datasets~$\mathcal{D}^2$--$\mathcal{D}^4$. This process enables systematic assessments of both the inference accuracy of the learned drift fields and their ability to generalize across datasets at each stage of the SPLIT-PINN framework.

\subsection{Prediction of joint PDF evolution}\label{subsec:results_jointPDF}

In the final stage of the SPLIT-PINN framework, we infer the correction drift term and thereby recover the complete three-dimensional drift field  $\mathbf{A}$. This yields the full 3D Liouville equation governing the temporal evolution of the joint PDF of von Mises stress, dislocation density, and equivalent plastic strain rate for dataset $\mathcal{D}^1$. Given the initial joint PDF from a new polycrystal realization $\mathcal{D}^2$--$\mathcal{D}^4$, we compute its time evolution by numerically solving the resulting Liouville PDE. To suppress nonphysical oscillations, we discretize all spatial dimensions using an upwind flux scheme~(\cite{leveque2002finite}), while time integration is performed with a fourth-order Runge-Kutta (RK4) method~(\cite{butcher2007runge}). To efficiently treat the multidimensional transport structure, we further employ the Strang operator-splitting method~(\cite{strang1968construction}). Additional details of the discretization are provided in \ref{app:discretization}. Zero Dirichlet boundary conditions are imposed at the end of each time step to ensure that the predicted probability distribution remains bounded and physically consistent.

\begin{figure}[H]
   \centering
\includegraphics[width=\linewidth]{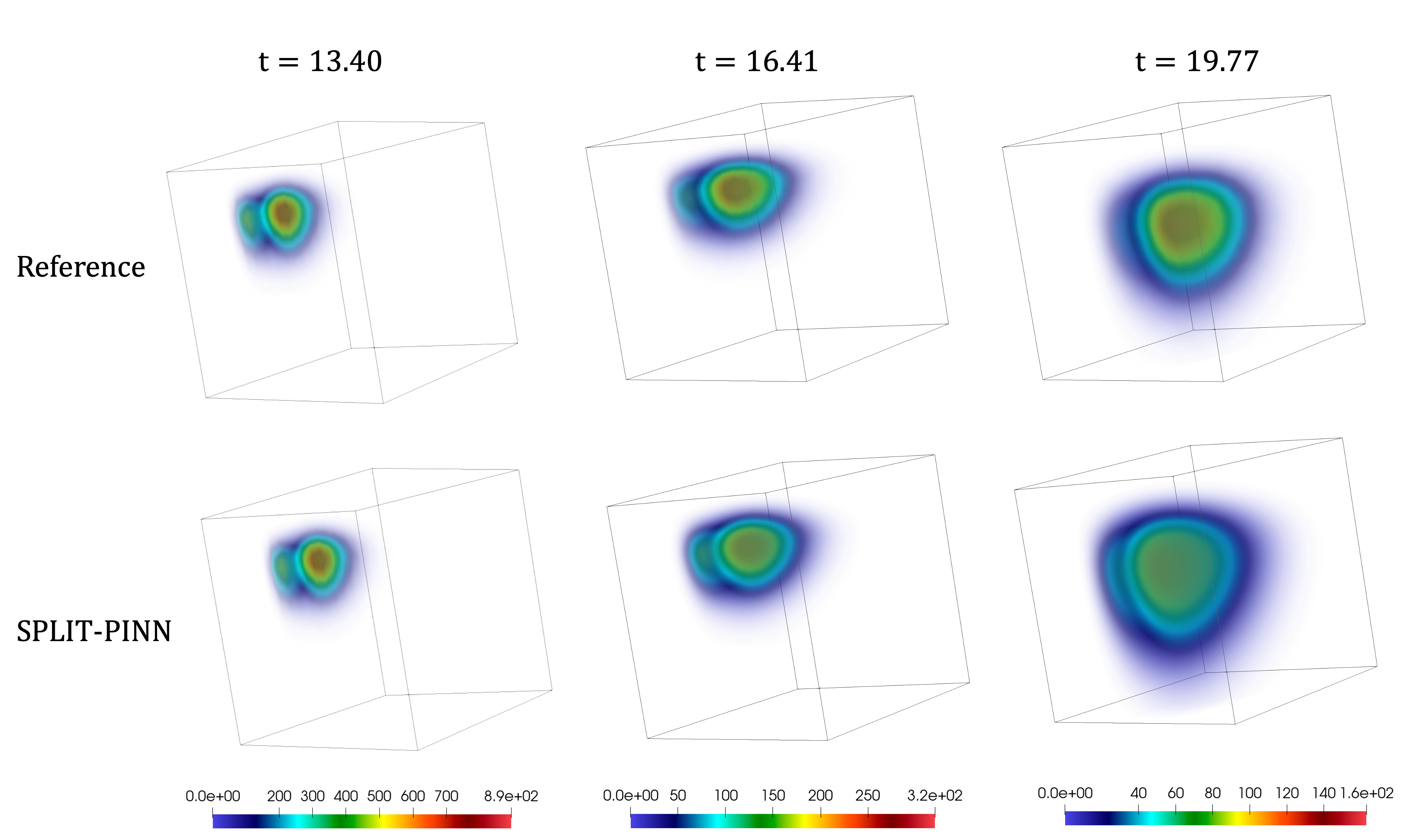}

                        \caption{Comparison of the reference and predicted 3D joint PDFs for Dataset~$\mathcal{D}^2$ at times $13.40$, $16.75$, and $19.77~\mu$s. The predicted PDFs are obtained by forward propagation using the Liouville model along with the drift field inferred from Dataset~$\mathcal{D}^1$.}
                        
                             \label{fig:3D_PDF}
\end{figure}

\begin{figure}[H]
   \centering
\includegraphics[width=\linewidth]{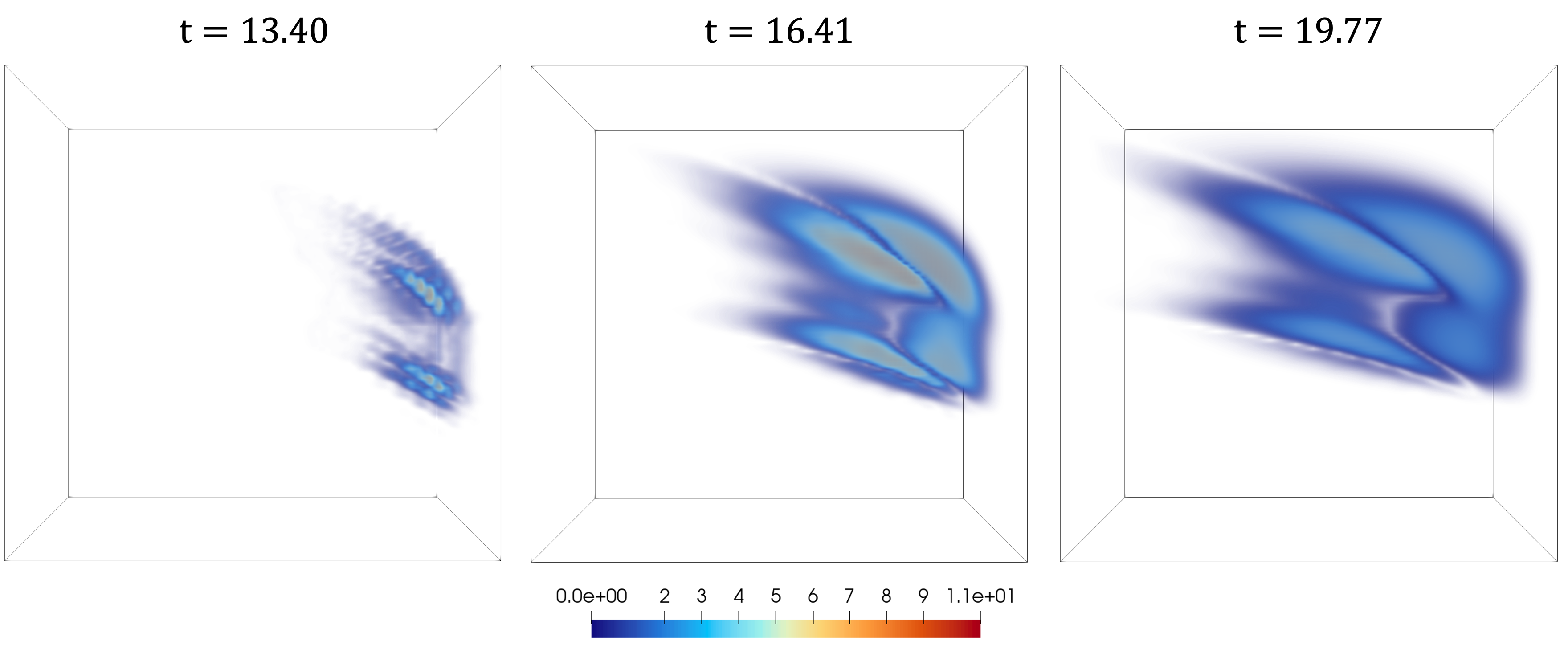}

                          \caption{Absolute pointwise error between the predicted and reference 3D joint PDFs for Dataset~$\mathcal{D}^2$ at times $13.40$, $16.75$, and $19.77~\mu$s. The bounded error values indicate that the learned drift field can effectively predict the evolution of the PDF for new polycrystal instances.}
                          
                             \label{fig:3D_PDF_err}
\end{figure}

\begin{figure}[H]
	\centering
	\includegraphics[width=\linewidth]{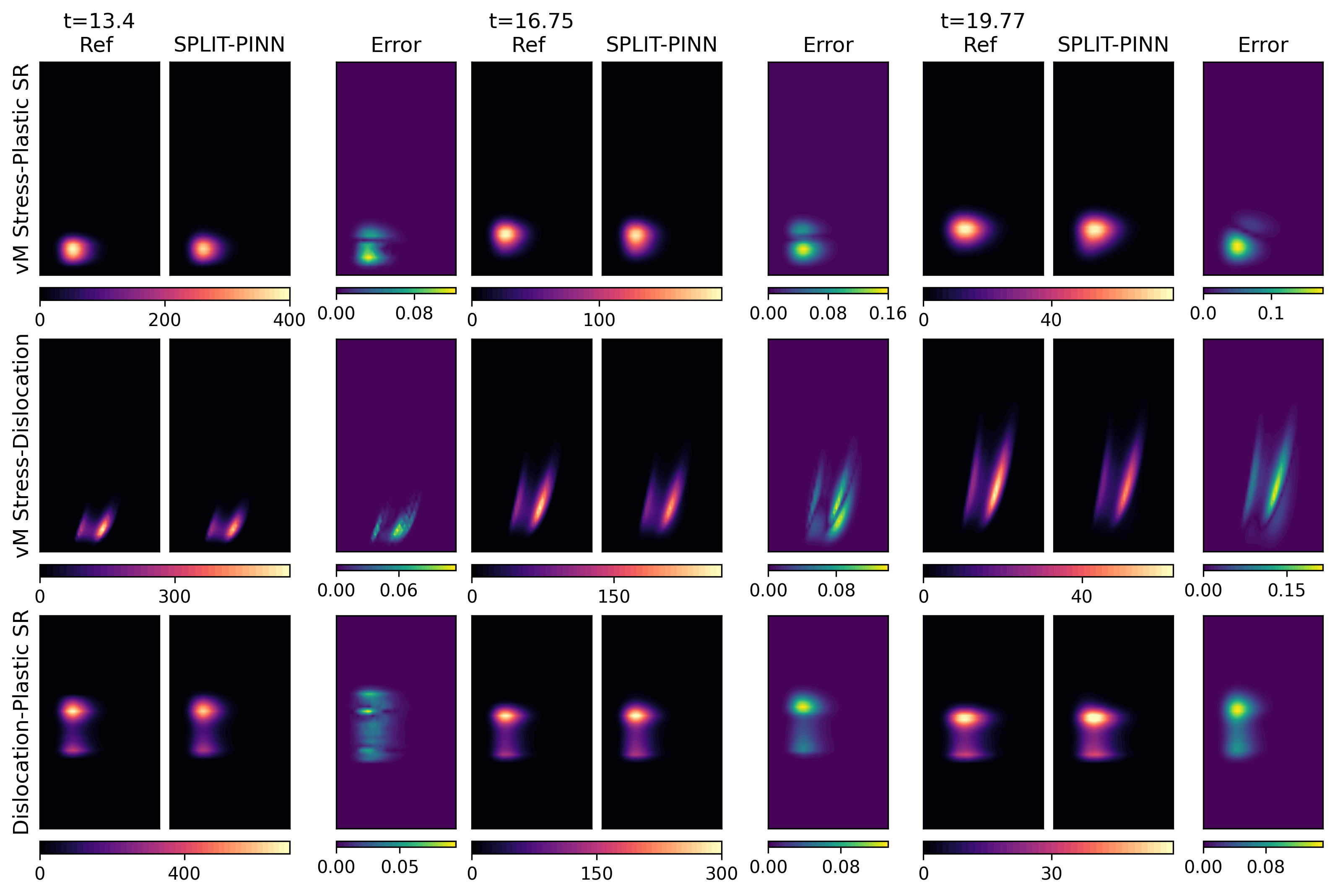}
	
\caption{2D cross-sectional views of the predicted joint PDF for Dataset~$\mathcal{D}^2$ at times $13.40$, $16.75$, and $19.77~\mu$s along with the corresponding relative pointwise error compared to the reference PDFs. The results highlight that the learned drift field along with the Liouville model accurately captures the main features of the PDF.}
	
	\label{fig:3D_cross_2}
\end{figure}

We compare the SPLIT-PINN results against the true PDF corresponding to $\mathcal{D}^2$ and present the absolute pointwise error in Figs.~\ref{fig:3D_PDF} and~\ref{fig:3D_PDF_err}, respectively. SPLIT-PINN can accurately capture the PDF's overall shape during prediction. During the early and intermediate time ranges, the SPLIT-PINN's predictions are very close to the ground truth. For a more detailed examination of the evolution of the estimated PDF, Fig.~\ref{fig:3D_cross_2} provides 2D cross sections of the reference and predicted PDF of $\mathcal{D}^2$ as well as the relative pointwise error. 

Bounding the maximum relative error approximately 20 \% even for longer times (here, 19.77 microsecond) shows the method's reliability in generalization given the error stems from several inherent challenges. First, we infer a high-dimensional (3D) drift field from a single training sample and then propagate it forward for a different polycrystal realization, which constitutes a form of operator learning with extrapolation. Approximation and generalization errors in PINNs can arise due to limited training data, finite network capacity, and optimization residuals, all of which contribute to discrepancies when the learned model is applied outside its training distribution. Theoretical analyses have shown that the generalization error of PINNs depends on the training error, the number of collocation points, and the stability of the underlying PDE, and that controlling the training loss can bound the solution error under suitable conditions (\cite{mishra2022estimates, mishra2023estimates}). Moreover, recent work on PINN extrapolation behavior highlights that predictive accuracy may degrade when the spectral characteristics of the solution shift beyond the training regime, a phenomenon that can occur in time-varying joint PDFs (\cite{fesser2023understanding}). In addition, numerical discretization error from solving the Liouville equation, even with upwind flux schemes and high-order time integrators, accumulates over time and interacts with the model error, particularly in multi-dimensional settings. Taken together, these factors make a maximum ~20 \% relative pointwise error acceptable in many practical scientific machine learning applications: it indicates that the learned drift captures the dominant dynamical behavior, while local discrepancies reflect intrinsic model transfer and numerical propagation limitations.

Following a similar approach, we extend the generality of the SPLIT-PINN framework by predicting the temporal evolution of the joint PDFs for Datasets~$\mathcal{D}^3$ and $\mathcal{D}^4$. In this case, the 3D Liouville equation with the drift field inferred from Dataset~$\mathcal{D}^1$ is solved, subject to the respective initial condition from each dataset. The results are presented as 2D cross-sectional slices of the joint PDF, enabling clearer visualization of the key correlations among von Mises stress, dislocation density, and equivalent plastic strain rate. Across all slices, the predicted PDFs show a good agreement with the ground truth, demonstrating that the Liouville model with the learned drift field generalizes effectively and can reliably capture the dominant probabilistic microstructural state evolution in new, unseen polycrystal realizations.

\begin{figure}[H]
	\centering
	\includegraphics[width=\linewidth]{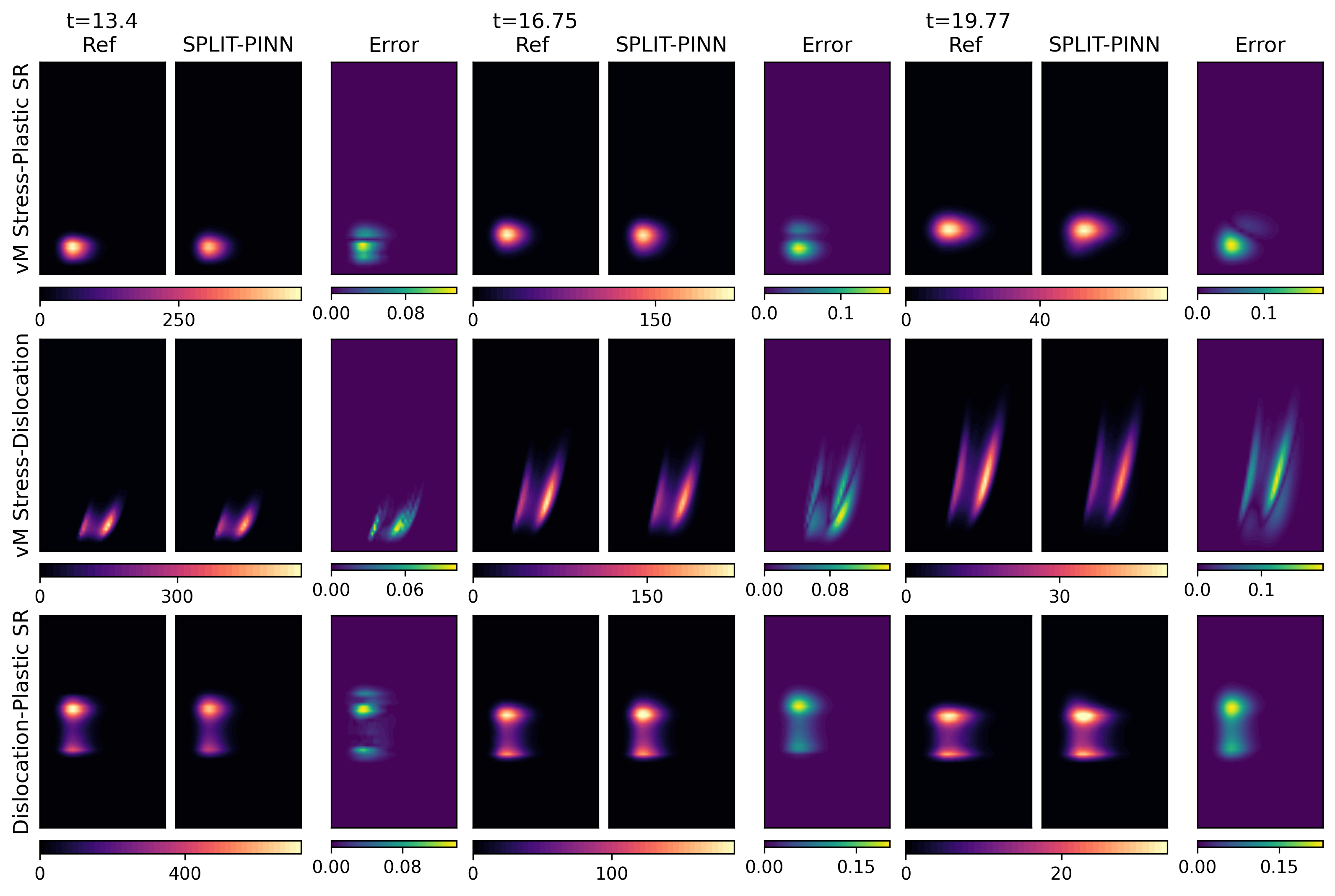}
	
\caption{2D cross-sectional slices of the predicted joint PDF for Dataset~$\mathcal{D}^3$ at times $13.40$, $16.75$, and $19.77~\mu$s, together with the relative pointwise error with respect to the reference PDFs. The predicted PDFs accurately reproduce the key probabilistic features, while the largest discrepancies remain confined.}

	\label{fig:3D_cross_3}
\end{figure}

In addition to pointwise errors, we quantify the discrepancy between predicted and reference joint PDFs using the Kullback--Leibler (KL) divergence, defined as
\begin{equation}
	\mathrm{KL}(P_\mathrm{ref} \,||\, P_\mathrm{pred}) = \sum_i P_\mathrm{ref}(x_i) \, \log \frac{P_\mathrm{ref}(x_i)}{P_\mathrm{pred}(x_i)},
\end{equation}
where $P_\mathrm{ref}$ and $P_\mathrm{pred}$ denote the reference and predicted PDFs, respectively, evaluated at the discretized grid points $x_i$. The KL divergence quantifies the overall difference in the probability mass between the predicted and reference distributions and is particularly informative for high-dimensional PDFs.  

For Dataset~$\mathcal{D}^2$, the KL divergence increases from 0.043 at the early time ($13.4\,\mu \,s$) to 0.243 at the intermediate time ($16.75\,\mu \,s$), and 0.432 at the latest time. Similarly, for Dataset~$\mathcal{D}^3$, it grows from 0.041 to 0.239 and 0.450, while for Dataset~$\mathcal{D}^4$, the corresponding values are 0.035, 0.218, and 0.429. These trends indicate robust learning for the diffusion-free, three-dimensional Liouville equation, in which small local errors in the inferred drift field accumulate during forward propagation. Importantly, even at later times, the KL values remain moderate, indicating that the dominant probability mass is accurately captured for all three datasets. The consistency of these results across datasets demonstrates that the drift field learned from Dataset~$\mathcal{D}^1$ generalizes well to new polycrystal realizations, supporting the robustness and reliability of the SPLIT-PINN framework.
%%%%%%%%%%%%%%%%%%%%%%%%
\begin{figure}[H]
	\centering
	\includegraphics[width=\linewidth]{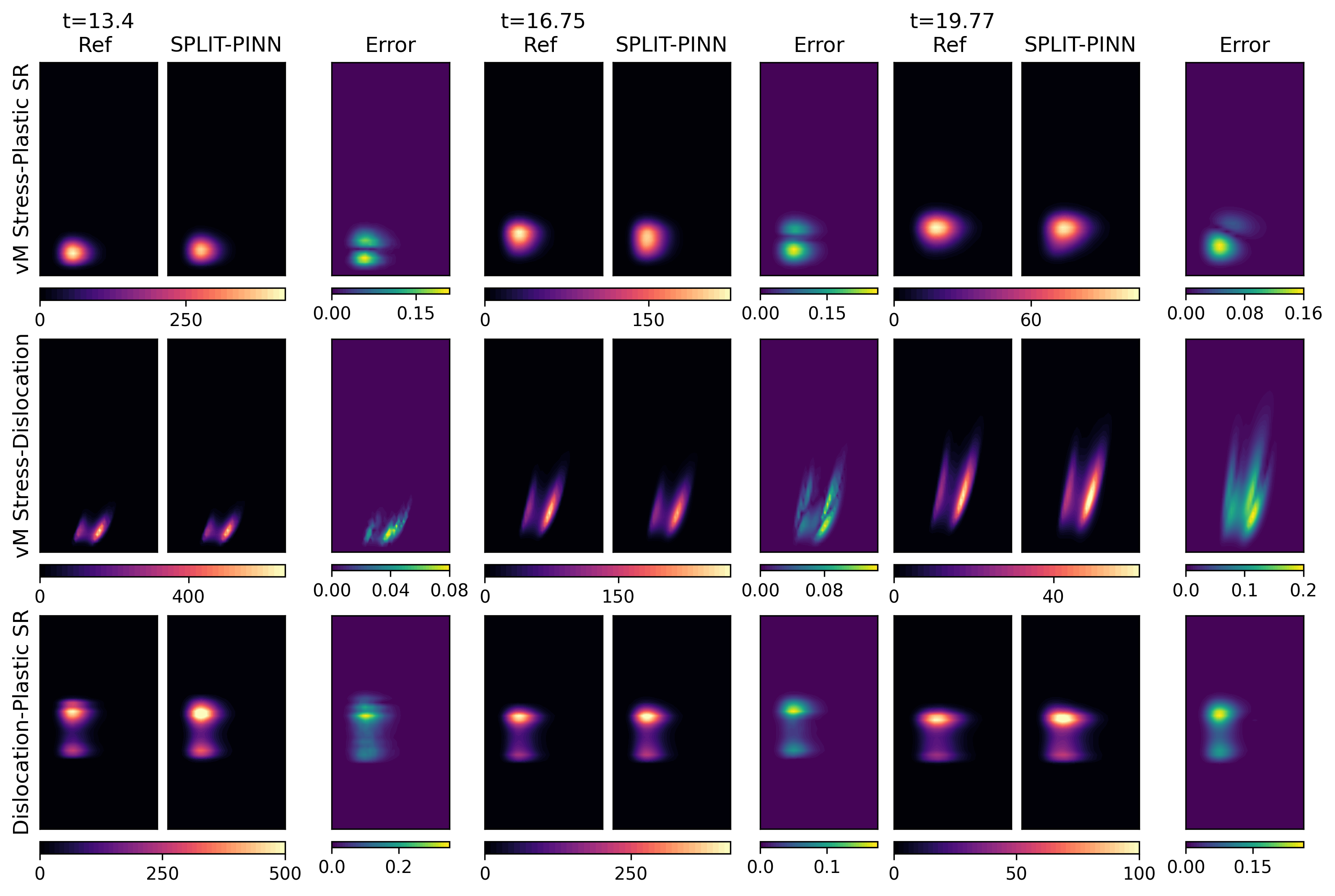}
	
	\caption{2D cross-sectional slices of the predicted joint PDF for Dataset~$\mathcal{D}^4$ at times $13.40$, $16.75$, and $19.77~\mu$s, together with the relative pointwise error compared to the reference PDFs. The results demonstrate that the predicted PDFs capture the dominant probabilistic microstructural state evolution for different polycrystal realizations.}

	\label{fig:3D_cross_4}
\end{figure}
\begin{remark}
	The error in the SPLIT-PINN's results is consistent and overperforming with previously reported PINN generalization behavior for transport-dominated problems.  For example, Mishra and Molinaro report relative generalization errors of approximately 11\% for Burgers’ equation with small viscosity, increasing to about 23\% in the inviscid (pure transport) limit, where diffusion is absent and the problem becomes more challenging (\cite{mishra2023estimates}). The Liouville equation considered here is likewise a hyperbolic conservation law without diffusion, and the present setting further involves a three-dimensional joint PDF and cross-dataset generalization. In this context, a maximum pointwise error on the order of 20\% is consistent with, and even favorable relative to, reported PINN performance for lower-dimensional inviscid problems, indicating that the learned drift field captures the dominant transport dynamics despite the increased dimensionality and transfer across realizations.	
\end{remark}

\begin{figure}[hbt!]
	\centering
    \begin{subfigure}{0.32\textwidth}
        \includegraphics[width=\textwidth]{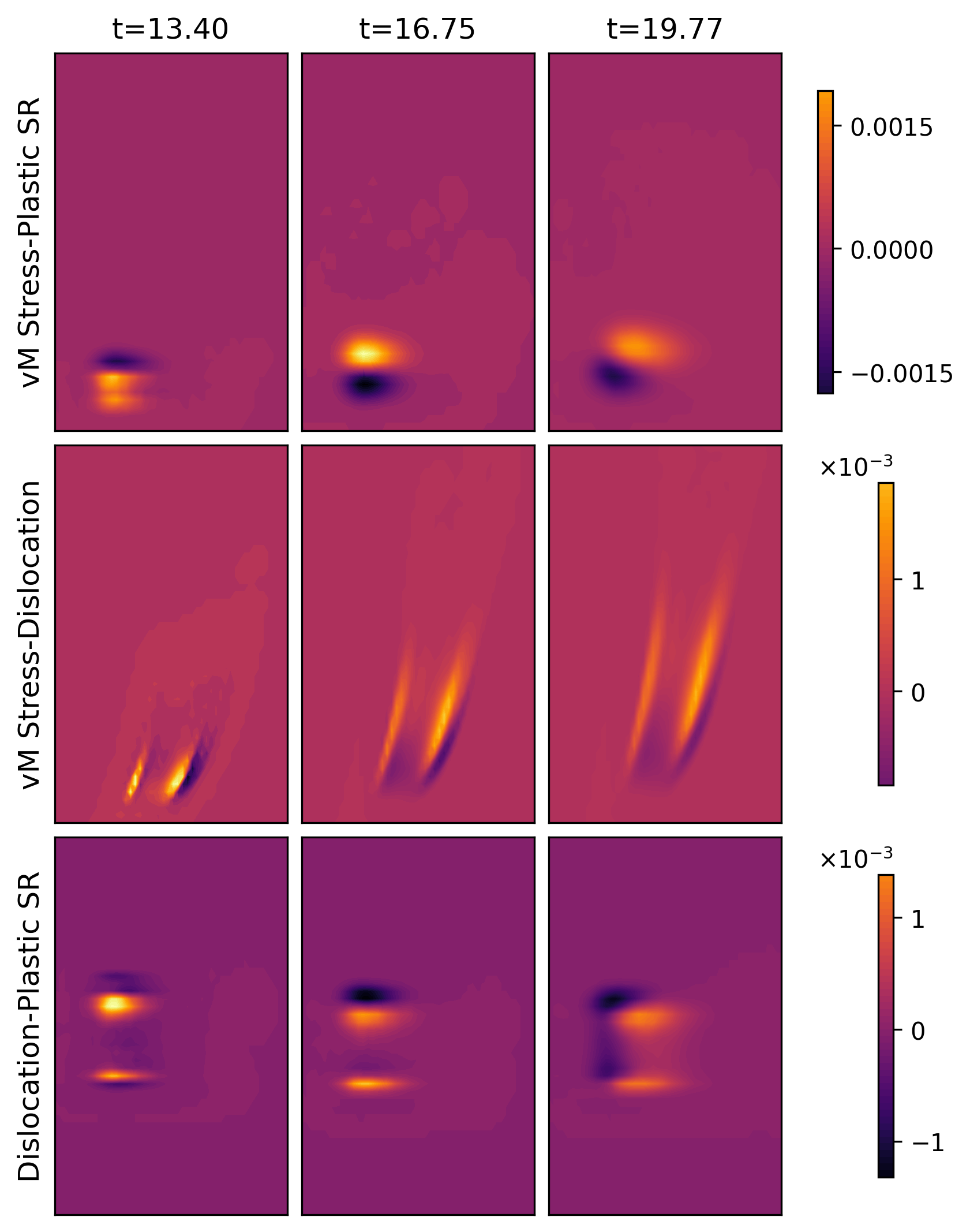}
        \caption{$\mathcal{D}^2$}
    \end{subfigure}
    \hfill
    \begin{subfigure}{0.32\textwidth}
        \includegraphics[width=\textwidth]{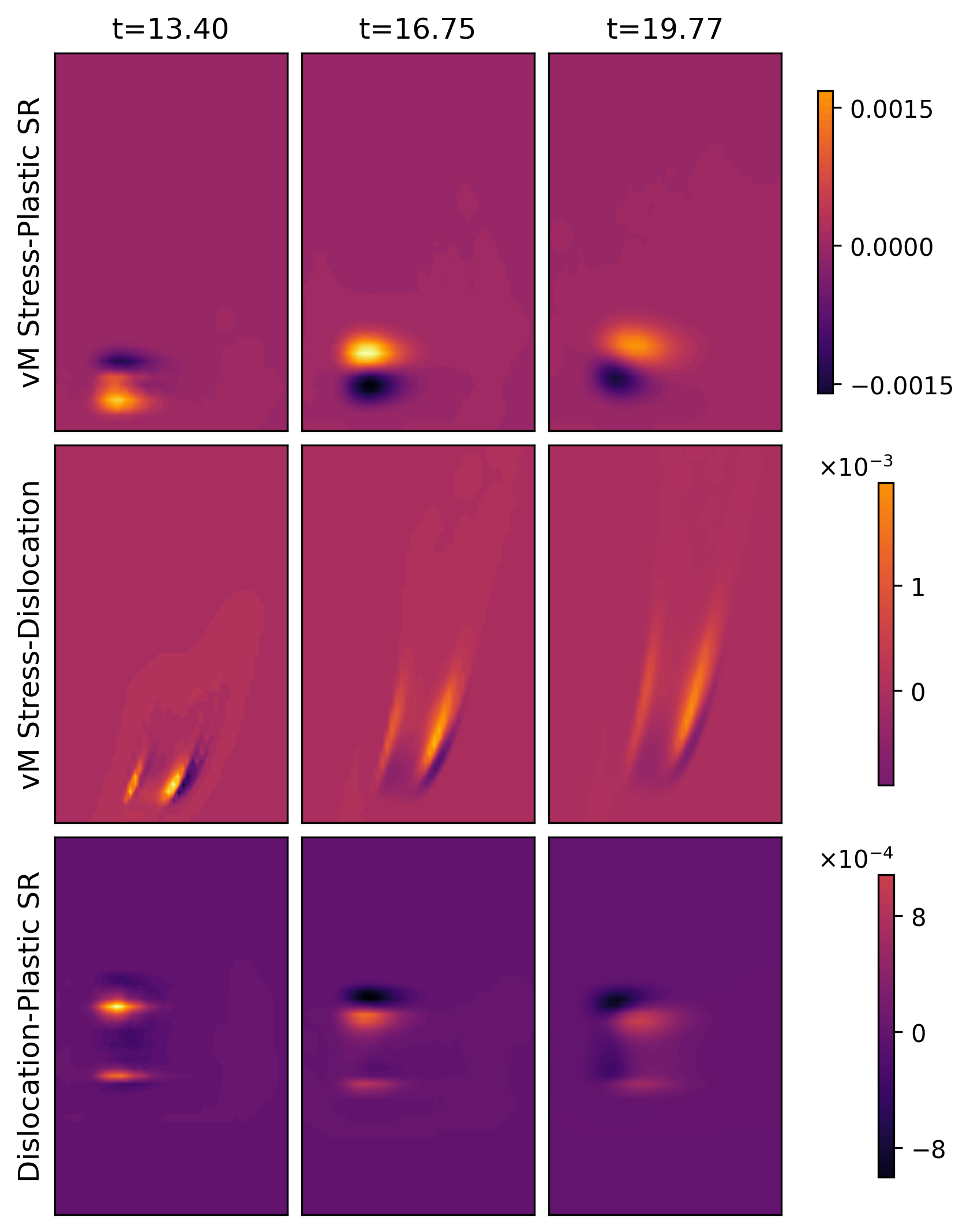}
        \caption{$\mathcal{D}^3$}
    \end{subfigure}
    \hfill
	 \begin{subfigure}{0.32\textwidth}
        \includegraphics[width=\textwidth]{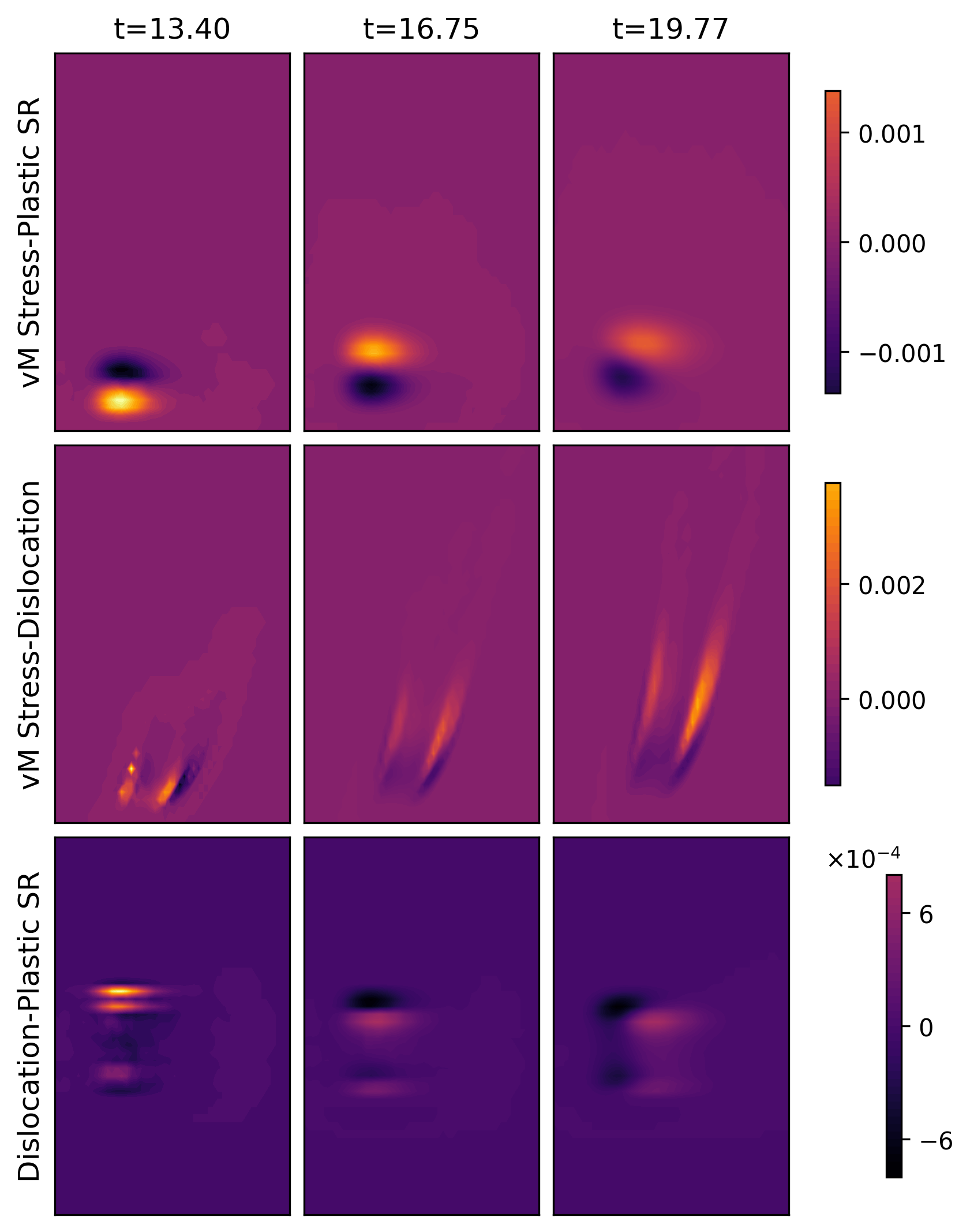}
        \caption{$\mathcal{D}^4$}
    \end{subfigure}
	\caption{\revv{Normalized pointwise KL divergence contours for datasets $\mathcal{D}^2-\mathcal{D}^4$, respectively. The uniformly low magnitude demonstrates small discrepancies at both the distribution modes and tails.}}
	\label{fig:pointKLdiv}
\end{figure}

\revv{
Furthermore, to rigorously quantify and localize the distributional discrepancies, we compute the normalized pointwise KL divergence between the reference distributions and the SPLIT-PINN approximations as:
\begin{equation}
	\mathrm{KL}(P_\mathrm{ref} \,||\, P_\mathrm{pred})_{\text{point-wise}} = P_\mathrm{ref}(x_i) \, \log \frac{P_\mathrm{ref}(x_i)}{P_\mathrm{pred}(x_i)}
\end{equation}
To ensure a mathematically consistent comparison, all 2D probability density slices are normalized to have unit integral prior to evaluation. Fig. \ref{fig:pointKLdiv} shows that the discrepancy magnitude is low across the entire domain for all datasets $\mathcal{D}^2$- $\mathcal{D}^4$. Moreover, these results emphasize that there is no structural bias or localized error spikes. The divergence is negligible both at the distribution modes (the high-density peaks) and within the tail regions of low probability density. This confirms that the SPLIT-PINN framework captures the entire joint probability topology with high spatial and density fidelity, without mischaracterizing the dominant physical states. Since these regions dominate the second-order statistics, the result indicates that discrepancies are not confined to low-probability tails and do not significantly affect the main inter-variable dependency structure.}
%%%%%%%%%%%%%%%
%%%%%%%%%%%%%%%%%%%%%%%

\begin{figure}[H]
	\centering
	\includegraphics[width=.7\linewidth]{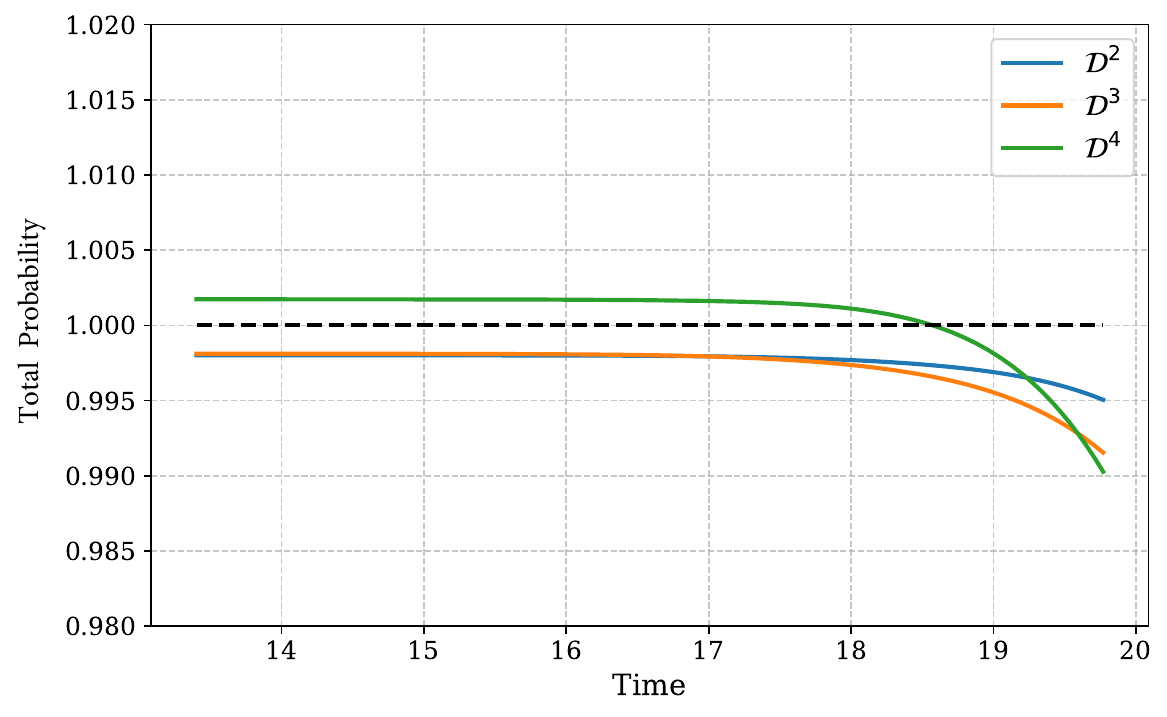}
	
	\caption{\revv{The total probability $M(t) = \int_\Omega P(\mathbf{x},t) d\mathbf{x}$ plotted over time for datasets 
$\mathcal{D}^2$, $\mathcal{D}^3$, and $\mathcal{D}^4$. Values remain close to unity 
throughout the time range considered, confirming that the predicted PDF preserves total probability.}}
\label{fig:conserv}
\end{figure}

\revv{To verify the conservation property of SPLIT-PINN, we evaluate the total probability $M(t) = \int_\Omega P(\mathbf{x},t) d\mathbf{x}$ over time for datasets $\mathcal{D}^2$ through $\mathcal{D}^4$, where a value of $M \approx 1$ indicates 
that the predicted joint PDF integrates to unity. Figure~\ref{fig:conserv} shows 
the evolution of $M$ over time for each dataset. The time-averaged values of $M$ remain close to unity across all datasets, with 
mean values of $0.9976$, $0.9973$, and $1.0005$ and standard deviations of 
$6.63 \times 10^{-4}$, $1.50 \times 10^{-3}$, and $2.44 \times 10^{-3}$ with respect to time for 
$\mathcal{D}^2$, $\mathcal{D}^3$, and $\mathcal{D}^4$, respectively.  These results 
confirm that SPLIT-PINN preserves the total probability with a 
high degree of accuracy.}

\subsection{Comparison of marginal PDFs}

While the joint 3D PDFs provide a complete description of the system's probabilistic state, examining the marginal distributions of each state variable offers additional insight into the performance of the SPLIT-PINN framework. In this section, we extract the 1D marginal PDFs for von Mises stress, dislocation density, and equivalent plastic strain rate from both the predicted and reference 3D joint PDFs. 

\begin{figure}[H]
    \centering
     \begin{subfigure}{\linewidth}
               \includegraphics[width=1\linewidth]{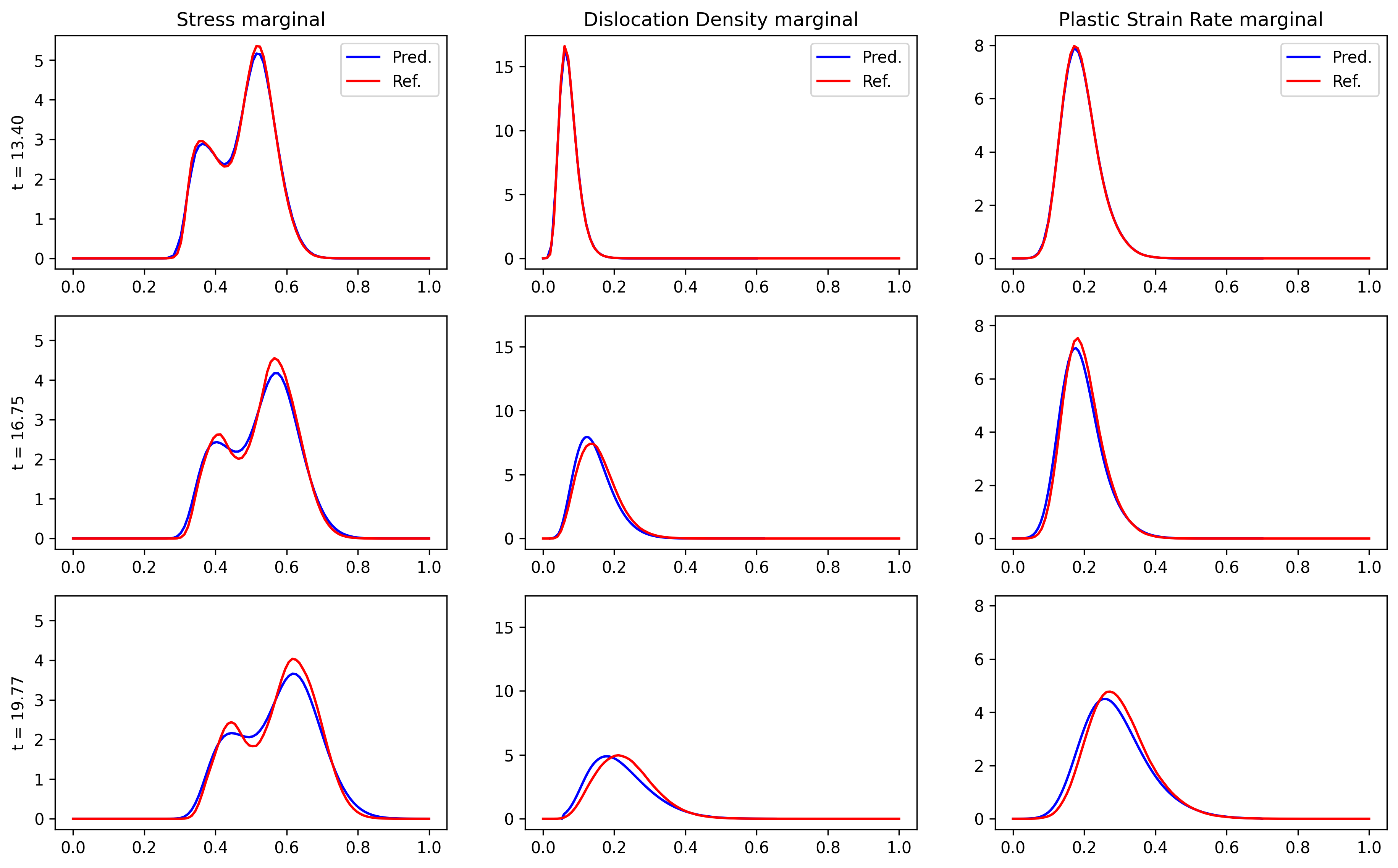}
                         \caption{Temporal evolution of the marginal PDFs for Dataset $\mathcal{D}^1$}
                             \end{subfigure}

    \begin{subfigure}{\linewidth}
               \includegraphics[width=1\linewidth]{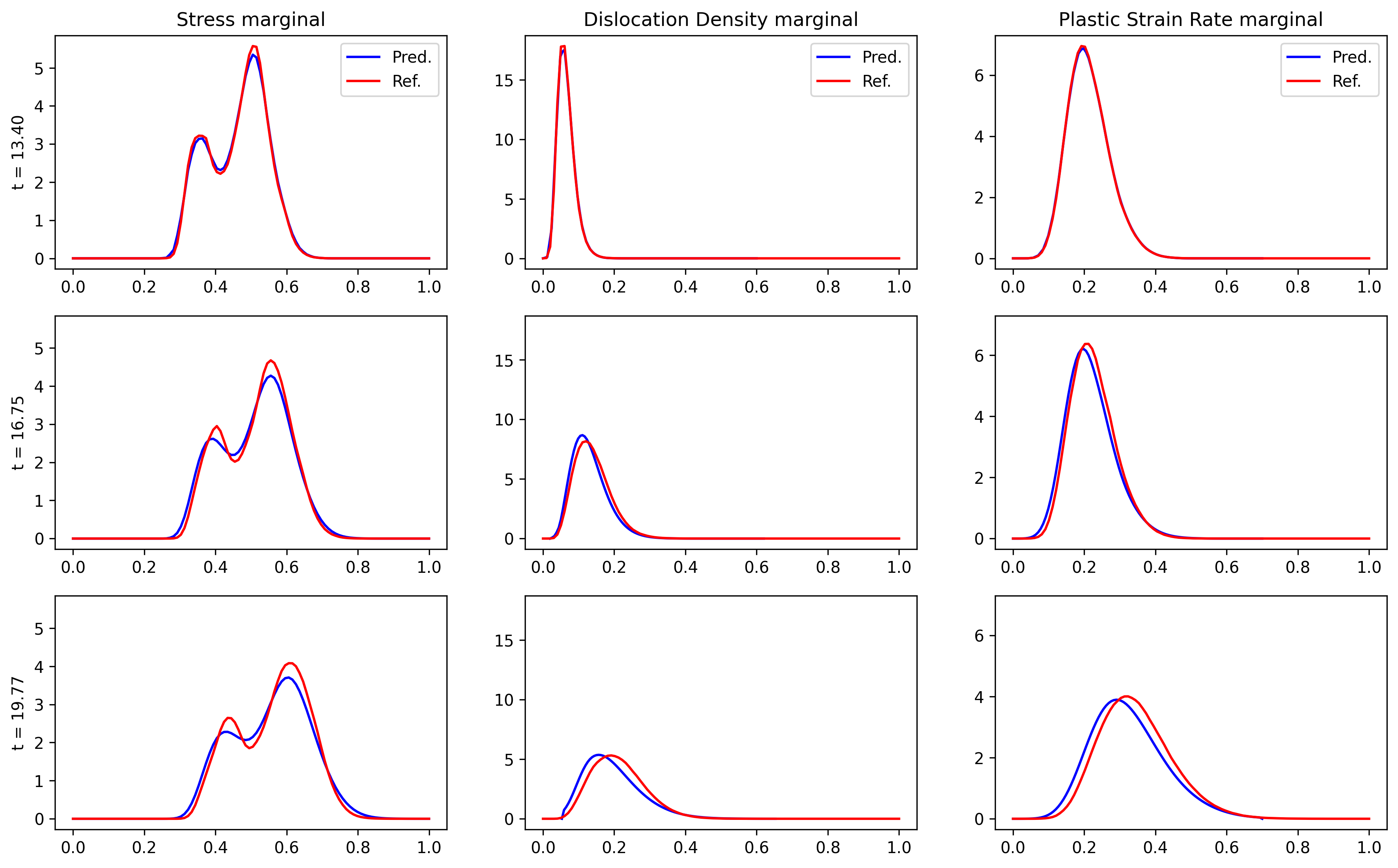}
                         \caption{Temporal evolution of the marginal PDFs for Dataset $\mathcal{D}^2$}
                             \end{subfigure}
                                       
                             \caption{Temporal evolution of one-dimensional marginal PDFs for Datasets $\mathcal{D}^1$ and $\mathcal{D}^2$. Each subfigure contains the three marginal PDFs obtained by integrating the full PDF over the remaining two dimensions. Results from SPLIT-PINN closely follow the corresponding reference PDFs, illustrating the method’s ability to reproduce the underlying probabilistic dynamics across all datasets. Time is reported in milliseconds $\mu s$.}
                             \label{fig:3D_PDF_joint1}
\end{figure}

\begin{figure}
                          \begin{subfigure}{\linewidth}
               \includegraphics[width=1\linewidth]{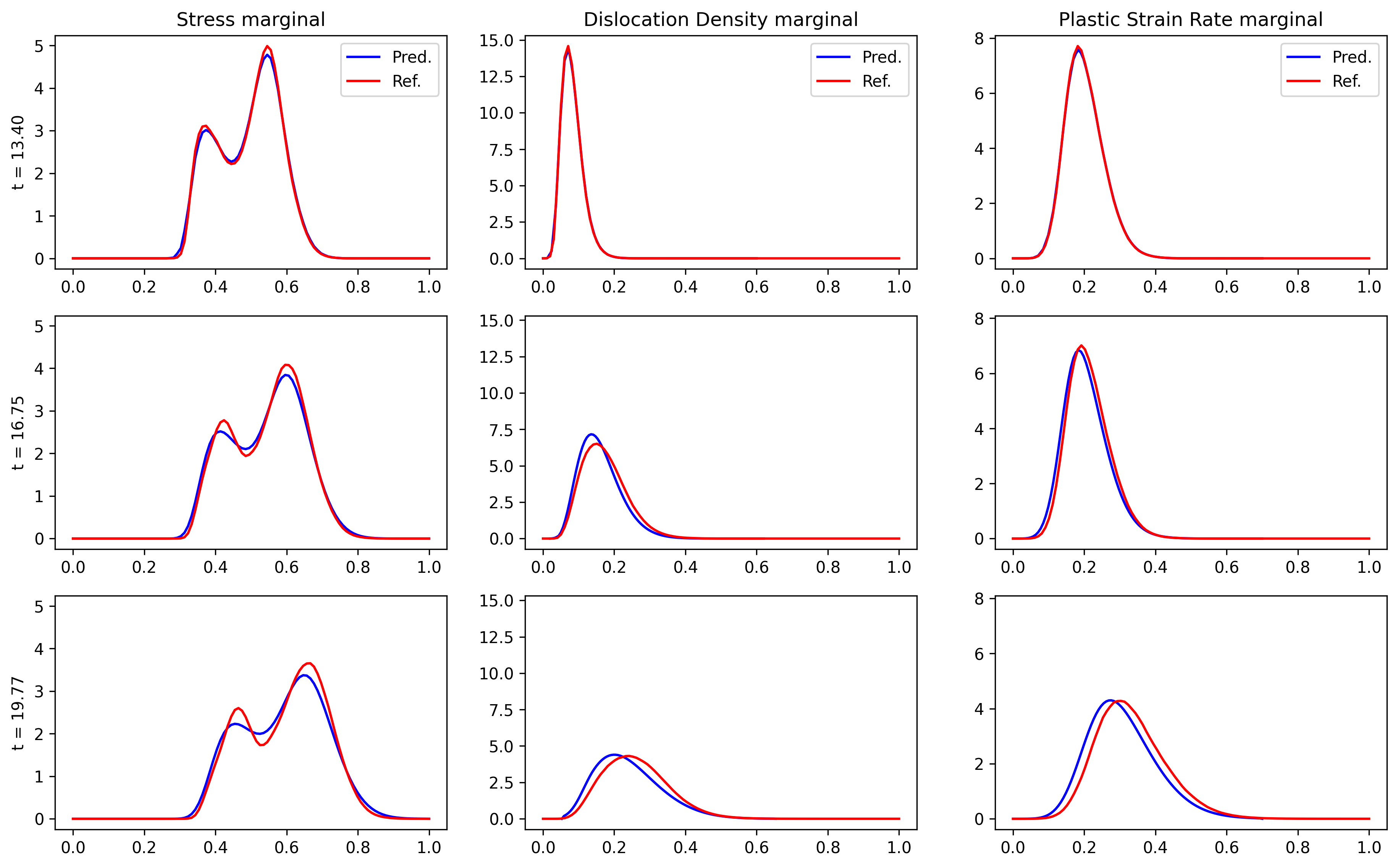}
                         \caption{Temporal evolution of the marginal PDFs for Dataset $\mathcal{D}^3$}
                             \end{subfigure}
                                 \begin{subfigure}{\linewidth}
               \includegraphics[width=1\linewidth]{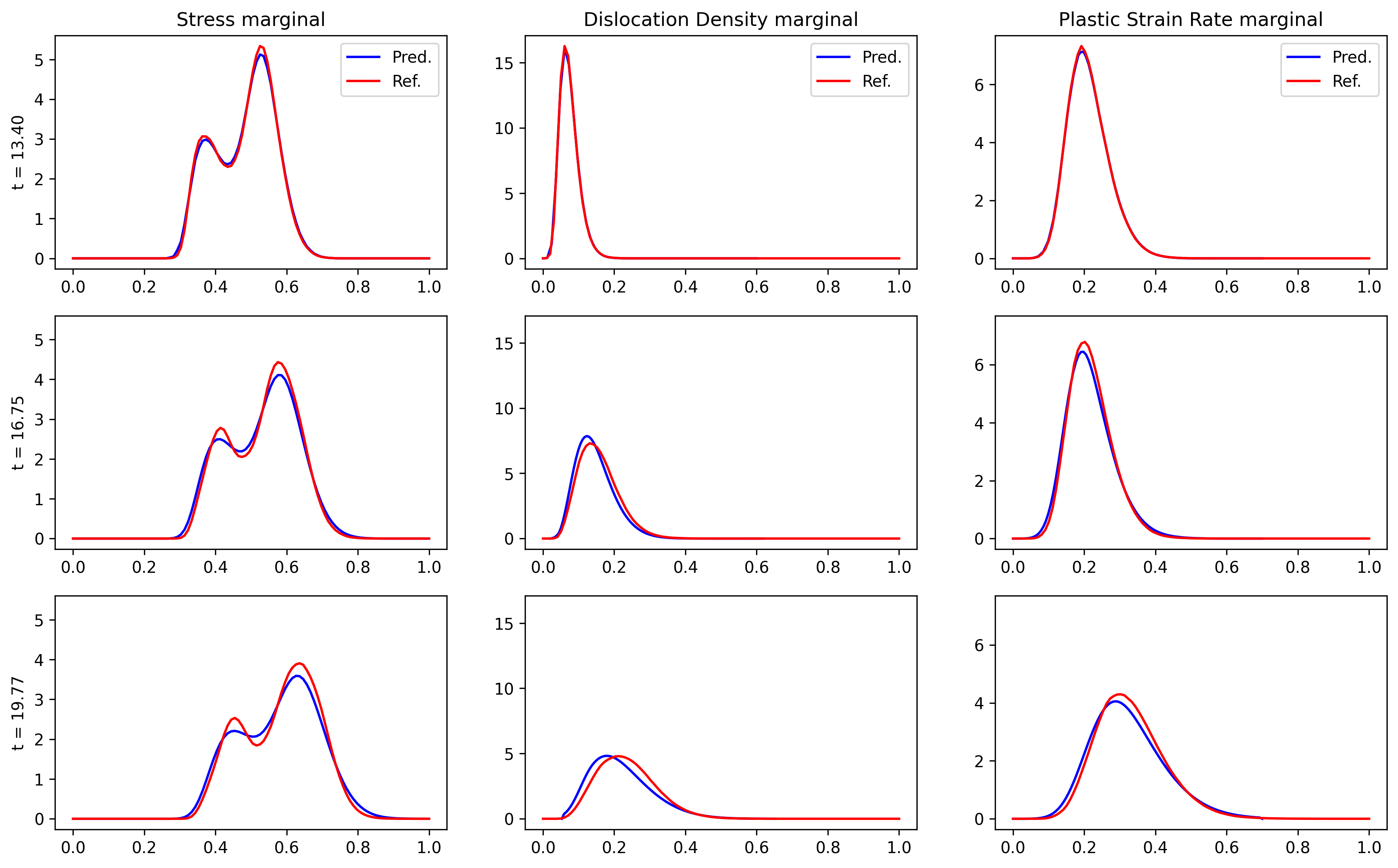}
                         \caption{Temporal evolution of the marginal PDFs for Dataset $\mathcal{D}^4$}
                             \end{subfigure}
                             \caption{Temporal evolution of one-dimensional marginal PDFs for Datasets $\mathcal{D}^3$ and $\mathcal{D}^4$. Similar to Fig.~\ref{fig:3D_PDF_joint1}, the SPLIT-PINN's predictions closely align with the reference PDFs. Time is reported in milliseconds $\mu s$. }
                             \label{fig:3D_PDF_joint2}
\end{figure}

 As shown in Figs.~\ref{fig:3D_PDF_joint1}--\ref{fig:3D_PDF_joint2}, the SPLIT-PINN predictions accurately reproduce both the central regions and the tails of the marginal distributions for all three datasets. This indicates that the learned drift field not only preserves the dominant probability mass in the bulk of the distribution but also faithfully captures the low-probability regions, which are critical for assessing extreme events in polycrystalline deformation behaviors.


\begin{figure}[H]
	\centering
	\includegraphics[width=.32\linewidth]{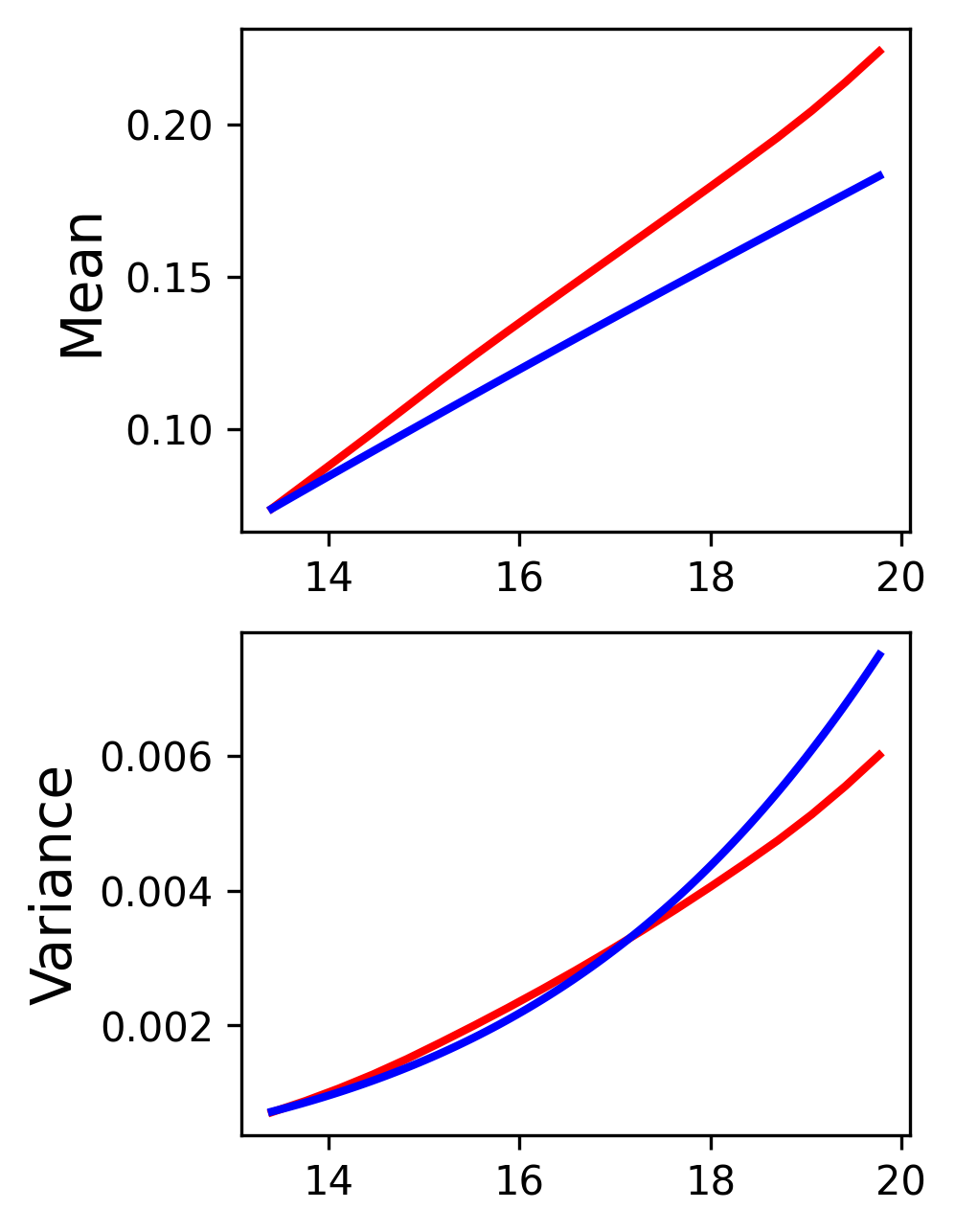}
	\includegraphics[width=.32\linewidth]{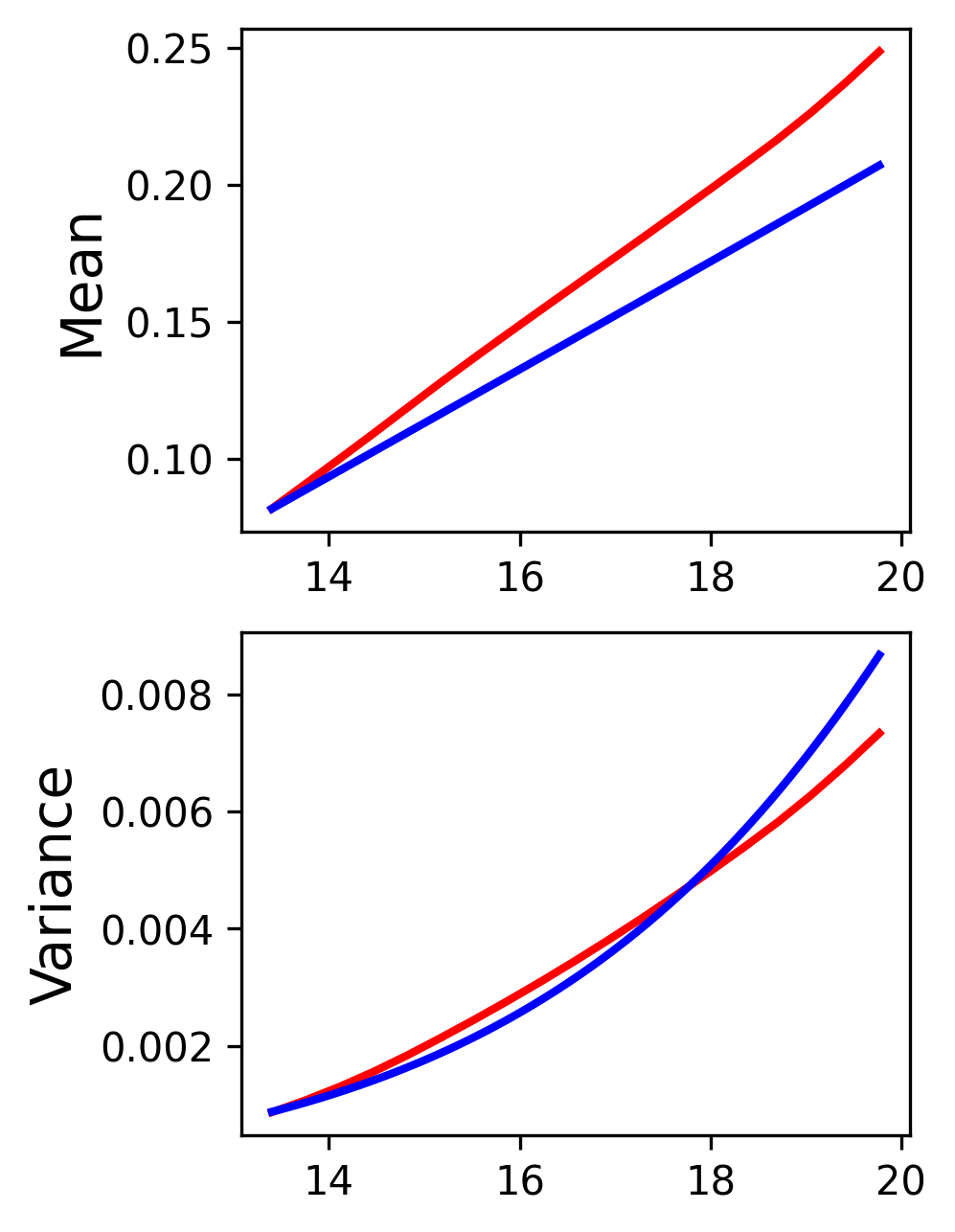}
	\includegraphics[width=.32\linewidth]{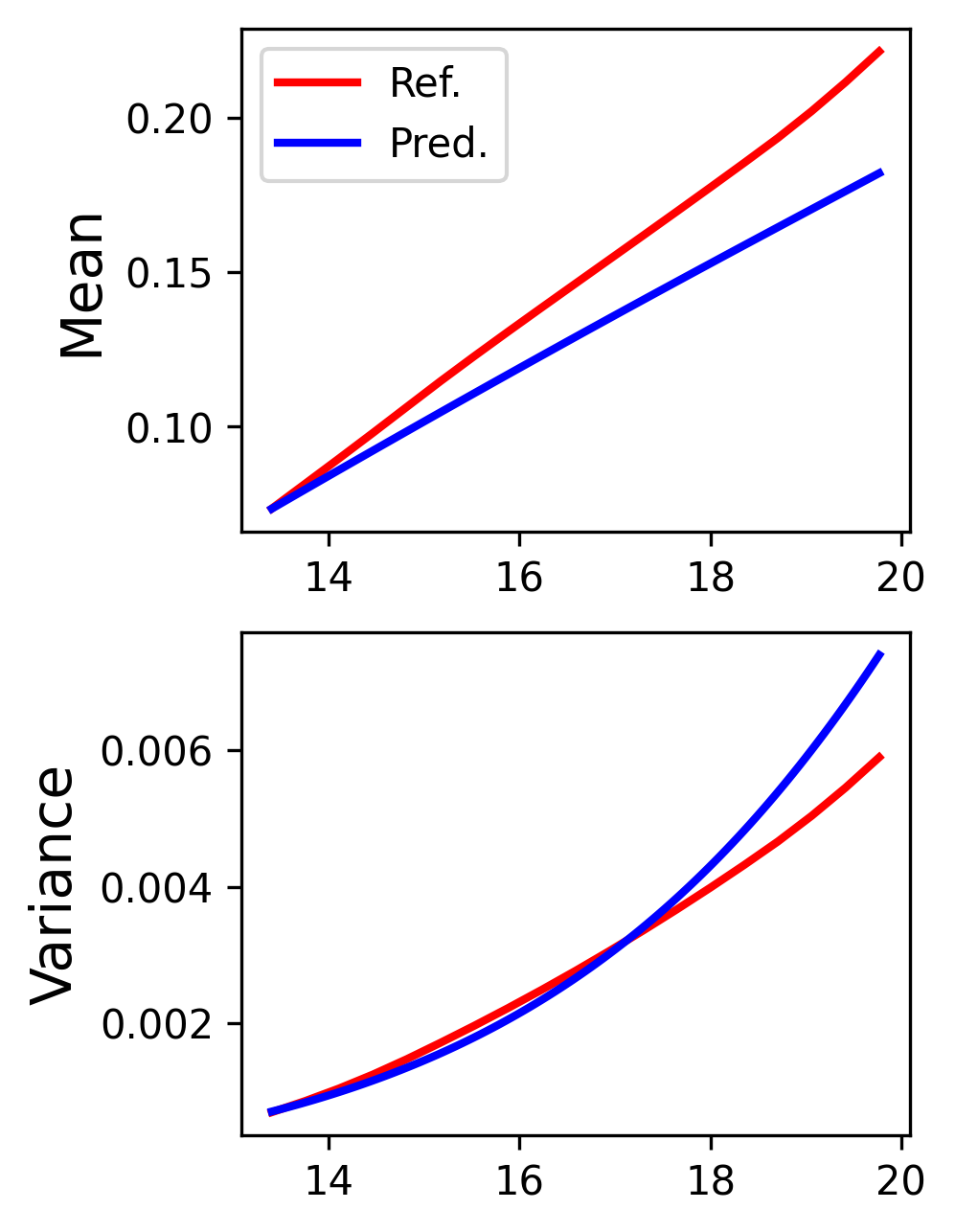}
	
	\caption{Comparison of the first two moments of the predicted PDFs for dislocation density on Datasets $\mathcal{D}^2$, $\mathcal{D}^3$, and $\mathcal{D}^4$.}
	\label{fig:3d:mom_dis}
\end{figure}

\begin{figure}[H]
    \centering
   
         \includegraphics[width=.32\linewidth]{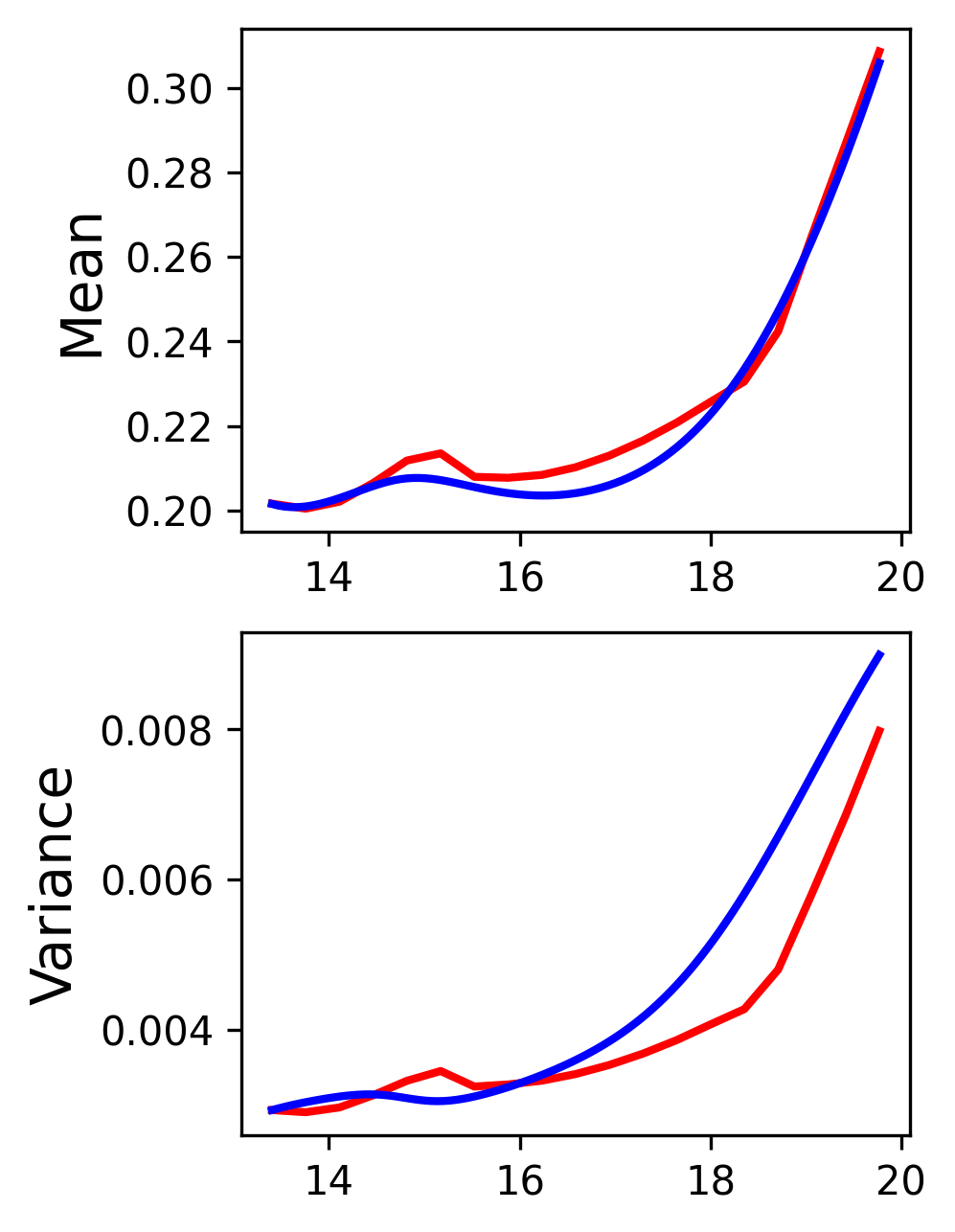}
         \includegraphics[width=.32\linewidth]{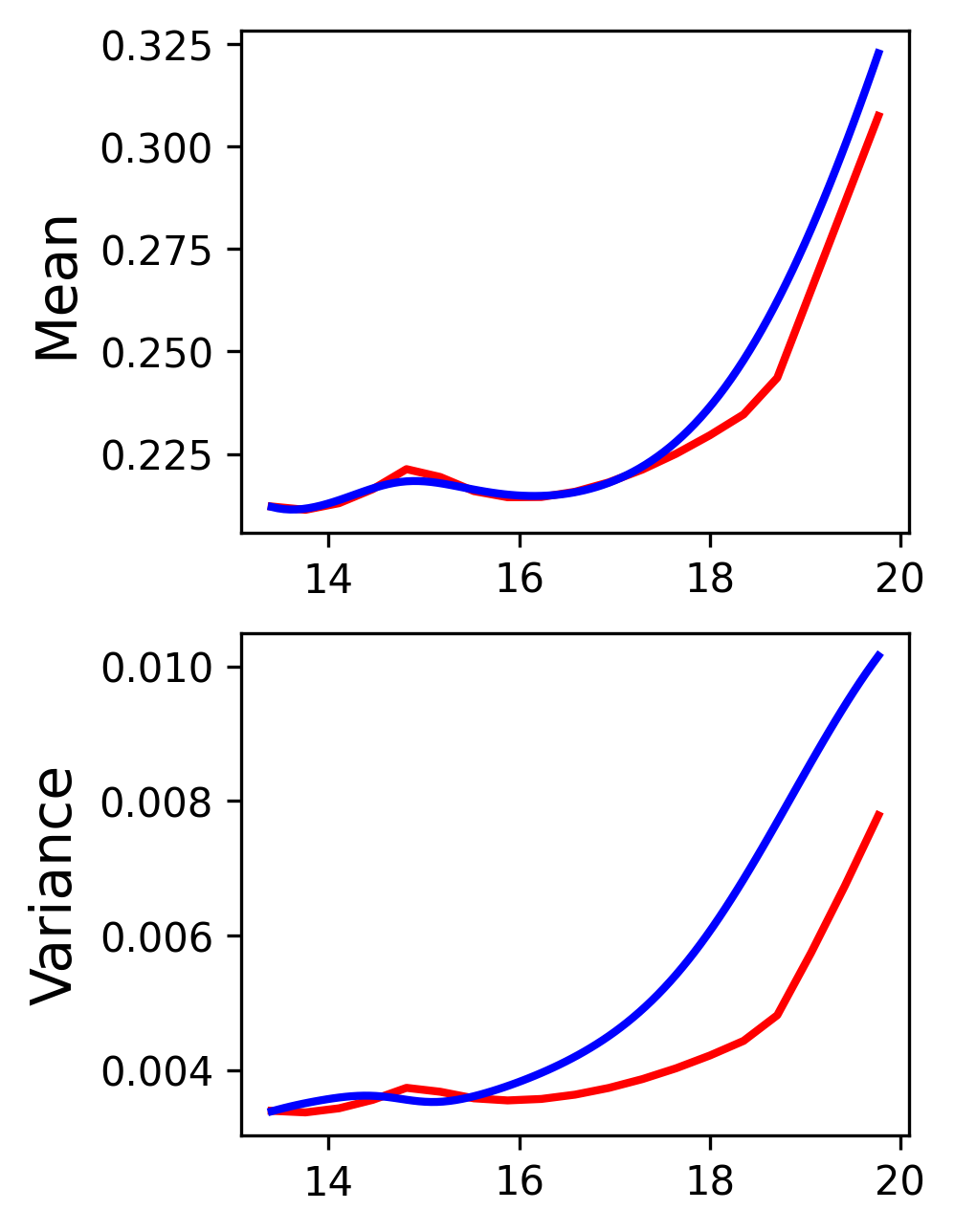}
\includegraphics[width=.32\linewidth]{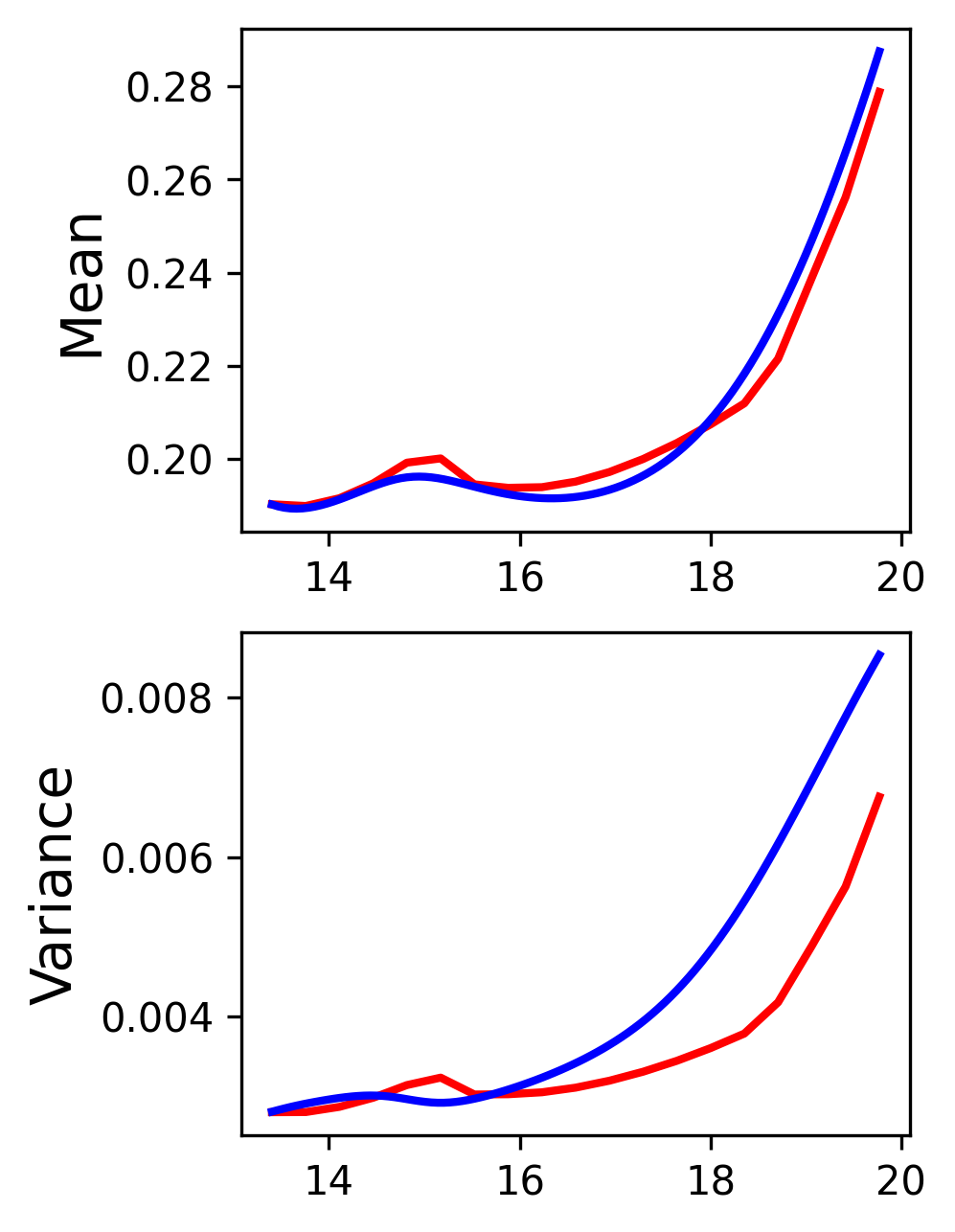}
         \caption{Comparison of the first two moments of the predicted PDFs for equivalent plastic strain rate on Datasets $\mathcal{D}^2$, $\mathcal{D}^3$, and $\mathcal{D}^4$.}
         \label{fig:3d:mom_plastic}
\end{figure}

\begin{figure}[H]
    \centering
             \includegraphics[width=1\linewidth]{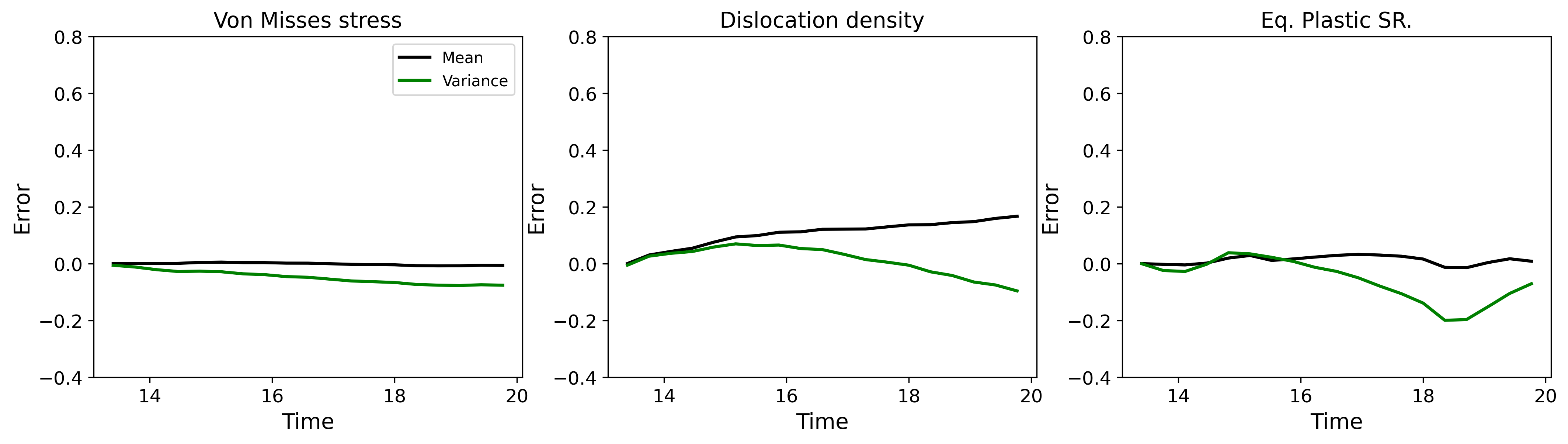}
               \includegraphics[width=1\linewidth]{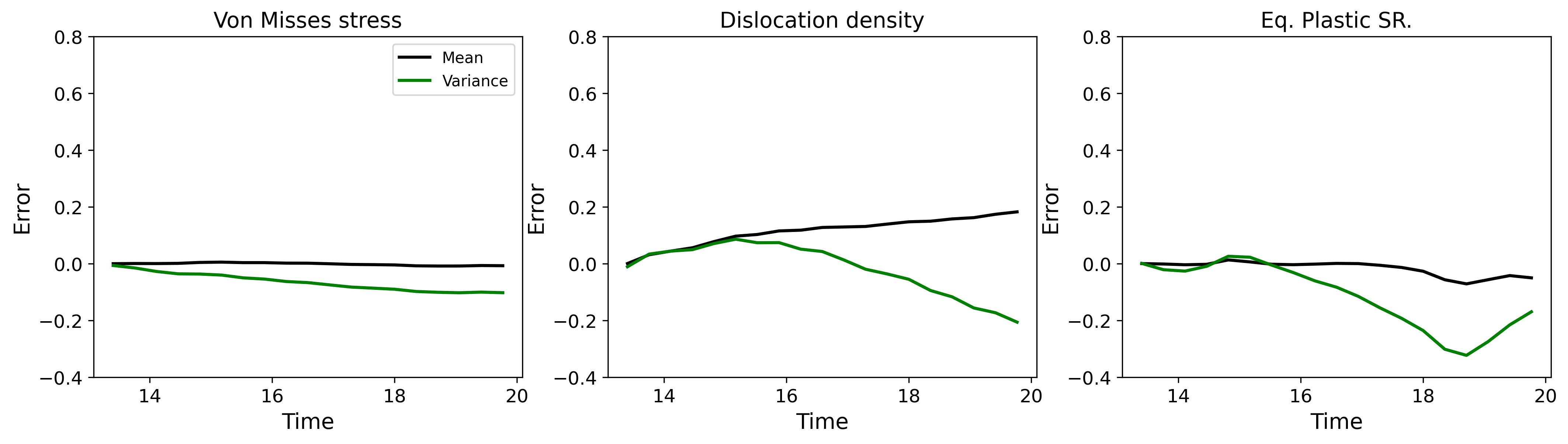}
                     \includegraphics[width=1\linewidth]{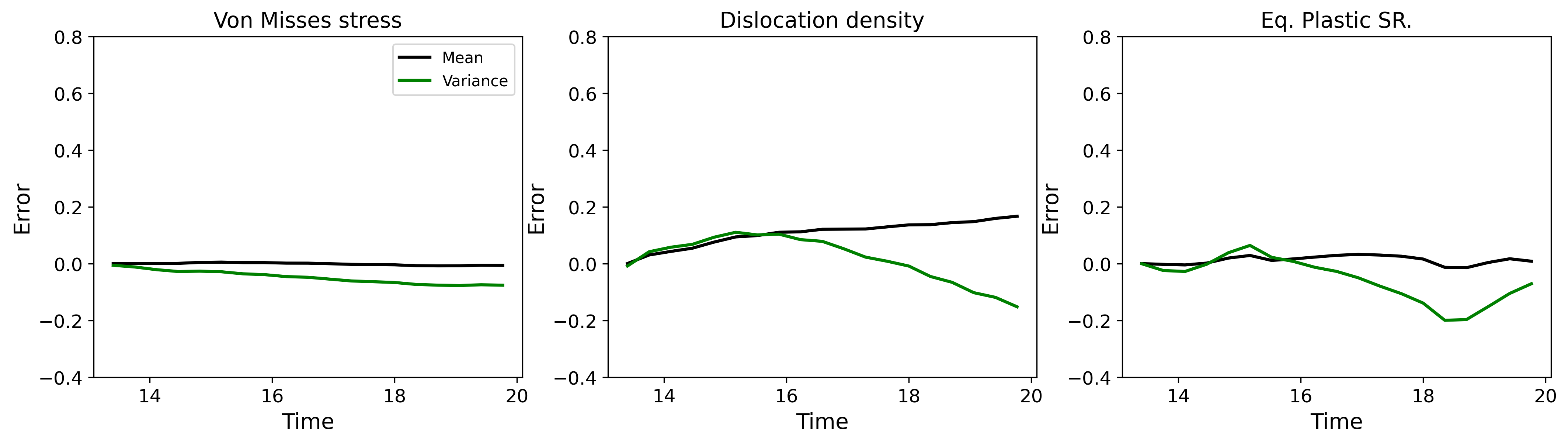}
               \caption{Relative errors on the moments of the predicted PDF compared to the reference ones. From top: Datasets $\mathcal{D}^2$, $\mathcal{D}^3$, and $\mathcal{D}^4$. Time is reported in milliseconds $\mu s$.}
               \label{fig:moment_err}
\end{figure}

To quantify the accuracy, we further analyze the first two statistical moments of the predicted PDFs across the three datasets and present the corresponding results in Figs.~\ref{fig:3d:mom_dis}--\ref{fig:3d:mom_plastic}. Each figure focuses on a different state variable, including von Mises stress, dislocation density, and equivalent plastic strain rate, and shows the temporal evolution of the moments alongside the corresponding reference results. The relative errors between the moments of the predicted and reference PDFs are also computed and presented in Fig.~\ref{fig:moment_err}. This comparative analysis demonstrates that the SPLIT-PINN framework successfully captures not only the overall shape of the marginal PDFs but also their key statistical signatures, such as mean and variance, across datasets with varying underlying microstructural conditions. These results, along with those discussed in \S\ref{subsec:results_jointPDF}, reinforce the robustness of SPLIT-PINN in learning the probabilistic model that generalizes effectively across unseen datasets, ensuring accurate predictions of both joint and marginal probabilistic features.
%%%%%%%%%%%%%%%%%
\begin{figure}[H]
    \centering
             \includegraphics[width=.32\linewidth]{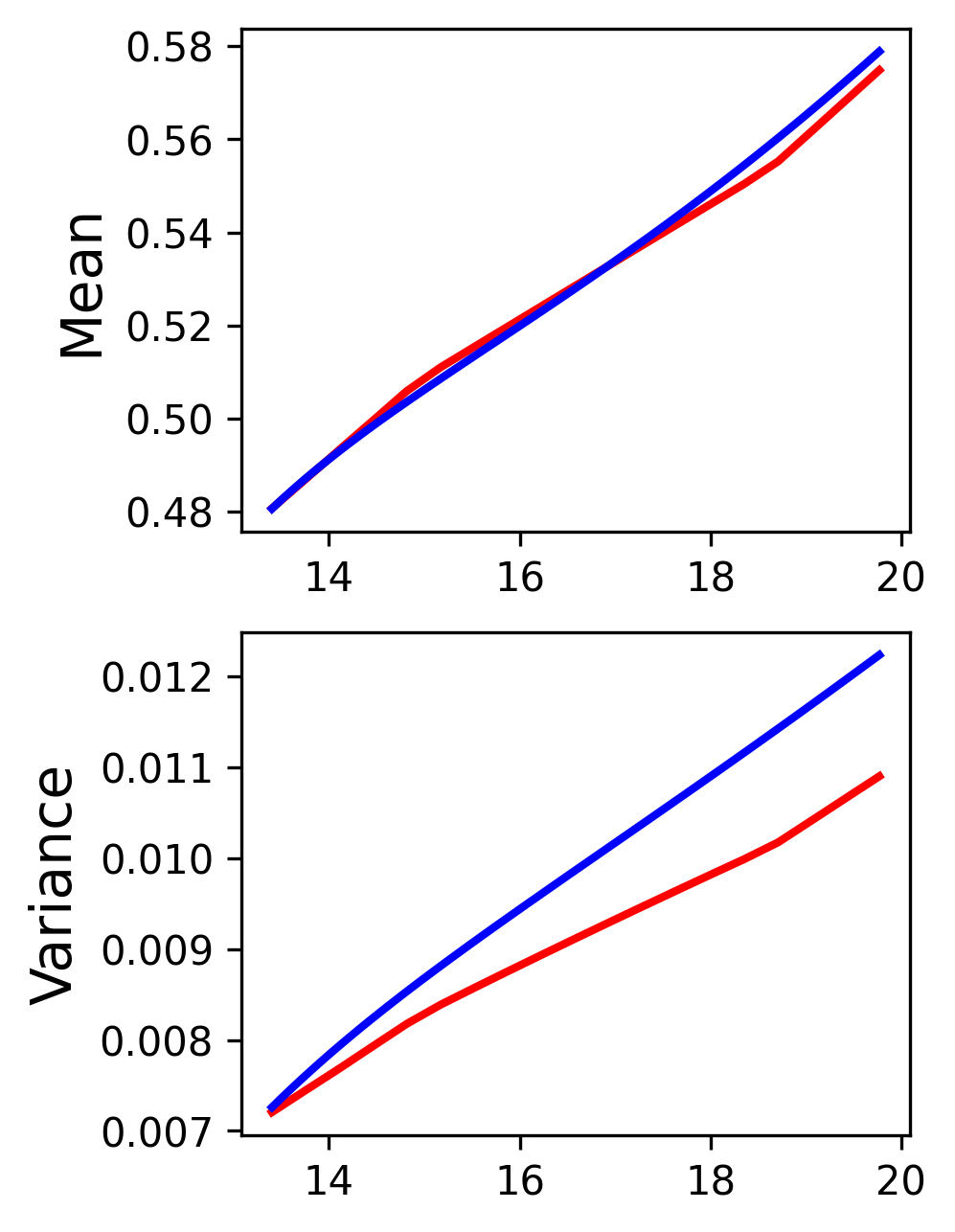}
               \includegraphics[width=.32\linewidth]{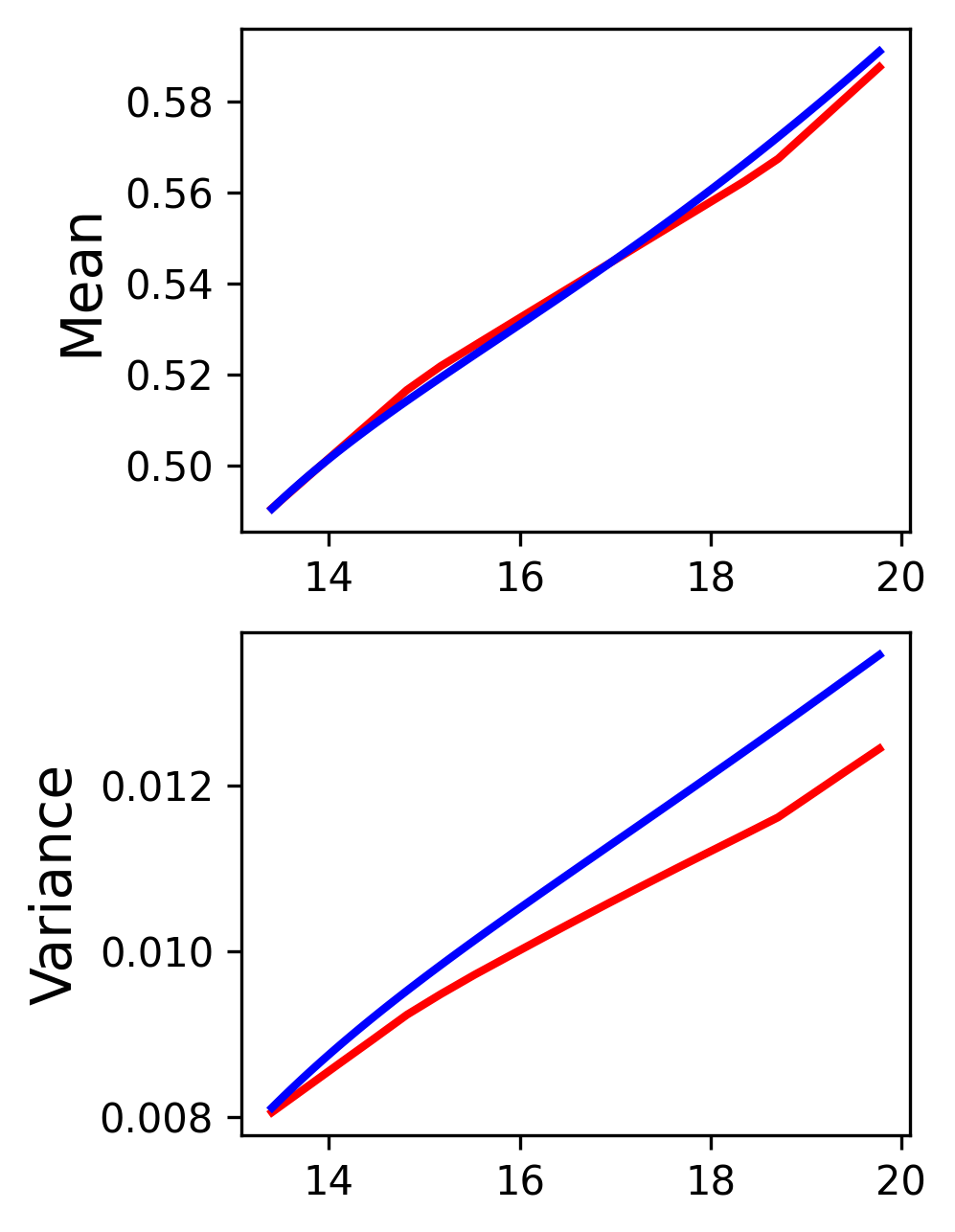}
         \includegraphics[width=.32\linewidth]{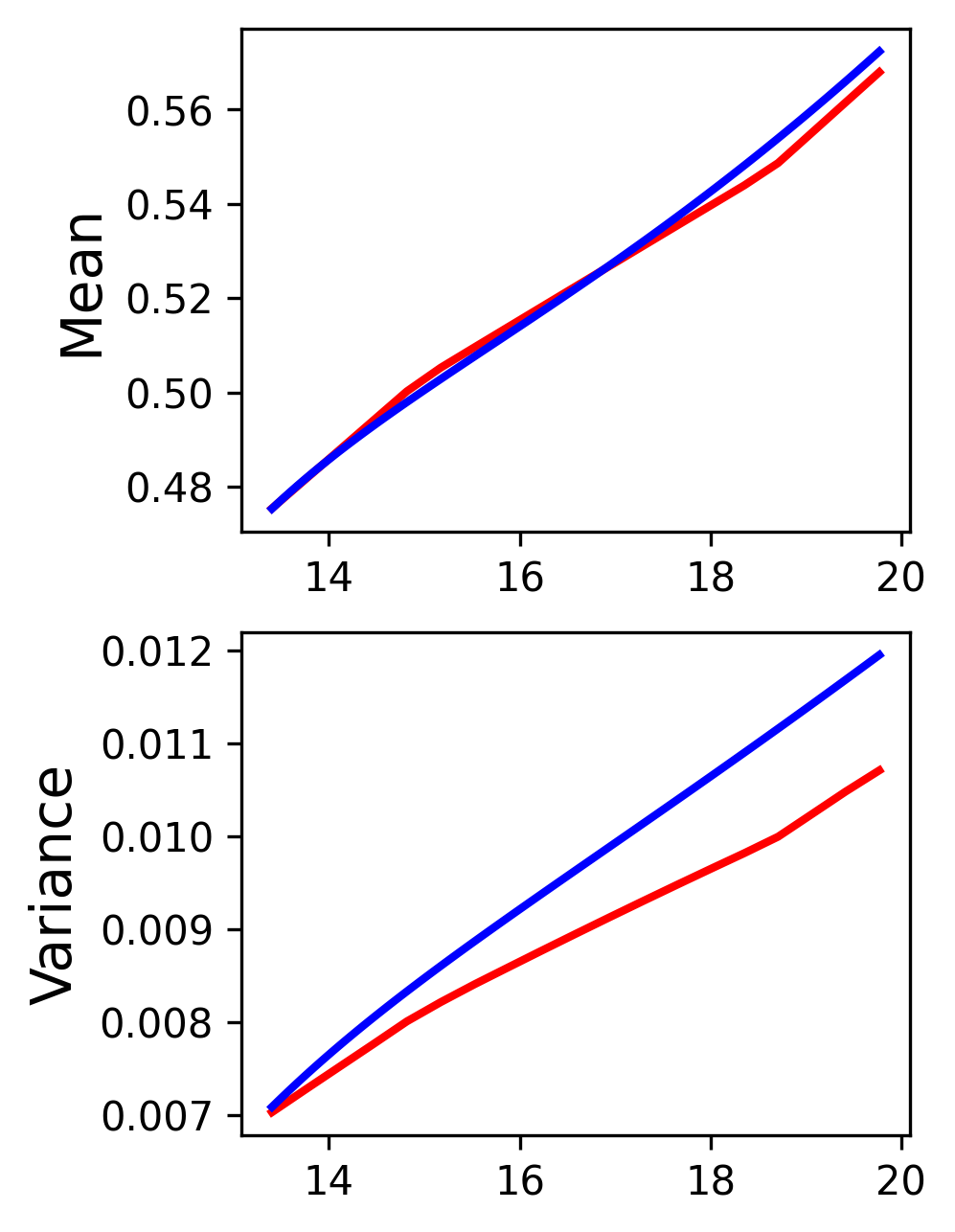}
         \caption{Comparison of the first two moments of the predicted PDFs for von Mises stress across the three datasets (from top: Datasets $\mathcal{D}^2$, $\mathcal{D}^3$, and $\mathcal{D}^4$). Each subplot illustrates the temporal evolution of the moments (reported in $\mu s$), highlighting the model’s ability to reproduce consistent statistical trends across various microstructural conditions.}
         \label{fig:3d:mom_stress}
\end{figure}
%%%%%%%%%%%%%%
%%%%%%%%%%%%%%%%%%%%%

\section{Conclusion}
\label{sec:conclude}

In this work, we have introduced a machine learning–based probabilistic modeling framework that enables the effective incorporation of small-scale spatial heterogeneity into large-scale descriptions of material behavior in polycrystalline metallic materials. By adopting PDFs to represent spatially heterogeneous material state fields, the proposed approach provides a principled statistical description of microstructural variability across different polycrystal realizations.

Central to this framework is the inverse identification of a probabilistic transport model, formulated as a Liouville equation with an unknown drift term. Using the SPLIT-PINN methodology, the drift field governing the temporal evolution of joint state PDFs is inferred directly from data. SPLIT-PINN is designed to address the challenges associated with high-dimensional, probabilistic, transport-dominated inverse problems, incorporating a marginal-correction drift decomposition, orthogonality constraints, and residual-based adaptive training, to stabilize learning and improve physical consistency. These features collectively enhance the well-posedness and robustness of drift reconstruction without imposing restrictive parametric assumptions.

The performance of SPLIT-PINN was systematically validated through a benchmark problem, enabling careful assessment of both inference accuracy and generalization capability at each stage of the framework. These studies highlight the severe ill-posedness of inverse drift identification when approached using standard PINN formulations and demonstrate that the hierarchical constraints embedded in SPLIT-PINN are essential for accurate and stable recovery of vector-valued drift fields.

Following benchmark validation, the framework was applied to real datasets describing the evolution of polycrystalline material states, including von Mises stress, dislocation density, and equivalent plastic strain rate. A Liouville model inferred from a single dataset was subsequently employed in forward simulations to predict the temporal evolution of joint and marginal PDFs across multiple unseen polycrystal statistical volume elements. Quantitative comparisons against ground-truth PDFs, using pointwise relative errors, KL divergence, and statistical moments, demonstrate that SPLIT-PINN learns a predictive probabilistic model that generalizes effectively across datasets and accurately captures both joint and marginal probabilistic features.

The observed error levels are consistent with, and in some cases favorable relative to, previously reported generalization behavior of PINNs for transport-dominated problems. In particular, while prior studies report relative errors of approximately 11–23\% for lower-dimensional inviscid Burgers-type equations, the present work addresses a three-dimensional Liouville equation without diffusion and requires generalization across distinct material realizations. Achieving maximum pointwise errors on the order of 20\% in this more challenging setting indicates that the learned drift field successfully captures the dominant transport dynamics underlying the evolution of high-dimensional state PDFs.

\rev{The scalability of the proposed framework to higher-dimensional joint PDFs remains an important consideration. While the SPLIT-PINN formulation mitigates the curse of dimensionality through its separable structure and marginal-correction strategy, increasing the number of state variables inevitably raises computational cost and data requirements, and may impact accuracy when higher-order dependencies are insufficiently sampled. In practice, the method is expected to remain tractable for moderately higher-dimensional systems, though this trade-off must be carefully managed. Moreover, for moderately high-dimensional systems (e.g., in the order of $O(10)$), solving the Liouville equation in the forward simulations can be more efficiently handled through Lagrangian models with Monte Carlo (LM-MC) (detailed formulations are provided in \ref{app:Lagrangian}), which can be further combined with variance reduction techniques to improve accuracy and reduce solution fluctuations (\cite{dimarco2014numerical}). Beyond this regime, expansion approaches such as BBGKY type approximation or ANOVA series expansion (\cite{cho2016numerical}), as well as machine learning methods (\cite{han2018solving,weinan2021algorithms}) can be leveraged. It should be noted, however, that for high-dimensional joint PDFs, the curse of dimensionality is unavoidable when the full distribution must be resolved. This is not merely a numerical limitation, but an informational one: a full joint PDF contains exponentially more information. In this regard, the Probability Density Evolution Method (PDEM)~(\cite{li2016probability}) has established arbitrary lower-dimensional density evolution equations, providing a viable alternative to the Liouville equation. 

Additionally, the present framework is based on a Liouville equation and therefore assumes deterministic system dynamics with uncertainty arising from initial conditions and/or system parameters. In the presence of stochastic external forcing, if the randomness can be represented through a finite expansion parameterized by a finite set of random variables, the formulation reduces to a direct extension of the Liouville equation, where the PDF becomes the joint distribution of the state variables and the random parameters that characterize the randomness coming from external forcing, as discussed by \cite{li2008principle}. If the stochastic forcing cannot be represented using a decomposition and truncation but assumed to be a process with idealized features (e.g., a Wiener process), a Fokker–Planck–type equation is required for probabilistic modeling. Extending the proposed inverse learning framework to generalized density evolution equations (\cite{li2008principle,li2016probability}) or to the identification of diffusion operators, constitutes an important direction for future work.}

Overall, this study establishes SPLIT-PINN as a systematic and robust framework for data-driven reconstruction of high-dimensional probabilistic transport models in materials physics. The proposed methodology offers a promising pathway for multiscale modeling of polycrystalline metallic materials, enabling predictive statistical descriptions of complex microstructural evolution beyond the reach of traditional deterministic approaches and unaffordable concurrent computational strategies.

\section*{Acknowledgment}

This work was funded by the U.S. Army Research Laboratory contract number W911NF2320073 within the University of Wisconsin - Madison, Center for Extreme Events in Structurally Evolving Materials.

\bibliographystyle{elsarticle-harv}

\bibliography{references}

% \printbibliography

\counterwithin{figure}{section}
\appendix
\section{Neural Network Architecture}\label{app:networks}

SPLIT-PINN employs three neural network components: the first data-driven network learns to predict the PDF from available computed PDFs; the second infers the one-dimensional drift term by incorporating a physics-informed loss based on the Liouville equation; and the third leverages these learned components to estimate the high-dimensional residual of the drift term.
    
	\subsection{Data-Driven Network for PDFs}

    We consider the joint PDF
$P(\boldsymbol{x},t)$, where $\boldsymbol{x} \in \mathbb{R}^d$ denotes the state vector.
The input domain $(\boldsymbol{x},t)$ is discretized on a space--time tensor grid.
Each spatial component $x_j \in [x_j^{\min}, x_j^{\max}]$, $j=1,\ldots,d$, is resolved by
$n_{x_j}$ points, and time $t \in [T_0,T_f]$ is sampled at discrete instances $\{t_k\}$.

The resulting training dataset is constructed as
\begin{equation}
X = \{(\boldsymbol{x}_i, t_k)\}_{i,k},
\qquad
Y = \{D(\boldsymbol{x}_i, t_k)\},
\end{equation}
where $D(\boldsymbol{x},t)$ denotes the measured or numerically evaluated
high-dimensional PDF. All target values are normalized
to lie in $[0,1]$.

The data-driven network is a fully connected feed-forward model
\[
(\boldsymbol{x},t) \longmapsto \widehat{P}(\boldsymbol{x},t),
\]
with the following architecture:
\begin{itemize}
\item Input layer of dimension $d+1$ (state variables and time).
\item Three hidden layers of width $100$ with $\tanh$ activation.
\item One hidden layer of width $100$ with $\mathrm{ReLU}$ activation.
\item Output layer of width $1$ with sigmoid activation to enforce positivity
and boundedness of the predicted PDF.
\end{itemize}

The dataset is randomly split into training ($90\%$) and validation ($10\%$) subsets.

	\subsection{Physics-Informed Network for the Drift Field $A(x,t)$ of marginal PDfs}
	
	To estimate 1D drift fields, we employ a 
	PINN with residual connections.
	The model receives the pair $(x,t)$ as input and predicts the drift value 
	$A(x,t)$. The architecture is as follows:
	\begin{itemize}
		\item Input layer of dimension $2$.
		\item One dense layer of width $50$ with $\tanh$ activation.
		\item A sequence of residual blocks, each defined by
		\[
		z \mapsto \sigma\!\big( W_2 \, \sigma(W_1 z + b_1) + b_2 + z \big),
		\qquad \sigma = \tanh,
		\]
		where $W$ and $b$ are weights and biases of the layer with output $z$. 
		We employ one initial block followed by three additional residual blocks,
		all with width $65$.
		\item Final linear output layer producing the drift field $A(x,t)$.
	\end{itemize}
	
	The use of residual connections improves gradient flow and stabilizes the 
	solution of the PDE-constrained optimization problem underlying PINN training~(\cite{xie2017aggregated}).

\subsection{Neural Network for High-Dimensional Residual Dynamics}

We construct a NN to learn a physics-informed model estimating the remaining high-dimensional residual term accounting for dependencies that are not captured by the of marginal drift contributions alone. Similar to the PINN’s structure in step 2, we use a 
neural network with a residual connection. The architecture of the residual network consists of:
\begin{itemize}
\item An input layer of dimension $d+1$.
\item one initial block followed by three additional residual blocks,
		all with width $100$ neurons and $\tanh$ activation function.

\item An output layer of dimension $d$, producing the residual drift components.

\end{itemize}

\section{Sequential learning vs. joint learning}
%\section{Joint network vs. single networks}
\label{app:joint}
We present a numerical experiment to demonstrate the importance of separating the learning of $P$ and $\mathbf{A}$, i.e., the proposed sequential learning approach. To this end, we compare it with the joint-learning approach, where a single neural network is trained to simultaneously approximate both $P$ and $\mathbf{A}$. In this example, the spatial dimension is $n_d=1$, and hence $\mathbf{A}$ can be reduced to a scalar operator $A$. 

In the joint-learning approach, the network takes $(x,t)$ as the input and outputs $(P,\,A)$, and is trained using a composite loss function that combines the data loss (based on the available observations of $P$) in Eq. \eqref{eq:data_loss}, positivity and conservation constraints on $P$, and a physics-informed loss derived from the Liouville equation. In contrast, the proposed sequential-learning approach decouples these learning tasks. First, a data-driven neural network is trained to approximate $P$ continuously using only observations of $P$ and its associated constraints. Next, a second physics-informed neural network is trained to infer $A$ using automatic differentiation on the trained first neural network of $P$ and the physics-informed loss derived from the Liouville equation.

\begin{figure}[H]
    \centering
    \includegraphics[width=1.05\linewidth]{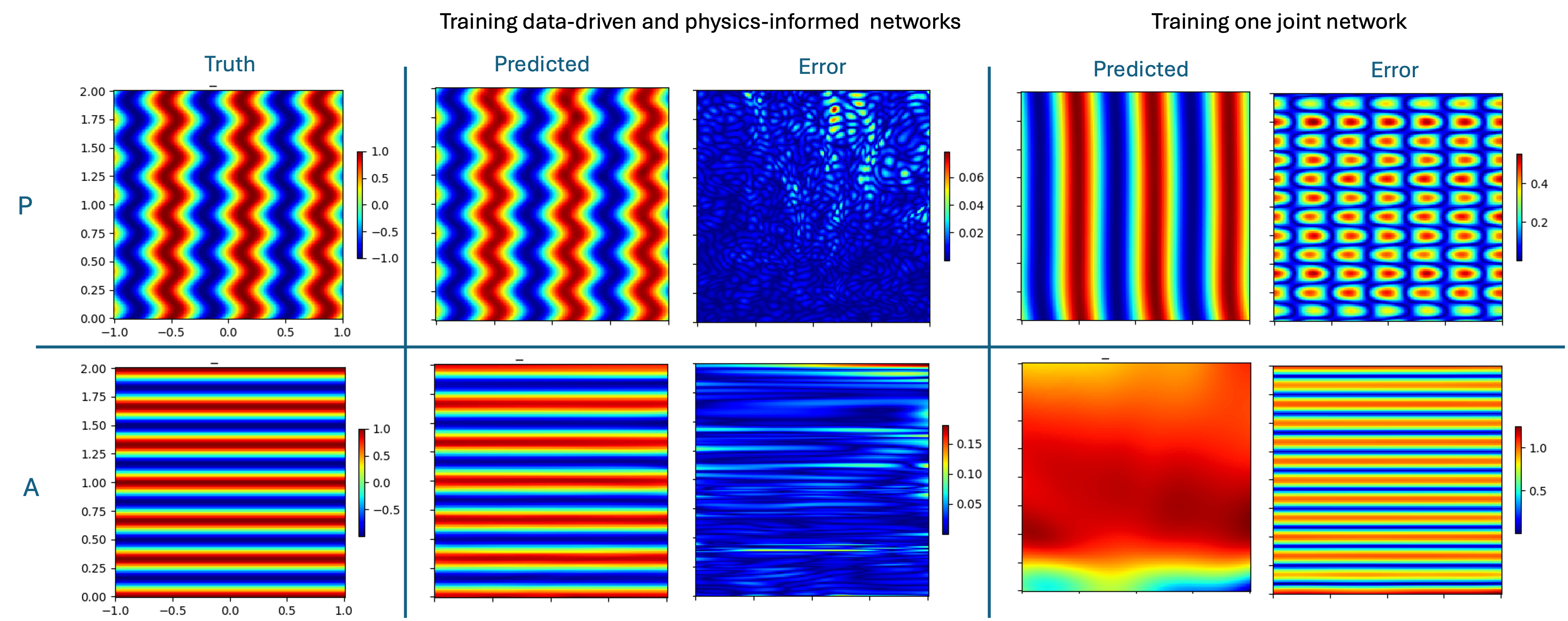}
    \caption{Comparison between the proposed sequential-learning approach (where $P$ is first learned in a data-driven manner and then used to train a physics-informed network for $A$) and the joint-learning approach (which learns $P$ and $A$ simultaneously).}
\label{fig:joint_seq}
\end{figure}

The numerical example considered has the following analytical solution: 
\begin{equation}
        A(t) = \cos(mt),\quad   P(x,t) = \sin\!\left( k \left( x - \frac{1}{m} \sin(mt) \right) \right)\;,
\end{equation}
with $m=6\pi$ and k = $3\pi$ over the computational domain defined by $\Omega = [-1,1]$ and $T= [0,2]$. We employ a similar network structure, collocation points, activation function, and learning rate for both joint-learning and sequential-learning approaches. The results for both are shown in Figure~\ref{fig:joint_seq}. The joint-learning approach struggled to resolve the high-frequency behavior of $P$. As a result, errors accumulated in the estimation of $A$, leading to a poor overall performance. In contrast, the proposed sequential-learning approach significantly improves the prediction accuracy for both $P$ and~$A$. 
%%%%%%%%%%%%%%%%%%

\section{Ensemble-size convergence study for the benchmark problem}\label{app:benchmark}
\begin{figure}[H]
    \centering
    \includegraphics[width=1\linewidth]{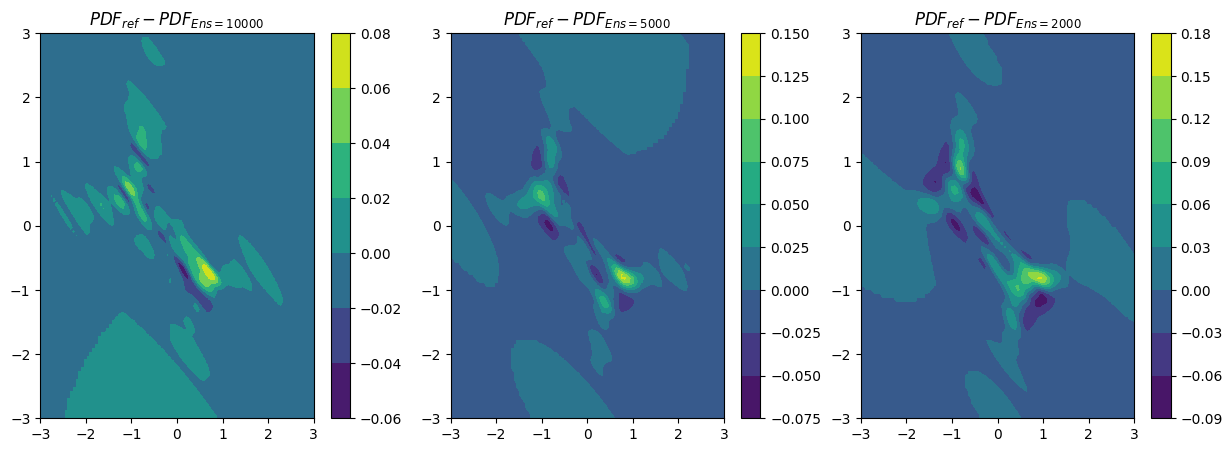}
    \caption{Pointwise error of the PDF estimates obtained with varying ensemble sizes, measured against the reference one computed using 15,000 ensembles.}
    \label{fig:ensemle_comp}
\end{figure}

\begin{figure}[H]
    \centering
     \includegraphics[width=.5\linewidth]{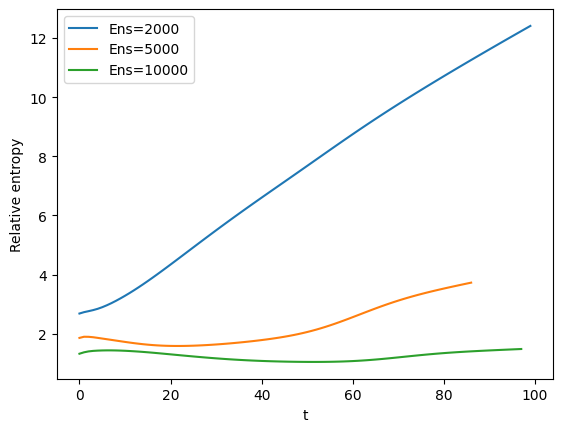}
    \caption{Relative entropy between PDF estimates computed with varying ensemble sizes and the reference PDF obtained using 15,000 ensembles.}
    \label{fig:ensemle_RE}
\end{figure}

In this section, we present a convergence study to determine the ensemble size required to accurately approximate the PDF examined in \S\ref{sec:bench}. The PDF obtained from 15,000 ensembles is used as the reference and compared with those computed from 2,000, 5,000, and 10,000 ensembles. The differences among these PDF estimates are quantified using pointwise error and relative entropy, as illustrated in Figs.~\ref{fig:ensemle_comp} and \ref{fig:ensemle_RE}. The relative entropy is calculated as 
\begin{equation}
\mathcal{D}_{\mathrm{KL}}\!\left(P_{\mathrm{ref}} \,\|\, P_{\mathrm{Ens}}\right)
=
\sum_{i=1}^{N}
P_{\mathrm{ref}}^{(i)}
\log\!\left(
\frac{P_{\mathrm{ref}}^{(i)}}{P_{\mathrm{Ens}}^{(i)}}
\right)\;,
\end{equation}
where $P_{\mathrm{Ens}}$ is the PDF calculated using sampling from the initial distribution, and $N$ denotes the number of discrete spatial points. 
These results indicate that the estimated PDFs converge with increasing ensemble size, and that 15,000 ensembles are sufficient to produce an accurate PDF.
%%%%%%%%%%%%%%%%%%

\section{\rev{Effect of Collocation Point Sampling and Orthogonality Constraint on SPLIT-PINN Performance}}
\label{sec:collocation}

\begin{figure}[H]
    \centering
     \includegraphics[width=\linewidth]{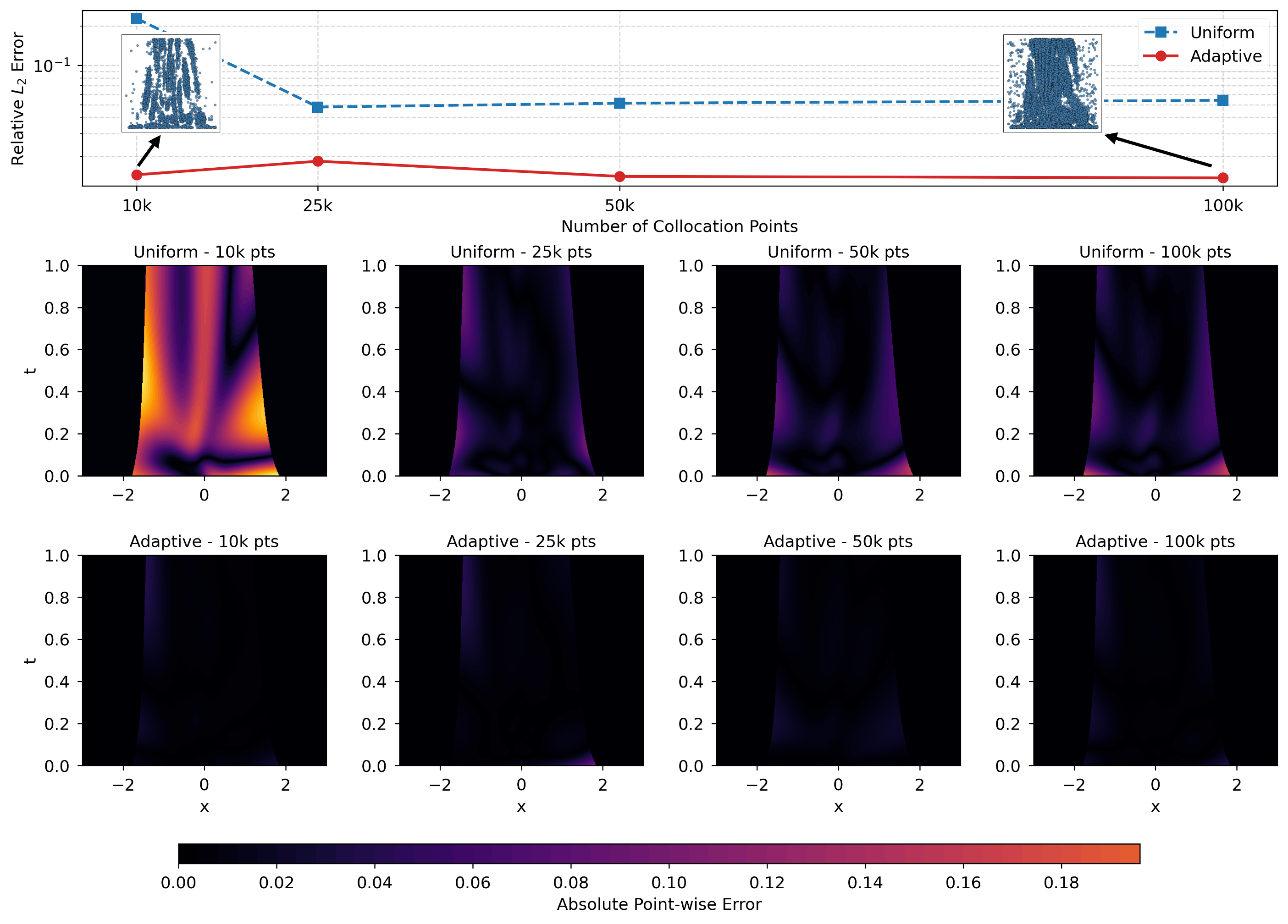}
    \caption{\rev{Sensitivity of the inferred drift term to the number of collocation points and sampling strategy. Top: Relative $L_2$ error of $A_x$ as a function of the number of collocation points for uniform random and adaptive sampling; insets show representative point distributions at 10k and 100k points for adaptive sampling. Bottom: Absolute point-wise errors in a slice of $A_x$ over the $(x,t)$ domain at $y=0$, for uniform and adaptive sampling, respectively, with varying numbers of collocation points. Adaptive sampling achieves consistently lower errors with fewer points by concentrating samples in regions of high probability density.}}
    \label{fig:conv}
\end{figure}

\rev{To assess the sensitivity of our SPLIT-PINN framework to the number of collocation points used in the learning of $R_i(\mathbf{x},t)$, we conduct a systematic study of the benchmark problem presented in \S\ref{sec:bench}, varying the number of collocation points across four levels (10k, 25k, 50k, and 100k), and compare two sampling strategies: uniform random sampling and the adaptive sampling scheme employed in this work. The results are summarized in Fig.~\ref{fig:conv}, where the top panel reports the relative $L_2$ error in the inferred drift term $A_x$ as a function of the number of collocation points for both sampling strategies, and the two rows in the bottom panel show the corresponding absolute point-wise error in one slice of $A_x$ over the $(x,t)$ domain at $y=0$. 

For both sampling strategies, the errors are largely insensitive to the number of collocation points, especially in the case of adaptive sampling. With uniform random sampling, the relative $L_2$ error is large with 10k points and decreases as more collocation points are added; however, even with 100k points, it remains larger than that obtained with adaptive sampling, due to the sharp interfaces in the drift field. This performance results from the ability of the adaptive scheme to concentrate collocation points in the regions of high probability density, precisely where the prediction of $R_i(\mathbf{x},t)$ has the greatest impact on the final joint PDF. These results demonstrate that the SPLIT-PINN framework is robust to the number of collocation points once a sufficient budget is reached, and the adaptive sampling strategy is more efficient and accurate than uniform random sampling, achieving convergence with far fewer points.}
%%%%%%%%%%%%%%%%%%%%%%%

\revv{
   Furthermore, in our SPLIT-PINN framework, the macroscopic marginal drifts ($\bar{A}_i$) are first learned to capture 1D, particularly, over the sharp interfaces between zero and non-zero marginal PDF values. Next, the model learns the interaction term ($R_i$) to capture the entire joint PDF dynamics via the Liouville equation. We evaluate the impact of orthogonality penalization (in Eq. \eqref{eq:loss_corr}) enforced in Step 3 of SPLIT-PINN (see Fig. \ref{fig:over_nd}) on the same problem. Specifically, we extract 1D cross-sectional profiles of the total drift components and mask it over the regions with non-zero PDF values, given that the PDE residual is automatically zero (or very small) outside of these regions. These regions are usually characterized by sharp spatial transitions. As illustrated below in Fig. \ref{fig:ortho_ablation}, training $R_i$ without the orthogonality constraint (blue line) fundamentally corrupts the macroscopic dynamics. Because the unconstrained $R_i$ network is driven purely by the joint PDE loss, it attempts to capture complex, high-dimensional dynamics. However, due to the inherent spectral bias of neural networks~(\cite{rahaman2019spectral}), it struggles with localized sharp gradients. Consequently, $R_i$ learns spurious, smooth, non-zero mean biases that unphysically overwrite the sharp macroscopic interfaces already resolved by the pre-learned $\bar{A}_i$. Conversely, enforcing the orthogonality constraint (Eq. \eqref{eq:ri_constraint}) during the third-step PINN training (red line) acts as a strict mathematical guardrail. }
	\begin{figure}[H]
		\centering
		 \includegraphics[width=0.8\textwidth]{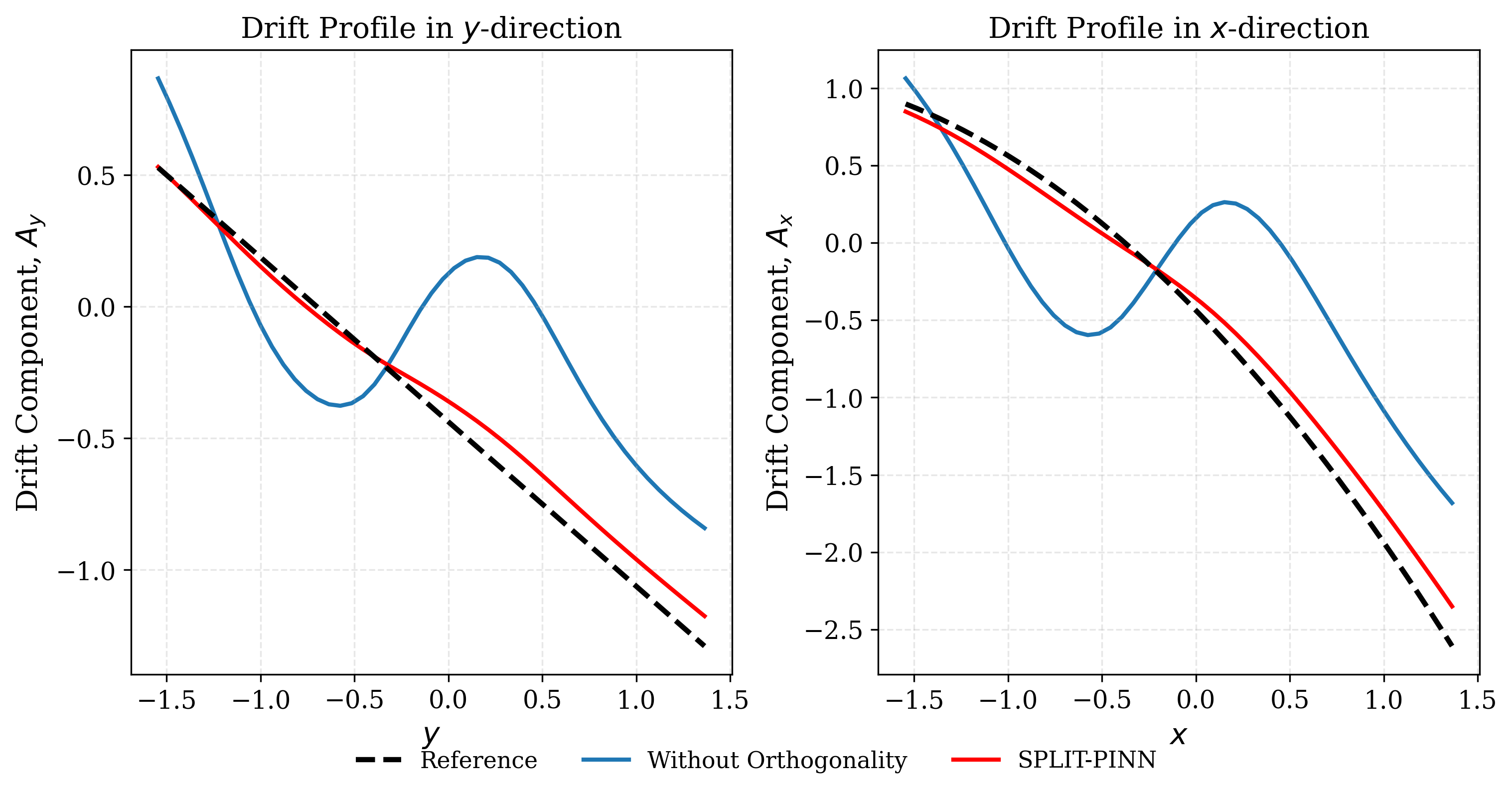}
		\caption{\revv{Effect of the orthogonality constraint on the inference of drift components for the benchmark problem in Section 4. 1D cross-sectional profiles of the drift in the $y$-direction (left) and $x$-direction (right) are shown. The model without the orthogonality constraint suffers from gradient interference and spectral bias, resulting in nonphysical smearing of sharp interfaces.}}
		\label{fig:ortho_ablation}
	\end{figure}
%%%%%%%%%%%%%%%%%

\section{Numerical solver for the Liouville equation}
\label{app:discretization}

With the drift term $\mathbf{A}$ inferred by SPLIT-PINN, the Liouville equation is fully determined, enabling the prediction of the temporal evolution of the state variables' joint PDF. Given the initial PDF for a new dataset, the time evolution of the joint PDF is computed by numerically solving the Liouville equation. In this section, we provide a detailed description of the numerical solver employed for this purpose.

\subsection{Spatial discretization: upwind flux}

Upwind discretization is essential for stabilizing hyperbolic transport equations, as it prevents nonphysical oscillations near steep gradients and ensures that information propagates along the characteristic directions (\cite{leveque2002finite}). 
%This approach has long been a cornerstone of numerical advection schemes; see, for example, \cite{leveque2002finite}. 
Thus, we discretize the flux term using an upwind scheme determined by the sign of the 
drift term $A$:
\[
F_{i+\frac{1}{2}} =
\begin{cases}
	A_{i+\frac{1}{2}} P_i, & A_{i+\frac{1}{2}} > 0, \\
	A_{i+\frac{1}{2}} P_{i+1}, & A_{i+\frac{1}{2}} < 0,
\end{cases}
\]
where $i$ is the space discretization index and $\cdot_{i+\frac{1}{2}} =(\cdot_{i}+\cdot_{i+1})/2 $; $F_{i+\frac{1}{2}}$ denotes the numerical flux at the discrete element interface; and $A_{i+\frac{1}{2}}$ denotes the drift term. The semi-discrete form of the Liouville equation then becomes:
\begin{equation}
    \left(\frac{dP}{dt}\right)_i = -\frac{F_{i+1/2} - F_{i-1/2}}{\Delta x} \;.
    \label{eq:semi-discrete}
\end{equation}

\subsection{Time integration: Runge--Kutta~4}
The semi-discrete form in Eq. \eqref{eq:semi-discrete} is solved using the classical fourth-order Runge-Kutta (RK4) temporal integration scheme~(\cite{butcher2007runge}):
\[
\begin{aligned}
	k_1 &= f(P^n), \\
	k_2 &= f\!\left(P^n + \tfrac{1}{2} \Delta t k_1\right), \\
	k_3 &= f\!\left(P^n + \tfrac{1}{2} \Delta t k_2\right), \\
	k_4 &= f\!\left(P^n + \Delta t k_3\right), \\
	P^{n+1} &= P^n + \frac{\Delta t}{6} (k_1 + 2k_2 + 2k_3 + k_4),
\end{aligned}
\]
where $f(P)$ denotes the spatial flux operator defined above.

\subsection{Operator splitting: Strang method}

To efficiently handle multi-dimensional PDFs, we employ the Strang operator splitting
(\cite{strang1968construction}), which decomposes the multi-dimensional advection
equation into a sequence of one-dimensional subproblems. For a three-dimensional
system, the probability density function \(P(x,y,z,t)\) satisfies
\[
\frac{\partial P}{\partial t}
+ \frac{\partial}{\partial x}(A_x P)
+ \frac{\partial}{\partial y}(A_y P)
+ \frac{\partial}{\partial z}(A_z P) = 0 .
\]

Instead of solving this equation in three dimensions simultaneously, the solution
is advanced over one time step \(\Delta t\) by sequentially evolving the PDF along
each spatial direction. One full time step consists of the following sub-steps:
\begin{enumerate}
	\item solve
	\[
	\frac{\partial P}{\partial t}
	+ \frac{\partial}{\partial x}(A_x P) = 0
	\]
	for a duration \(\Delta t/2\),
	\item solve
	\[
	\frac{\partial P}{\partial t}
	+ \frac{\partial}{\partial y}(A_y P) = 0
	\]
	for a duration \(\Delta t/2\),
	\item solve
	\[
	\frac{\partial P}{\partial t}
	+ \frac{\partial}{\partial z}(A_z P) = 0
	\]
	for a duration \(\Delta t\),
	\item solve
	\[
	\frac{\partial P}{\partial t}
	+ \frac{\partial}{\partial y}(A_y P) = 0
	\]
	for a duration \(\Delta t/2\),
	\item solve
	\[
	\frac{\partial P}{\partial t}
	+ \frac{\partial}{\partial x}(A_x P) = 0
	\]
	for a duration \(\Delta t/2\).
\end{enumerate}

This symmetric ordering is the defining feature of Strang splitting and ensures
second-order accuracy in time for the full three-dimensional advection problem,
provided each one-dimensional subproblem is solved with sufficient accuracy.
The method reduces the multi-dimensional PDF evolution to a sequence of
one-dimensional advection solves, enabling dimension-by-dimension integration
while preserving the advective structure of the Liouville equation and the total
probability up to the accuracy of the underlying numerical scheme.

\revv{The numerical stability of the solver is governed by a Courant--Friedrichs--Lewy (CFL) condition associated with the upwind discretization. Strang operator splitting reduces each substep to a one-dimensional advection problem~(\cite{strang1968construction}), leading to a directional CFL number of the form~(\cite{hairer1993solving}):
\[
\text{CFL}_d = \frac{|A_d| \, \Delta t}{\Delta x}, \quad d \in \{x,y,z\},
\]
where $d= x,\,y,\,z$. We use time-step size $\Delta t = 0.04$ and uniform space discretization $\Delta x = 0.01$ in each direction, yielding to
\[
\text{CFL}_d = 4 |A_d|.
\]
In the simulations considered, the maximum magnitude of the drift components satisfies $\max |A_d| \approx 0.2$ corresponding to the disclocation density variable, resulting in $\text{CFL}_d \approx 0.8 < 1$, which ensures stability of the upwind scheme. The use of RK4 time integration does not significantly alter this stability bound, and the overall scheme remains stable under the chosen discretization.}

\rev{
\section{Liouville Equation in Lagrangian Point of View}\label{app:Lagrangian}
Instead of evaluating $P(\cdot, t)$ at a fixed Eulerian point, if we follow the trajectory of a particle $\mathbf{x}(t)$ (Lagrangian point of view), we obtain:
\begin{equation}
\begin{aligned}
    \frac{d P(\mathbf{x}(t), t)}{d t} &= \frac{\partial P(\mathbf{x}(t), t)}{\partial \mathbf{x}(t)}\frac{d \mathbf{x}(t)}{d t} + \frac{\partial P(\mathbf{x}(t), t)}{\partial t} \\
&= \cancel{\sum_{i=1}^{n_d}\frac{\partial P(\mathbf{x}(t), t)}{\partial x_i(t)}\frac{d x_i(t)}{d t}}
- \sum_{i=1}^{n_d} \frac{\partial A_i(\mathbf{x}(t), t)}{\partial x_i} P(\mathbf{x}(t), t)
- \cancel{\sum_{i=1}^{n_d} A_i(\mathbf{x}(t), t) \frac{\partial P(\mathbf{x}(t), t)}{\partial x_i(t)}} \\
&= - \sum_{i=1}^{n_d} \frac{\partial A_i(\mathbf{x}(t), t)}{\partial x_i} P(\mathbf{x}(t), t)
\end{aligned}
\label{eq:Lagrangian1}
\end{equation}
By taking $\log$ on $P$ and substituting Eq. \eqref{eq:Lagrangian1}, we arrive:
\begin{equation}\label{eq:Lagrangian2}
\frac{d \log P(\mathbf{x}(t), t)}{d t} = \frac{1}{P(\mathbf{x}(t), t)} \frac{d P(\mathbf{x}(t), t)}{d t} = - \sum_{i=1}^{n_d} \frac{\partial A_i(\mathbf{x}(t), t)}{\partial x_i}
\end{equation}
\eqref{eq:Lagrangian2} can be combined with the dynamical equation of $\mathbf{x}(t)$ to form a set of ODEs of size $n_d+1$, as:
\begin{equation}
\frac{d}{dt} \begin{bmatrix}
\mathbf{x}(t) \\ \log P(\mathbf{x}(t), t)
\end{bmatrix} = \begin{bmatrix}
\mathbf{A}(\mathbf{x}(t), t) \\ -\sum_{i=1}^{n_d} \frac{\partial A_i(\mathbf{x}(t), t)}{\partial x_i}
\end{bmatrix}
\end{equation}
Note that, when solving this system of ODEs, the right-hand side of Eq. \eqref{eq:Lagrangian2} can be efficiently evaluated via automatic differentiation, since $A_i(\mathbf{x}(t), t)$ is represented by a neural network (see Fig. \ref{fig:over_nd}); therefore, no numerical approximation (e.g., finite difference) is required herein.

Thus, by leveraging Lagrangian point of view, we can utilize the trajectories of $\mathbf{x}(t)$ to effectively sample the $n_d$-dimensional state space and predict the probability density $P(\mathbf{x}(t), t)$.
}

\end{document}